\documentclass[11pt,a4paper]{article}
\pdfoutput=1
\usepackage[pdftex]{graphics}
\usepackage{jheppub}
\usepackage{amsmath,amssymb,amsfonts,mathtools}
\usepackage{enumitem}
\usepackage{multirow}
\usepackage{array,booktabs}
\usepackage{mathrsfs,bbm}

\usepackage[dvipsnames]{xcolor}
\definecolor{labelcolor}{RGB}{194, 175, 116}

\definecolor{skyblue}{RGB}{34,139,230}
\definecolor{indigo}{RGB}{31,31,180}

\definecolor{schutz}{RGB}{65,113,251}
\definecolor{bosoniclight}{RGB}{183,171,165}

\usepackage{bold-extra}

\def\enter{\medskip}

\setlist[itemize]{
    label=\adjustbox{scale=0.7}{$\bullet$}, itemsep=-3pt,topsep=0px
}

\setlist[enumerate]{
	label=(\alph*),      
	itemsep=0em,         
	topsep=.3em,           
}

\hypersetup{
    citecolor=schutz,
    linkcolor=schutz,
    urlcolor=indigo
}

\usepackage[outline]{contour}

\usepackage[final]{showlabels} 

\usepackage{hyphenat}
\let\oldeqref\eqref
\renewcommand{\eqref}[1]{Eq.\,\smash{\oldeqref{#1}}}
\newcommand{\eqrefs}[2]{Eqs.\,\smash{\oldeqref{#1}} and \smash{\oldeqref{#2}}}
\newcommand{\eqrefss}[3]{Eqs.\,\smash{\oldeqref{#1}}, \smash{\oldeqref{#2}}, and \smash{\oldeqref{#3}}}

\newcommand{\rcite}[1]{Ref.\,\cite{#1}}
\newcommand{\rrcite}[1]{Refs.\,\cite{#1}}

\newcommand{\fref}[1]{Fig.\,\ref{#1}}

\newcommand{\Sec}[1]{Section\:\ref{#1}}
\newcommand{\Secs}[2]{Sections\:\ref{#1} and \ref{#2}}
\newcommand{\Secss}[3]{Secs.\,\ref{#1}, \ref{#2}, and \ref{#3}}
\newcommand{\App}[1]{Appendix\:\ref{#1}}

\usepackage{tikz}
\usetikzlibrary{calc} 
\usetikzlibrary{shapes.geometric} 
\usetikzlibrary{positioning} 
\usetikzlibrary{fit} 
\usepackage[a]{esvect} 
\tikzset{empty/.style = {inner sep = 0pt, outer sep = 0, minimum size = 0}}
\tikzset{b/.style = {inner sep = 2pt, outer sep = 4pt, minimum size = 12pt}}
\tikzset{c/.style = {inner sep = 2pt, outer sep = 4pt, minimum size = 12pt}}
\tikzset{w/.style = {inner sep = 1pt, outer sep = 2pt, minimum size = 12pt, anchor = west}}
\tikzset{lin/.style = {draw, line width = 0.5pt}}
\usepackage[export]{adjustbox}

\tikzset{dot/.style = {circle, draw=black, fill=black, inner sep=0pt, outer sep=0pt, minimum size=3pt, line width=1.2pt}}

\definecolor{lgray}{RGB}{150,150,150}
\tikzset{
	dprop/.style = {
		draw, line width=0.8pt,
		dotted, 
		line cap=round,
		dash pattern=on 0pt off 2.53pt,
		color=lgray
	}
}

\tikzset{l/.style = {draw, line width = 1.2pt}}
\tikzset{s/.style = {inner sep = 2.5pt, outer sep =2.5pt, minimum size = 1pt, font = \small}}

\let\oldc\c
\let\oldi\i

\def\mem{\hspace{0.1em}}
\def\hem{\hspace{0.05em}}
\def\nem{\hspace{-0.1em}}
\def\hnem{\hspace{-0.05em}}
\def\hhem{\hspace{0.025em}}
\def\hhnem{\hspace{-0.025em}}
\def\hhhem{\hspace{0.0125em}}

\def\blank{{\,\,\,\,\,}}

\def\qiq{{\quad\implies\quad}}
\def\qfq{{\quad\iff\quad}}

\newcommand{\bigbig}[1]{\big(\mem{#1}\mem\big)}
\newcommand{\BB}[1]{\Big(\,{#1}\,\Big)}
\newcommand{\bb}[1]{\bigg(\,{#1}\,\bigg)}

\newcommand{\lrp}[1]{\left(\mem{#1}\mem\right)}

\DeclareMathOperator{\diag}{diag}

\DeclareMathOperator{\Sym}{Sym}

\def\id{\mathrm{id}}

\def\a{\alpha}
\def\b{\beta}
\def\c{\gamma}
\def\d{\delta}
\def\e{\epsilon}

\def\m{\mu}
\def\n{\nu}
\def\r{\rho}
\def\s{\sigma}
\def\k{\kappa}

\def\t{\tau}

\def\vth{\vartheta}

\def\bpsi{{\smash{\bar{\psi}}\kern0.02em}\vphantom{\psi}}

\def\bchi{{\bar{\chi}}}

\def\bvth{{\bar{\vartheta}}}

\def\tgamma{\tilde{\gamma}}   

\def\bQ{{\bar{Q}}}

\def\tD{\bar{D}}
\def\tT{\bar{T}}
\def\tR{\bar{R}}
\def\tgamma{\bar{\gamma}}

\def\O{{\mathcal{O}}}

\def\mplus{{\mem+\mem}}
\def\mminus{{\mem-\mem}}

\def\mdot{{\mem\cdot\mem}}

\def\swedge{{\mem{\wedge}\,}}

\def\modot{{\mem\odot\mem}}
\def\moplus{{\,\oplus\,}}
\def\mtensor{{\,\otimes\,}}

\def\mlra{{\mem\leftrightarrow\mem}}

\def\inn{{\:\in\:}}
\def\eqq{{\:=\:}}
\def\too{{\:\to\:}}

\newcommand{\wrap}[1]{{\smash{#1}\vphantom{\b}}}
\newcommand{\Wrap}[1]{{\smash{#1}\vphantom{0^0}}}

\def\i{\iota}

\def\lsq{{
    \kern-0.037em
    \adjustbox{scale=0.99,valign=c}{$
        {\lfloor \llap{\reflectbox{\rotatebox[origin=c]{180}{$\lfloor$}}}}
    $}
    \kern-0.04em
}}
\def\rsq{{
    \kern-0.04em
    \adjustbox{scale=0.99,valign=c}{$
        {\rlap{\reflectbox{\rotatebox[origin=c]{180}{$\rfloor$}}} \rfloor}
    $}
    \kern-0.037em
}}

\def\mpad{\kern0.525em}
\def\hpad{\kern0.8em}
\def\wpad{\kern0.95em}

\definecolor{paracolor}{RGB}{27,54,126}
\newcommand{\para}[1]{\paragraph{\color{schutz}\scshape
	#1%
}}

\def\kerr{{\smash{\text{$\kern-0.075em\sqrt{\text{Kerr\hem}}$}}}}

\def\cov{{\text{cov}}}
 
\def\cyc{{\text{cyc}}}

\def\top{{\text{top}}}

\newcommand{\dbar}{
    d\kern-.20em\makebox[0pt][l]{$\bar{}$}\kern.20em
}
\newcommand{\deltabar}{
    \delta\kern-.20em\makebox[0pt][l]{$\bar{}$}\kern.20em
}

\newcommand{\expval}[1]{
    \big\langle\hem{
        #1
    }\hem\big\rangle
}

\let\oldexp\exp
\renewcommand{\exp}{\oldexp\nem}
\def\Pexp{\mathrm{P}\kern-0.1em\exp}

\def\Pexp{\mathrm{P}\kern-0.1em\exp}

\def\R{\mathbb{R}}
\def\C{\mathbb{C}}

\def\V{\mathbb{V}}

\def\ps{\mathcal{P}}

\def\A{\mathcal{A}}
\def\Aflat{\mathbb{A}}

\def\M{{\mathcal{M}}}

\def\mflat{{\mathbb{M}}}

\def\CX{{\mathcal{X}}}

 \def\Xvec{{\mathbbm{X}\kern-.03em}}

\def\U{\mathrm{U}}
\def\su{\mathfrak{su}}
\def\SU{\mathrm{SU}}
\def\so{\mathfrak{so}}

\def\g{\mathfrak{g}}

\def\Cinfty{C^\infty\hnem}
\newcommand{\act}[1]{[\,{#1}\,]}
\newcommand{\cont}[2]{\big\langle\hem{#1},\hem{#2}\hem\big\rangle}

\def\Tstar{T^*\nem}

\def\tpartial{\tilde{\partial}}
\def\tE{\tilde{E}}

\def\blob{{\color[RGB]{168,175,194}\bullet}}
\def\crc{{\color[RGB]{121,126,140}{}\circ{}}}
\def\crctight{{\color[RGB]{121,126,140}{\circ}}}

\def\psp{{\partial/\partial}}

\newcommand{\pb}[2]{\{\hem{#1},{#2}\hem\}}

\newcommand{\Jac}[3]{\mathrm{Jac}\hem\big(\hem{#1},\mem{#2},\mem{#3}\hem\big)}
\def\jac{{\mathrm{Jac}}}

\newcommand{\JJ}[3]{J\hhem\big(\hem{#1},\mem{#2},\mem{#3}\hem\big)}

\newcommand{\comm}[2]{[\hem{#1},{#2}\hem]}

\usepackage{simpler-wick}

\newcommand{\pif}[2]{{\omega^\circ{}^{-1}\hnem
    \bigbig{
        \hem{d#1},
        {d#2}\hem
    }
}}

\newcommand{\pib}[2]{{\omega^\bullet{}^{-1}\hnem
    \bigbig{
        \hem{#1},
        {#2}\hem
    }
}}
\newcommand{\pia}[2]{{\omega^{-1}\hhhem
    \bigbig{
        \hem{#1},
        {#2}\hem
    }
}}
\newcommand{\piA}[2]{{\omega^{-1}\hhhem
    \BB{
        \hem{#1}
        \mem\hhem,\mem
        {#2}\hem
    }
}}

\def\E{\mathbf{E}}
\def\bme{\mathbf{e}}
\def\Om{\mathbf{\Omega}}

\def\EE{\smash{\mathbf{\acute{E}}}{}}
\def\ee{\smash{\mathbf{\acute{e}}}{}}
\def\OOm{\smash{\mathbf{\acute{\Omega}}}{}}

\def\EE{\smash{\acute\E}{}}
\def\ee{\smash{\acute\bme}{}}
\def\OOm{\smash{\acute\Om}{}}

\def\XX{\smash{\acute{\mathbf{X}}}{}}
\def\pp{\smash{\acute{\mathbf{p}}}{}}

\def\XQ{X}
\def\XQcan{{\scalebox{1.25}[1.02]{$\mathtt{X}$}}}
\def\YQcan{{\scalebox{1.25}[1.02]{$\mathtt{Y}$}}}

\def\pcan{{\kern.02em\mathtt{p}}}
\def\qcan{{\kern.02em\mathtt{q}}}
\def\xcan{{\kern.02em\mathtt{x}}}

\def\ham{{\hhem\mathsf{H}\hhem}}

\def\dX{{{\delta}X}}

\contourlength{0.15pt}
\renewcommand{\emph}[1]{\textcolor{schutz}{\contour{schutz}{{\textit{#1}}}{\kern0.01em}}}

\usepackage{titlecaps}

\definecolor{defcolor}{RGB}{85,94,119}
\newcommand{\define}[1]{{\color{defcolor}\bfseries\scshape \titlecap{#1}}}

\begin{document}

\setlength{\parindent}{1.38em}

\begin{titlepage}

\begin{flushright}\scriptsize
    {
        CALT-TH 2026-012
    }
\end{flushright}
\vskip 172pt
\begin{center}
    {\huge\textmd{\scshape\bfseries{%
		Covariant Symplectic Geometry of
        \\[0.7\baselineskip]
		Classical Particles
    }}}
    \vskip30pt
\end{center}
\begin{center}
    {\large
        \scshape\bfseries{%
            Joon-Hwi Kim%
        }\rlap{\textsuperscript{\textsc{%
            \kern1pt%
            a%
        }}}
    }\\[0.6\baselineskip]
    {\footnotesize
        \textsuperscript{\textsc{a}}\kern-1pt\textit{
            Walter Burke Institute for Theoretical Physics,\\
            California Institute of Technology, Pasadena, CA 91125
        }\\
        \vphantom{.}
    }
\end{center}
\vspace{10pt}
\noindent\parbox{\linewidth}{
    \noindent\textsc{Abstract}:\,
        We investigate the tension between 
        symplecticity and gauge covariance
        in classical Hamiltonian mechanics.
        The pursuit of manifest covariance
        over manifest symplecticity
        results in 
        a unique
        geometric formulation.
		Firstly,
		covariant yet non-canonical coordinates
		are employed
        by adopting
        Souriau's approach to minimal coupling.
		Secondly,
		covariant yet non-coordinate frames
		arise from
        Ehresmann connections
        in phase space.
    	Thirdly,
    	the concept of covariant Poisson bracket is introduced,
    	facilitating direct derivations of covariant equations of motion.
        In this way,
        we establish
		manifestly covariant Hamiltonian formulations of
		particles coupled to background gauge and gravitational fields,
		with or without spin.
        The variational principle and path integral origins of our framework
        are also explicated.
}

\end{titlepage}

\bibliographystyle{utphys-modified}
\renewcommand*{\bibfont}{\footnotesize}

\beforetochook
\hrule
\tableofcontents
\afterTocSpace
\hrule
\afterTocRuleSpace
\setcounter{footnote}{0}
\pagestyle{myplain}\pagenumbering{arabic}

\newpage
\section{Introduction}

\begin{figure}[t]
	\centering
	\begin{tikzpicture}
		\node[empty] (O) at (0,0) {};
		\node[empty] (n) at (0,0.19) {};
		\node[empty] (A) at ($(O)+(  90:0.7)$) {};
		\node[empty] (L) at ($(O)+(-150:0.7)$) {};
		\node[empty] (R) at ($(O)+(- 30:0.7)$) {};
		\draw[-] (A)--(L)--(R)--(A);
		\node[s, align=center] (a) at ($1.05*(A) + (n)$) {\scshape Determinism};
		\node[s, anchor=east] (l) at ($0.8*(L) - 1.42*(n)$)
		{\scshape Symplecticity};
		\node[s, anchor=west] (r) at ($0.8*(R) - 1.42*(n)$) {\rlap{\scshape Gauge Invariance}\hphantom{\scshape Symplecticity}};
	\end{tikzpicture}
	\caption{
		Three principles in tension.
	}
	\label{tension}
\end{figure}
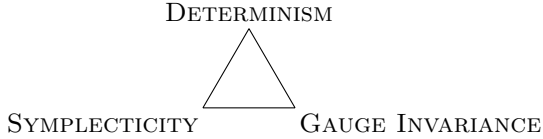

In quantum field theory,
there lies a tension
between
locality, unitarity, and gauge invariance.
A classic example is the Aharonov-Bohm effect
\cite{aharonov1959significance},
which illustrates
the tension between locality and gauge invariance.
%
A more contemporary example is
the on-shell program for scattering amplitudes
\cite{Britto:2005fq,Trnka:2013the,Arkani-Hamed:2016rak},
which pursues
modern formulations that replace Feynman diagrams
by
prioritizing gauge invariance
at the cost of
manifest locality or unitarity.

In fact,
there exists
an analogous tension 
in classical mechanics.
As illustrated in \fref{tension},
the conflicting principles are
determinism, symplecticity, and gauge covariance.
The various formulations of classical mechanics---%
Newtonian, Lagrangian, Hamiltonian, etc.---%
can be viewed as the result of navigating the tradeoffs among these axes in different ways.
Each formulation highlights different aspects of classical mechanics,
offering distinct advantages.

The term ``symplecticity'' refers to
the conservation of classical probability
under time evolution,
which may serve as the classical counterpart of unitarity.
In fact, it can be seen that
its equivalent definition is the existence of an action principle:
symplecticity means the system is Lagrangian.

In Hamiltonian mechanics,
symplecticity is represented as the Liouville theorem:
time evolution preserves 
the phase space measure
by leaving the symplectic form $\omega$ invariant.
The Liouville property is mathematically equivalent to the closure $d\omega = 0$ of the symplectic form.
This closure is trivialized when the phase space
is described with canonical (Darboux \cite{darboux1882probleme}) coordinates,
in which case the symplectic form takes the form
$\omega = d\pcan \wedge d\qcan$.
Hence
the rationale for canonical coordinates
is manifest symplecticity.
This explains why
the use of canonical coordinates
is portrayed as a sacred doctrine
in typical textbooks.

Meanwhile, it is well-known to physicists
that the canonical momentum $\pcan_\m$
of a charged particle in external electromagnetic field
is not gauge invariant.
It is instead the kinetic momentum $p_\m$
that is gauge invariant,
which is famously related to the canonical momentum
by the relation $p_\m = \pcan_\m - qA_\m(x)$.
Here, $q$ is the electric charge, and $A_\m(x)$ is the electromagnetic gauge potential.
The symplectic form is
$\omega = d\pcan_\m \swedge dx^\m = 
    dp_\m \swedge dx^\m + qF
$,
where
$F = dA$
is the field strength two-form.

In our view, 
this well-known discourse on
the canonical and kinetic momenta
illustrates the tension between symplecticity and gauge invariance
in classical mechanics.

First,
symplecticity is manifested in
the description by 
the canonical momentum.
The symplectic form is identically closed as
$d\omega = d(dP_\m \swedge dx^\m) = 0$.
However, the Hamiltonian
explicitly involves the gauge potential as
$H = \frac{1}{2}\, ( P \mminus qA(x) )^2
= \frac{1}{2}\mem P^2 - qA(x) \mdot P + \frac{1}{2}\, q^2 A^2(x)$,
obscuring gauge invariance.
In perturbation theory
\cite{Feynman:1948ur,Schubert:2001he,wlf-ps},
the $\frac{1}{2}\, q^2 A^2(x)$ term
translates to the very spurious four-point interaction vertex
in the Feynman rules of scalar quantum electrodynamics,
whose sole role is to restore gauge invariance for scattering amplitudes.

In contrast,
gauge invariance is manifest in
the description
by the kinetic momentum.
The symplectic form
$\omega \eqq  dp_\m \swedge dx^\m + qF$
involves only the gauge-invariant field strength
while the Hamiltonian is simply $H \eqq \frac{1}{2}\mem p^2$
\cite{souriau1970structure}.
Moreover, the Feynman rules describe a cubic perturbation theory
in which the photon coupling is only linear
\cite{Feynman:1948ur,Schubert:2001he,wlf-ps}.
However,
$(x^\m,p_\mu)$ is a non-canonical coordinate system on the phase space
because
$\omega =  dp_\m \swedge dx^\m + qF$
is not Darboux.
Therefore, symplecticity is not trivialized in this description.
In fact,
demanding the closure $d\omega = 0$ yields $dF = 0$
as a consistency condition on the background field,
which derives a half of the Maxwell's equations
\cite{dyson1990feynman}.
In this way,
the gauge potential $A$ such that $F = dA$
is discovered \textit{a posteriori}
via the Poincar\'e lemma.

\begin{figure}
    \centering
    \begin{tabular}{ccc}
        \includegraphics[scale=0.86,valign=c]{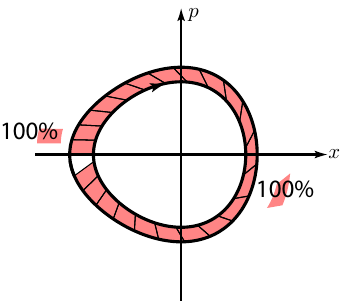}
        &
        \includegraphics[scale=0.18,raise=-7.92pt]{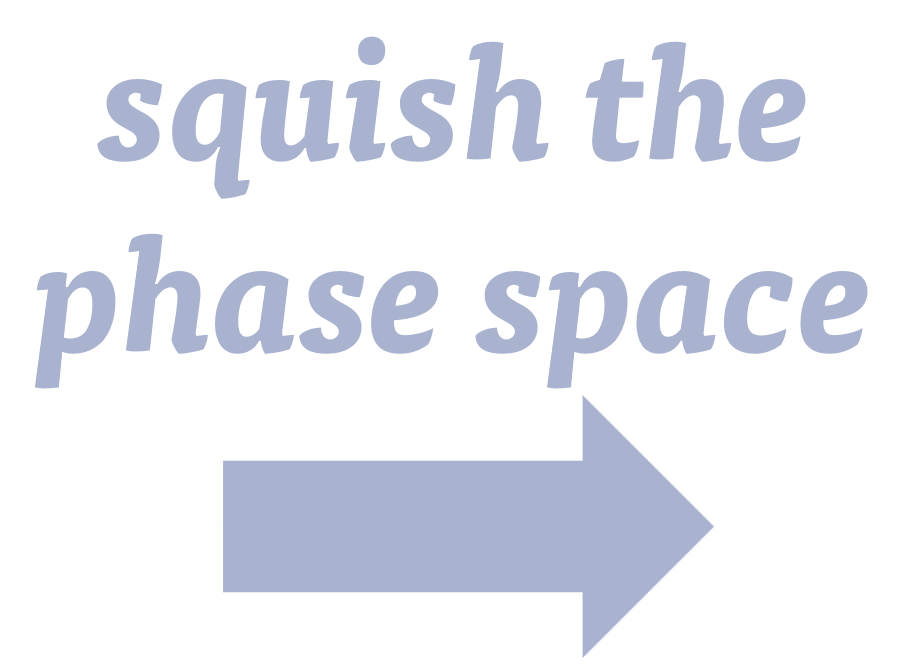}
        &
        \includegraphics[scale=0.86,valign=c]{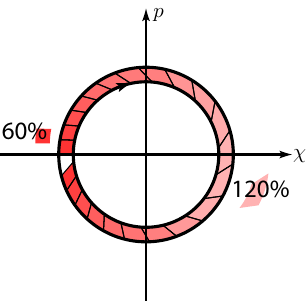}
    \\
        {\small\scshape Canonical Coordinates}
        &
        {}
        &
        {\small\scshape Non-Canonical Coordinates}
    \\[0.1\baselineskip]
        {}
    \end{tabular}
    \caption{
        Anharmonic oscillator in canonical and non-canonical coordinates.
        In a canonical coordinate system (\textit{left}),
        the apparent area element in phase space is preserved
        under time evolution.
        In a non-canonical coordinate system (\textit{right}),
        the apparent area element
        shrinks or expands by
        60\% or 120\%
        of the original scale.
        Such a non-canonical coordinate system
        arises by squishing the phase space by a generic diffeomorphism.
    }
    \label{spt-anharmonic}
\end{figure}


The coordinate transformation
$\pcan_\m = p_\m + qA_\m(x)$
that relates between the manifestly symplectic and gauge-invariant
formulations
is depicted in \fref{spt-anharmonic}.
Intuitively, it describes a squishing of the phase space
such that the contour surfaces of the Hamiltonian
are brought back to the free-theory form.
This results in apparent variations in the sizes of area parcels
under time evolution,
visualized as varying color densities in the right panel of \fref{spt-anharmonic}.
In perturbation theory,
$\pcan_\m \eqq p_\m \mplus qA_\m(x)$
describes a worldline field redefinition 
that cubicizes the Feynman rules.

To sum up,
gauge invariance is manifested 
in Hamiltonian mechanics
by sacrificing symplecticity,
which means to use non-canonical coordinates.
This idea is formulated in a coordinate-free fashion
by the statement that
the minimal coupling can be implemented 
in a gauge-invariant fashion
by a \textit{sole modification of the symplectic structure}.
This observation is due to
Souriau \cite{souriau1970structure},
while 
Feynman \cite{dyson1990feynman}
had also provided a physicist's perspective
in the language of Poisson geometry.
Here, we have rephrased
the Souriau-Feynman discourse
in our own perspective
that draws a dichotomy between
symplecticity and gauge invariance.

The purpose of this paper is to
extend, abstract, and generalize
this discussion
to achieve manifestly covariant Hamiltonian formulations of
particles coupled to gauge theories and gravity,
with or without spin.
Despite previous works
\cite{%
	souriau1974modele,%
	woodhouse1997geometric,guillemin1990symplectic,%
	torrence1973gauge,
	sniatycki1974prequantization,guillemin1978equations,
	soni1992classical,
	lee1990feynman,tanimura1992relativistic,stern1993deformed,
	chou1994dynamical,berard1999dirac,
	Gibbons:1993ap%
},
our modern treatment
introduces many new concepts
and explicates their precise geometrical identities
and physical origins.

\begin{figure}[t]
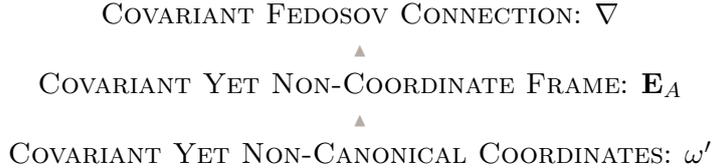

	\centering
	\begin{center}
		\textsc{Covariant Fedosov Connection}: $\nabla$
		\\[-0.18\baselineskip]
		\adjustbox{valign=c,scale=0.65}{\color{bosoniclight}$\blacktriangle$}
		\\[-0.05\baselineskip]
		\textsc{Covariant Yet Non-Coordinate Frame}: $\E_A$
		\\[-0.18\baselineskip]
		\adjustbox{valign=c,scale=0.65}{\color{bosoniclight}$\blacktriangle$}
		\\[-0.05\baselineskip]
		\textsc{Covariant Yet Non-Canonical Coordinates}: $\omega'$
	\end{center}
	\caption{
		The scaffolding of covariant symplectic geometry.
		The first two levels are sufficient for characterizing classical geometry,
		while the third level characterizes quantum geometry.
	}
	\label{tower}
\end{figure}

\Sec{NONCAN}
reviews the Souriau \cite{souriau1970structure} method
and provides its generalizations to non-abelian gauge theory and gravity,
demonstrating that gauge-covariant coordinates are inevitably non-canonical.
For non-abelian interactions,
it will be observed that
the occurrence of bare connection terms
is unavoidable
even in gauge-covariant coordinates,
mandating a more elaborate approach.
A coordinate-free characterization of these ideas are presented in \Sec{NONCAN>SPT}
under the name of ``symplectic perturbation theory.''

\Sec{NONCOORD}
provides the crucial refinement for non-abelian interactions,
based on gauge-covariant yet non-coordinate frames due to the Ehresmann \cite{ehresmann1948connexions} notion of connections.
This refinement leads to the definition of covariant Poisson bracket in \Sec{CPB},
which examines the components of the Poisson bivector with the gauge-covariant frames.
We elaborate in detail
how the check of symplecticity becomes nontrivial
in terms of the covariant Poisson bracket,
which introduces the idea of covariant Jacobi identity.

\Sec{CEOM} shows how the covariant Poisson bracket facilitates direct derivations of the classical equations of motion
in which gauge covariance is perfectly manifest at all intermediate steps.
\Sec{PI} explicates the variational principle or path integral origins
of the covariant Poisson bracket and equations of motion.
This will also offer a glimpse into the quantum theory,
which we plan to present in forthcoming work \cite{CSG-qu}.
\fref{tower} illustrates our big plan;
the third level \cite{fedosov1994simple} will not be entered in this paper.

We shall emphasize that
our direct derivations of covariant Hamiltonian equations of motion
are of a great practical value.
Computations and derivations with bare connection terms
are often bulky and inefficient.
They are not only conceptually uninsightful
but also practically discouraged.
To our knowledge, however,
such manifestly covariant methods
have not been widely recognized nor practiced
in the current physics literature.

We present our ideas via concrete examples,
which are various particles
in general spacetime dimensions.
We work in the relativistic context
and assume zero rest-mass for simplicity.
We presume basic familiarity in 
the geometric treatment of Hamiltonian mechanics
\cite{arnold1989mathematical,woodhouse1997geometric,guillemin1990symplectic}.
Relevant mathematical details may be found in \App{GUISES},
where we clarify on the equivalent definitions of the term symplecticity
in the differential-geometric language.
Our spinning particle examples
are the counterparts of the worldsheet models of \rrcite{adamo-ym,adamo-grav},
tracing back to \rrcite{Brink:1976sz,Brink:1976uf,Balachandran:1976ya,Barducci:1976xq,Barducci:1976wc}.
When working in generic index notations,
we always presume proper resolutions of the orderings
for Grassmann-odd objects if any.

\newpage

\section{Covariant Yet Non-Canonical Coordinates}
\label{NONCAN}

\subsection{Electromagnetism}
\label{NONCAN>EM}

\para{Scalar Particle}

Let $\mflat = (\R^d,\eta)$ be flat spacetime,
which is a $d$-dimensional real vector space
equipped with a Lorentzian-signature flat metric $\eta$.
Let $x^m$ be its Cartesian coordinates,
where $m,n,r,s,\cdots$ are Lorentz indices running from $0$ to $d{\:-\mem}1$.
We work in the convention
$\eta_{mn} = \diag(-1,+1,{\cdots},+1)$.

Consider the cotangent bundle $T^*\mflat$
with base and fiber coordinates $x^m$ and $p_m$.
This serves as the phase space of a relativistic point particle
in flat spacetime,
with position $x^m$ and momentum $p_m$.
Along with 
the Hamiltonian $\frac{1}{2}\mem p^2$,
the canonical symplectic form
\begin{align}
	\label{scalar.omega0}
	\omega^\circ \,=\, dp_m \wedge dx^m
\end{align}
implements a free point particle  
that moves along straight-line trajectories.
\eqref{scalar.omega0} gives rise to
the canonical Poisson brackets,
\begin{align}
    \label{scalar.pb0}
    \pb{x^m}{x^n}^\circ
    \,=\,
    	0
    \,,\quad
    \pb{x^m}{p_n}^\circ
    \,=\,
        \delta^m{}_n
    \,,\quad
    \pb{p_m}{p_n}^\circ
    \,=\,
		0
    \,.
\end{align}

\para{Covariant Coordinates}
Souriau \cite{souriau1970structure}'s observation was that
this particle can be coupled to electromagnetism
via a sole modification of the symplectic form:
\begin{align}
	\label{EM.omega}
	\omega \,=\, 
		dp_m \wedge dx^m
		+ \frac{1}{2}\,
			qF_{mn}(x)\mem dx^m \swedge dx^n
	\,.
\end{align}
Here, $F \eqq \frac{1}{2}\, F_{mn}(x)\mem dx^m \swedge dx^n \inn \Omega^2(\mflat)$
is the electromagnetic field strength.
This modification
endows the particle with charge $q$,
which can be seen by deriving
the Hamiltonian equations of motion.

To elaborate,
the Poisson brackets due to \eqref{scalar.pb0} are
\begin{align}
    \label{EM.pb}
    \pb{x^m}{x^n}
    \,=\,
    	0
    \,,\quad
    \pb{x^m}{p_n}
    \,=\,
        \delta^m{}_n
    \,,\quad
    \pb{p_m}{p_n}
    \,=\,
		qF_{mn}(x)
    \,.
\end{align}
Clearly, $\pb{p_m}{p_n} = F_{mn}(x)$ describes a non-canonical Poisson bracket,
the intuition for which could be the cyclotron precession of the momentum $p_m$
in the external field.
The resulting Hamiltonian equations of motion
implement the Lorentz force law
precisely through this non-canonical bracket:
\begin{align}
\begin{split}
	\label{EM.eom}
	\dot{x}^m
	\,&=\,
	\pb{x^m}{p_n}\mem p^n
	\,=\,
		p^m
	\,,\\
	\dot{p}_m
	\,&=\,
		\pb{p_m}{p_n}\mem p^n
	\,=\,
		qF_{mn}(x)\mem p^n
	\,.
\end{split}
\end{align}
\eqref{EM.eom} confirms that $p_m$ is the \textit{kinetic momentum}:
up to index-raising,
it is equated with the particle's physical velocity $\dot{x}^m$ on equations of motion.

Souriau's method 
provides an elegantly gauge-invariant Hamiltonian formulation of the charged particle.
(See, e.g., \rcite{woodhouse1997geometric}, p.\,36.)
It is directly based on the gauge-invariant field strength, $F$.
There is no need to invoke the gauge connection $A \inn \Omega^1(\R^d)$ such that $F = dA$,
which is subject to the gauge redundancies $A \sim A + \lambda$
for closed one-forms $\lambda$.

However, to ensure that the modified phase space geometry remains symplectic,
one must not forget to impose the closure $d\omega = 0$.
Crucially, this symplecticity condition demands
the closure of the field strength:
$dF = 0$.
In this way,
the gauge connection $A$
arises \textit{a posteriori}
as a secondary construct
from its local existence
by Poincar\'e lemma.

\newpage

\para{Canonical Coordinates}

It is worth reiterating that
Souriau implements the coupling by 
changing only the symplectic structure
while \textit{fixing the Hamiltonian}.
In the meantime, one could have pursued the reverse approach
via the freedom to choose coordinates on the phase space.

To this end, suppose a gauge potential $A$ is provided.
It is easy to see that the symplectic form in \eqref{EM.omega} is equal to
\begin{align}
	\label{EM.omega.can}
	\omega \,=\, 
		d\pcan_m \wedge dx^m
	\,,
\end{align}
if one defines $\pcan_m = p_m + qA_m(x)$.
This $\pcan_m$ is the canonical momentum.
The closure $d\omega \eqq 0$ now holds identically
without any requirement on the background.
The trade-off is that
the Hamiltonian is deformed to
$\frac{1}{2}\mem (\pcan - qA(x))^2$
and makes explicit references to the gauge potential,
grossly obscuring gauge invariance.

In particular, consider how the Hamiltonian equations of motion are derived in this description.
By using the canonical Poisson brackets between $x^m$ and $\pcan_m$,
one derives
\begin{align}
\begin{split}
	\label{EM.eom.can}
	\dot{x}^m
	\,&=\,
		\pcan^m - qA^m(x)
	\,,\\
	\dot{\pcan}_m
	\,&=\,
		- qA_{n,m}(x)\,
		\bigbig{
			\pcan^n - qA^n(x)
		}
	\,.
\end{split}
\end{align}
This derivation has evident disadvantages
in both practical and conceptual aspects.
First of all,
gauge invariance is grossly obscured
by the appearances of bare gauge potentials.
Second of all,
deriving and identifying the Lorentz force law
becomes a complicated task,
which involves using several gymnastics about partial derivatives and index permutations.
Third of all,
\eqref{EM.eom.can} does not directly concern physical observables.
The canonical momentum is gauge-dependent and is not a physical observable:
$\pcan_m \sim \pcan_m + \lambda_m(x)$.
As a result, the above derivation in canonical coordinates
is neither useful, efficient, nor insightful.

\enter
To sum up, we have drawn a dichotomy between
manifest gauge invariance
and 
manifest symplecticity.
The former description modifies the symplectic structure,
while the latter description modifies the Hamiltonian.
The former employs the gauge-invariant yet non-canonical momentum $p_m$,
which is equated to the particle's physical velocity 
on equations of motion
and thus
referred to as kinetic momentum in the physics literature.
The latter employs the
canonical yet gauge-dependent
momentum $\pcan_m$.

\subsection{Gauge Theory}
\label{NONCAN>YM}

Let us expand our scope to non-abelian gauge theories.
The equations of motion of
color-charged particles in background non-abelian gauge fields
are known as Wong's equations
\cite{wong1970field},
whose Hamiltonian formulation
has been provided in the literature
\cite{lee1990feynman,tanimura1992relativistic,stern1993deformed}
(see also \rrcite{guillemin1978equations,guillemin1990symplectic}).
However, we find that
such previous approaches
might not be fully satisfactory
in light of manifest gauge covariance.

\para{Colored Scalar Particle}
For concreteness,
take $G = \SU(N)$ as the gauge group.
Let $\g = \su(N)$ be the Lie algebra
and $\g^*$ be its dual.
Let $q_a$ coordinatize $\g^*$, where $a,b,c,d,\cdots$ 
are adjoint indices running through $\dim G = (N^2 {\,-\,} 1)$ integers.
Provided the Jacobi identity
$f^d{}_{e[c}\mem f^e{}_{ab]} = 0$,
the dual Lie algebra $\g^*$ is a Poisson manifold with the Poisson brackets
$\pb{q_a}{q_b} = q_c\mem f^c{}_{ab}$.

\newpage

The phase space implementing the color charge
can be any symplectic realization of this Poisson manifold $\g^*$,
such as $T^*G$.
Our choice in this paper will be
$\Pi\V$
with $\V$ a complex or complexified representation space of $G = \SU(N)$,
such as $\C^N$ or $\g^\C$.
This is a K\"ahler vector space
featuring linear fermionic coordinates $\vth^i$
whose complex conjugates are 
identified as
$\bvth_i = \delta_{i\bar{j}}\mem [\vth^j]^*$
by the K\"ahler metric $\delta_{i\bar{j}}$,
where $i,j,\cdots$ run through $\dim\V$ integers.
The symplectic structure is $i\mem d\bvth_i \swedge d\vth^i$.
The color charge is
a composite variable
\begin{align}
	\label{q}
	q_a \,=\, i\mem \bvth_i\mem (t_a)^i{}_j\mem \vth^j
	\,,
\end{align}
where
$(t_a)^i{}_j$ are the generators of $G$
such that $(\comm{t_a}{t_b})^i{}_j = (t_c)^i{}_j\mem f^c{}_{ab}$.
Since $q_a$ is a bilinear, it Poisson-commutes with a $\U(1)$ generator
$\bvth_i\hem \vth^i$.
Hence a constraint $\bvth_i\hem \vth^i = w$
will be imposed,
where $w$ is a constant.

Finally,
let $E \too \mflat$ be a vector bundle over flat spacetime
whose typical fiber is $\Pi\V$.
The colored relativistic point particle
is a constrained Hamiltonian system
defined on
the phase space $T^*\mflat \moplus E$,
equipped with
the mass-shell constraint $\frac{1}{2}\mem p^2$
and
the $\U(1)$ constraint $\bvth_i\hem \vth^i - w$.
The symplectic structure is
\begin{align}
    \label{colored.struct0}
    \theta^\circ
    \,=\,
        p_m\mem dx^m
        + i\mem \bvth_i\mem d\vth^i
    \qiq
    \omega^\circ
    \,=\,
    	d\theta^\circ
    \,=\,
        dp_m \swedge dx^m
        + i\mem d\bvth_i \swedge d\vth^i
    \,,
\end{align}
which gives rise to the canonical (super-)Poisson brackets
\begin{align}
    \label{colored.pb0}
    \pb{x^m}{p_n}^\circ
    \,=\,
        \delta^m{}_n
    \,,\quad
    \pb{\vth^i}{\bvth_j}^\circ
    \,=\,
        -i\mem \delta^i{}_j
    \,.
\end{align}

\para{Covariant Coordinates}

To couple the colored scalar particle to external non-abelian gauge fields,
the standard recipe
covariantizes the symplectic potential $\theta^\circ$ in \eqref{colored.struct0}
by replacing the ordinary exterior derivative $d$
with the covariant exterior derivative $D$:
\begin{align}
    \label{YM.thetaA}
    \theta
    \,=\,
        p_m\mem dx^m
        + i\mem \bvth_i\mem D\vth^i
    \,=\,
        \theta^\circ
        + q_a\hem A^a
    \,.
\end{align}
Here, $A^a \eqq A^a{}_m(x)\mem dx^m$
describes the non-abelian gauge potential
$A \inn \Omega^1(\mflat;\g)$.
Consequently,
the symplectic form $\omega = d\theta$
is found as
\begin{align}
    \label{YM.omegaA}
    \omega
    \,=\,
        \omega^\circ
        \,+\,
        \BB{
            dq_a \wedge A^a
            + q_a\mem dA^a
        }
    \,.
\end{align}
\eqref{YM.omegaA} gives rise to the non-canonical Poisson brackets,
\begin{align}
\begin{split}
    \label{YM.pb}
    \pb{x^m}{p_n}
    \,=\,
        \delta^m{}_n
    &\,,\quad
    \pb{\vth^i}{p_r}
    \,=\,
        - A^i{}_{jr}(x)\mem \vth^j
    \,,\\
    \pb{\vth^i}{\bvth_j}
    \,=\,
        -i\mem \delta^i{}_j
    &\,,\quad
    \pb{p_m}{p_n}
    \,=\,
        q_a\hem F^a{}_{mn}(x)
    \,,
\end{split}
\end{align}
where we omit components related by complex conjugation.
Here, $F^a = dA^a + \frac{1}{2}\mem f^a{}_{bc}\mem A^b \swedge A^c
$ $	= \frac{1}{2}\, F^a{}_{mn}(x)\mem dx^m \swedge dx^n
$
describes
the non-abelian field strength $F \in \Omega^2(\mflat;\g)$.
Note that the Poisson brackets in \eqref{YM.pb}
are precisely the ones used in
\rrcite{lee1990feynman,tanimura1992relativistic,stern1993deformed}.

Finally,
the equations of motion 
are derived
by using
the time-dependent Hamiltonian
$\k^0(\t)\mem \bigbig{
	\frac{1}{2}\mem p^2
} + \k^1(\t)\mem \bigbig{
	\bvth_i\hem\vth^i - w
}$
that summarizes all the constraints
with Lagrange multipliers $\k^0(\t)$ and $\k^1(\t)$.
In the worldline parametrization such that $\k^0(\t) = 1$, one finds
\begin{align}
\begin{split}
    \label{YM.eom}
    \dot{x}^m
    \,&=\,
        p^m
    \,,\\
    \dot{p}_m
    \,&=\,
        q_a\hem F^a{}_{mn}(x)\mem p^n
	\,,\\
    \dot{\vth}^i
    \,&=\,
        - A^i{}_{jr}(x)\mem \vth^j\mem p^r
        - i\mem \k^1\mem \vth^i
    \,.
\end{split}
\end{align}
The time evolution of the color charge $q_a$
is uniquely determined even when
the Lagrange multiplier $\k^1(\t)$
is left agnostic,
as it is $\U(1)$-invariant:
$
	\dot{q}_a
	=
		q_b\mem f^b{}_{ca}\mem A^c{}_r(x)\mem p^r
$.
With this understanding,
\eqref{YM.eom} precisely reproduces
Wong's equations \cite{wong1970field}.

Let us explicate
the physical interpretations.
First of all,
$\dot{x}^m = p^m$ 
in \eqref{YM.eom}
shows that, up to index-raising,
$p_m$ is equated to the particle's actual velocity $\dot{x}^m$ in spacetime.
For this reason, $p_m$ must be gauge invariant.
Clearly, it deserves the name kinetic momentum.
Second of all,
$\dot{p}_m = q_a\hem F^a{}_{mn}(x)\mem p^n$ 
in \eqref{YM.eom}
describes the non-abelian Lorentz force law,
whose origin is precisely the non-canonical bracket $\pb{p_m}{p_n} = q_a\hem F^a{}_{mn}(x)$ in \eqref{YM.pb}.
Third of all,
$
	\dot{q}_a
	=
		q_b\mem f^b{}_{ca}\mem A^c{}_r(x)\mem p^r
$
describes that
the particle's color charge is covariantly constant along the worldline,
i.e., is parallel-transported
with respect to the non-abelian gauge connection $A \in \Omega^1(\mflat;\g)$.

Overall, we conclude that
$(x^m,p_m,\vth^i)$
together provides a non-canonical yet gauge-covariant coordinate system
on the phase space $T^*\mflat \moplus E$ of a colored scalar particle in background non-abelian gauge fields.

\enter

In this paper, our practical criterion for
a manifestly covariant Hamiltonian formulation
is whether
the derivation of the classical equations of motion is manifestly covariant.
Does the above derivation of Wong’s equations pass this test?
The answer is \textit{no}.
Despite the covariance of the coordinates 
$(x^m,p_m,\vth^i,\bvth_i)$,
\eqref{YM.eom} makes an explicit reference to the gauge potential $A$
while producing the bare, non-covariant time derivative $\dot{\vth}^i$ of the color variable.
Evidently,
this problem originates from the non-covariant nature of the Poisson brackets in \eqref{YM.pb},
yielding bare connection coefficients on the right-hand sides.
An ideal derivation of Wong's equations
would instead directly produce 
the gauge-covariant time derivative
$D\vth^i\nem/d\t = -i\mem \k^1\mem \vth^i$.

On a related note,
\eqref{YM.omegaA} understands the symplectic form $\omega$
as a modification of the free-theory's $\omega^\circ$
by a group of \textit{gauge-dependent} terms,
$dq_a \swedge A^a + q_a\mem dA^a$.
This makes a stark contrast with the abelian case in \Sec{NONCAN>EM},
where $\omega = \omega^\circ + qF$ with the gauge-invariant field strength $F$.
This issue can be traced back to
the non-abelian nature
of the gauge interaction;
the color charge $q_a$ is a dynamical variable.
In fact, we will see shortly that
the same problem plagues the case of gravitational couplings as well.

Therefore, it generally seems true that
another layer of refinement
is mandated
to achieve a fully gauge-covariant formulation
when the particle exhibits non-abelian interactions.
This provides the very motivation for \Sec{NONCOORD}.

\para{Canonical Coordinates}

It remains to remark on the canonical coordinates for the colored scalar particle.
Provided the gauge potential $A \in \Omega^1(\mflat;\g)$,
one defines $\pcan_m = p_m + q_a\hem A^a{}_m(x)$
to rewrite \eqref{YM.thetaA}
as $\theta = \pcan_m\mem dx^m$.
As a result, the symplectic form is $\omega = d\pcan_m \swedge dx^m$
and gives rise to the canonical Poisson brackets between $x^m$ and $\pcan_m$.

Of course, 
gauge covariance is grossly obscured
when one derives the equations of motion
with canonical coordinates.
The mass-shell constraint is deformed as
$\frac{1}{2}\mem (\pcan - q_a\hem A^a(x))^2$,
and its derivatives spawn messy partial derivatives of the gauge potential.
Repackaging them in terms of the nonabelian field strength
is a not-so-trivial task.
By the very same reasons
as in \Sec{NONCAN>EM},
using canonical coordinates
is neither useful, efficient, nor insightful.

\subsection{Gravity}
\label{NONCAN>BI}

\para{Canonical Coordinates}

Finally, let us concern gravitational interactions.
Let $\M = (\R^d,g)$ be general-relativistic spacetime,
which is a $d$-dimensional smooth manifold
equipped with a pseudo-Riemannian metric $g \inn \Gamma(\Sym^2\Tstar\M)$.
Let $x^\m$ be local coordinates on $\M$,
where $\m,\n,\r,\s,\cdots$ are the spacetime indices running from $0$ to $d{\:-\mem}1$.
Let $g^{-1} \inn \Gamma(\Sym^2T\hnem\M)$ be the inverse metric.

Consider the cotangent bundle $\Tstar\M$
with base and fiber coordinates $x^\m$ and $\pcan_\m$.
This is the phase space of a general-relativistic point particle.
For typical occasions,
this particle is well-described by
the canonical symplectic structure
\begin{align}
	\label{BI.struct-can}
	\theta \,=\, \pcan_\m\mem dx^\m
	\qiq
	\omega \,=\, d\pcan_\m \wedge dx^\m
	\,,
\end{align}
together with the Hamiltonian
$\frac{1}{2}\mem g^{-1}{}^{\m\n}(x)\mem p_\m\mem p_\n$.
The Poisson brackets due to \eqref{BI.struct-can} are
\begin{align}
	\label{BI.pb-can}
	\pb{x^\m}{x^\n}
	\,=\, 0
	\,,\quad
	\pb{x^\m}{\pcan_\n}
	\,=\, \delta^\m{}_\n
	\,,\quad
	\pb{\pcan_\m}{\pcan_\n}
	\,=\, 0
	\,.
\end{align}

This is a working formulation of the point particle in general relativity.
However, if we are to directly succeed the philosophies of the Souriau method in \Sec{NONCAN>EM},
two immediate issues will be pointed out.
First, the symplectic form in \eqref{BI.struct-can} employs a non-covariant differential $d\pcan_\m$ of a tensorial (covector) variable $\pcan_\m$,
obscuring general covariance.
Second, the Hamiltonian $\frac{1}{2}\mem g^{-1}{}^{\m\n}(x)\mem \pcan_\m\mem \pcan_\n$
is deformed away from the free-theory form.

Crucially,
this conventional approach
fails our practical test for a manifestly gauge-covariant formulation.
It derives the Hamiltonian equations of motion
in terms of partial derivatives of the inverse metric,
which are not generally covariant.

\para{Covariant Coordinates}

To this end, let us employ the vielbein formalism.
Let $e^m = e^m{}_\m(x)\mem dx^\m$ be the orthonormal coframe
such that $g = \eta_{mn}\mem e^m \mtensor e^n$.
Let $E_m = E^\m{}_m(x)\mem \partial_\m$ be the orthonormal frame
such that $\langle e^m , E_n \rangle = \delta^m{}_n$.
Again, $m,n,r,s,\cdots$ are Lorentz indices running from $0$ to $d{\:-\mem}1$,
yet local.

We propose that
the momentum components in the orthonormal frame,
$p_m = \pcan_\m\hem E^\m{}_m(x)$,
define the gauge-covariant ``kinetic momentum''
in the Souriau sense.
First of all,
it brings back the Hamiltonian 
to the free-theory form:
$\frac{1}{2}\mem \eta^{-1}{}^{mn} p_m\hem p_n$
with the \textit{flat} inverse metric $\eta^{-1}{}^{mn}$.
Second of all, $p_m$ transforms covariantly under local Lorentz transformations.
Third of all,
it encodes the physical velocity of the particle in orthonormal frames
as will be shown shortly.
Consequently, we identify $(x^\m,p_m)$ as the gauge-covariant coordinates
for the phase space $\Tstar\M$.

The symplectic structure in \eqref{BI.struct-can} now appears as
\begin{align}
	\label{BI.structA}
	\theta \,=\, p_m\mem e^m
	\qiq
	\omega \,=\, dp_m \wedge e^m \,+\, p_m\mem de^m
	\,,
\end{align}
which describes the non-canonical Poisson brackets
\begin{align}
	\label{BI.pb}
	\pb{x^\m}{x^\n}
	\,=\,
		0
	\,,\quad
	\pb{x^\m}{p_n}
	\,=\,
		E^\m{}_n(x)
	\,,\quad
	\pb{p_m}{p_n}
	\,=\,
		- p_k\mem \Omega^k{}_{mn}(x)
	\,.
\end{align}
Here, $\Omega^k{}_{mn}(x)$ denote the anholonomy coefficients
of the orthonormal frame:
\begin{align}
	\Omega^k{}_{mn}
	\,=\,
		\cont{
			e^k
		}{
			\comm{E_m}{E_n}
		}
	\qfq
	de^k
	\,=\,	
		- \frac{1}{2}\,
			\Omega^k{}_{mn}\,
			e^m \swedge e^n
	\,.
\end{align}

Unfortunately, \eqref{BI.structA} still describes non-covariant objects
such as $dp_m$ and $de^m$,
despite the invariances of the symplectic potential $p_m\mem e^m$.
The anholonomy coefficients $\Omega^k{}_{mn}(x)$
in \eqref{BI.pb}
are also famously non-invariant
under local Lorentz transformations.

This alternative description also fails our practical test for manifest covariance.
The Hamiltonian equations of motion are derived as
\begin{align}
\begin{split}
	\label{BIEOM.A}
	\dot{x}^\m
	\,&=\,
		\pb{x^\m}{p_m}\mem p^m
	\,=\,
		E^\m{}_m(x)\mem p^m
	\,,\\
	\dot{p}_m
	\,&=\,
		\pb{p_m}{p_n}\mem p^n
	\,=\,
		- p_k\, \Omega^k{}_{mn}(x)\mem p^n
	\,.
\end{split}
\end{align}
Although \eqref{BIEOM.A} can be eventually boiled down to the geodesic equation,
it does not manifest local Lorentz covariance
due to the anholonomy coefficients $\Omega^k{}_{mn}(x)$.


\subsection{Theory of Symplectic Perturbations}
\label{NONCAN>SPT}

Before moving on to \Sec{NONCOORD},
let us provide a coordinate-free characterization of
the ideas presented and observed in \Secss{NONCAN>EM}{NONCAN>YM}{NONCAN>BI}.
This analysis should clarify
the invariant geometrical content of
the tension between symplecticity and gauge covariance.

First, we provide a mathematical abstraction of
the Souriau method.
Second, we provide its generalization 
based on the languages of almost-symplectic geometry.
To our knowledge, these constructions are new.

\para{Almost-Symplectic Geometry}

We begin with basic definitions and conventions.
An \define{Almost-Symplectic Structure}
is a nondegenerate two-form
$\omega \inn \Omega^2(\ps)$
on a smooth manifold $\ps$
\cite{gelfand1997fedosov}.
An \define{Almost-Symplectic Manifold}
is a smooth manifold $\ps$ equipped with an almost-symplectic structure $\omega \in \Omega^2(\ps)$,
whose dimension is necessarily even.

The \define{Pointwise Inverse}
of an almost-symplectic structure $\omega \inn \Omega^2(\ps)$
is the unique bivector
$\omega^{-1} \inn \Gamma(\wedge^2T\ps)$ 
that satisfies
$\omega^{-1}(\blank,\omega(\blank,v)) \eqq v$
for all $v \inn \Gamma(T\ps)$,
referred to as the \define{almost-Poisson bivector}.
It is convenient to regard
$\omega : T\ps \to \Tstar\ps : v \mapsto \omega(\blank,v)$
and
$\omega^{-1} : \Tstar\ps \to T\ps : \alpha \mapsto \omega^{-1}(\blank,\alpha)$
as bundle isomorphisms,
in which case
$\omega \crc \omega^{-1} = \id$
and
$\omega^{-1} \crc \omega = \id$.
Here, $\crctight$ denotes composition of maps.

In an almost-symplectic manifold $(\ps,\omega)$,
the \define{Hamiltonian Vector Field}
of $f \inn \Cinfty(\ps)$
is the unique vector field $\Xvec_f \inn \Gamma(T\ps)$
such that $\i_{\Xvec_f} \omega \eqq {-df}$,
i.e.,
$\Xvec_f \eqq \omega^{-1}(\blank,df)$.
(This is right-action convention, suitable for particle mechanics.)
The \define{Almost-Poisson Bracket}
is the bi-differential operator
$\pb{\blank}{\nem\hnem\blank} : \Cinfty(\ps) {\,\times\,} \Cinfty(\ps) \to \Cinfty(\ps) : (f,g) \mapsto \pb{f}{g}$,
where
\begin{align}
    \label{pb-def}
    \pb{f}{g}
    \,=\,
        \omega^{-1}(df,dg)
    \,.
\end{align}
Note that $
    \pb{f}{g} \eqq \Xvec_g\act{f} \eqq {-\Xvec_f\act{g} }
    \eqq {-\omega(\Xvec_f,\Xvec_g)}
$.
The \define{Jacobiator}
is the tri-differential operator
$\jac : \Cinfty(\ps) {\,\times\,} \Cinfty(\ps) {\,\times\,} \Cinfty(\ps) \to \Cinfty(\ps) : (f,g,h) 
\mapsto 
    \jac(f,g,h)
$, where
\begin{align}
    \label{jac-def}
    \Jac{f}{g}{h}
    \,=\,
        \pb{\pb{f}{g}}{h} + \pb{\pb{g}{h}}{f} + \pb{\pb{h}{f}}{g}
    \,=\,
        d\omega(\Xvec_f,\Xvec_g,\Xvec_h)
    \,.
\end{align}
See \App{GUISES} for more details.

A \define{Symplectic Structure}
is an almost-symplectic structure that is closed.
A \define{Symplectic Manifold}
is a smooth manifold $\ps$ equipped with a symplectic structure $\omega \in Z^2(\ps)$.
When $\omega$ is symplectic,
the almost-Poisson bivector (almost-Poisson bracket)
is called the \define{Poisson bivector} (\define{Poisson Bracket}).
The Jacobi identity refers to the property
$\jac(f,g,h) = 0$ $\forall f,g,h \in \Cinfty(\ps)$
of the Poisson bracket.


\para{Symplectic Perturbations}

A \define{Symplectic Perturbation}
on a symplectic manifold $(\ps,\omega^\circ)$
is a two-form $\omega' \inn \Omega^2(\ps)$
such that
$\omega = \omega^\circ + \omega' \in \Omega^2(\ps)$
is nondegenerate.
A symplectic perturbation is \define{Proper} iff $d\omega' = 0$.
A symplectic perturbation is \define{Improper} iff $d\omega' \neq 0$.

Proper symplectic perturbations
deform symplectic manifolds
to symplectic manifolds.
Improper symplectic perturbations
deform symplectic manifolds
to almost-symplectic manifolds.

The pointwise inverse of the modified symplectic form 
$\omega = \omega^\circ + \omega'$
admits
a power series expansion.
In the notation that regards $\omega$, $\omega^\circ$,
and $\omega'$
as bundle maps,
it reads
\begin{align}
    \label{spt}
    \omega^{-1} 
    \mem\,=\,\mem
	    \omega^\circ{}^{-1}
	    \,-\,
            \omega^\circ{}^{-1} \crc
            \omega' \crc
            \omega^\circ{}^{-1}
	    \,+\,
            \omega^\circ{}^{-1} \crc
            \omega' \crc
            \omega^\circ{}^{-1}
            \omega' \crc
            \omega^\circ{}^{-1}
	    \,-\, \cdots
    \,\mem.
\end{align}
Concretely, suppose local coordinates $\XQ^I$ on $\ps$,
where $I,J,K,L,\cdots$ run through $(\dim \ps)$ integers.
Denoting 
$\pb{f}{g} = \omega^{-1}(df,dg)$
and
$\pb{f}{g}^\circ = \omega^\circ{}^{-1}(df,dg)$,
one finds that
\eqref{spt} encodes
the series expansion for the modified almost-Poisson bracket:
\begin{align}
\begin{split}
    \label{spt.pb}
    \pb{f}{g}
    \mem\,=\,\mem
        \pb{f}{g}^\circ
        \,&-\,
        \pb{f}{\XQ^I}^\circ
            {\mem\,\omega'_{IJ}(\XQ)\,\hhem}
        \pb{\XQ^J}{g}^\circ
    \\
        \,&+\,
        \pb{f}{\XQ^I}^\circ
            {\mem\,\omega'_{IJ}(\XQ)\,\hhem}
        \pb{\XQ^J}{\XQ^K}^\circ
            {\mem\,\omega'_{KL}(\XQ)\,\hhem}
        \pb{\XQ^L}{g}^\circ
        \,-\,
        \cdots
    \,\mem.
\end{split}
\end{align}
Similarly,
the Hamiltonian vector fields
admit the expansion
\begin{align}
\begin{split}
    \label{spt.X}
    \Xvec_f
    \,\,=\,\,
        \Xvec_f^\circ
        \,&-\,
            \pb{\blank}{\XQ_I}^\circ
                {\mem\,\omega'_{IJ}(\XQ)\,\hhem}
            \pb{\XQ^J}{f}^\circ
    \\
        \,&+\,
            \pb{\blank}{\XQ_I}^\circ
                {\mem\,\omega'_{IJ}(\XQ)\,\hhem}
            \pb{\XQ^J}{\XQ^K}^\circ
                {\mem\,\omega'_{KL}(\XQ)\,\hhem}
            \pb{\XQ^L}{f}^\circ
        \,-\,
            \cdots
    \,\mem.
\end{split}
\end{align}
Here, we have denoted
$\Xvec_f = \omega^{-1}(\blank,df)$
and
$\Xvec_f^\circ = \omega^\circ{}^{-1}(\blank,df)$.

\enter
Let us elicit the intuitive physical interpretations of the above abstractions
by referencing the charged particle example in \Sec{NONCAN>EM}.
The symplectic perturbation $\omega' \inn \Omega^2(\ps)$
brings the \textit{free theory} $(\ps,\omega^\circ,H)$
to the \textit{interacting theory} $(\ps,\omega,H)$,
where $H \inn \Cinfty(\ps)$ is the time-evolution generator.
The expansion of the Hamiltonian vector field in \eqref{spt.X}
precisely describes how
the ``straight-line'' trajectories
in the free theory
are deflected
by the ``field strength'' $\omega'_{IJ}(\XQ)$.
As a result,
\eqref{spt.X}
describes the ``Lorentz force''
due to the interaction.
Indeed, \eqref{EM.eom} describes the time-evolution vector field
\begin{align}
    \label{EM.XH}
    \Xvec_H
    \,=\,
        \Xvec^\circ_H
    \,-\,
        \pb{\blank}{x^m}^\circ
            {\hem\,F_{mn}(x)\,\hhem}
        \pb{x^n}{H}^\circ
    \,=\,
        p^m\mem
        \frac{\partial}{\partial x^m}
        + F_{mn}(x)\mem p^n\mem
        \frac{\partial}{\partial p_m}
    \,,
\end{align}
where the expansion truncates at the leading order.
As remarked earlier, the symplectic perturbation
$\omega' = \frac{1}{2}\, F_{mn}(x)\mem dx^m \swedge dx^n$
generates the Lorentz force $F_{mn}(x)\mem p^n$ in \eqref{EM.XH}.

It is also insightful to see how the expansion in \eqref{spt.pb} is applied.
The $dx \swedge dx$ perturbation on the symplectic form
induces the change in the $\pb{p_m}{p_n}$ bracket
via the free theory's
$\pb{x^m}{p_n}^\circ \neq 0$:
\begin{align}
	\label{EM.seesaw}
    \pb{p_m}{p_n}
    \,=\,
        \pb{p_m}{p_n}^\circ
        -
        \pb{p_m}{x^r}^\circ
            {\hem\,F_{rs}(x)\,\hhem}
        \pb{x^s}{p_n}^\circ
    \,=\,
        F_{mn}(x)
    \,.
\end{align}
This Poisson non-commutativity
leads to $\dot{p}_m = \pb{p_m}{H} \neq 0$,
giving rise to the orbital precession frequency proportional to $F_{mn}(x)$.

In sum, symplectic perturbations are ``field strengths''
that deflect particle trajectories
by ``Lorentz forces''
while inducing non-canonical Poisson brackets
encoding ``precession behaviors.''

\newpage

From physical grounds,
we will always be interested in proper symplectic perturbations
in this paper.
This means to impose $d\omega' = 0$,
or equivalently the Jacobi identity
$\jac(f,g,h) = 0$
in the interacting theory.
Based on \eqref{jac-def}
and \eqref{spt.X},
we find
\begin{align}
    \label{spt.jacs}
    d\omega'(\Xvec_f^\circ,\Xvec_g^\circ,\Xvec_h^\circ)
    \,=\,0
    \,,\quad
    \smash{\sum_\cyc}\,\mem
    d\omega'(\Xvec_{\XQ^I}^\circ,\Xvec_g^\circ,\Xvec_h^\circ)
        {\mem\,\omega'_{IJ}(\XQ)\,\hhem}
    \Xvec_f\act{\XQ^J}
    \,=\, 0
    \,,\quad
    \cdots
    \,,
    \vphantom{\big|_0}
\end{align}
at each order in $\omega'$.
Here, $\sum_\cyc$ sums over the three cyclic permutations of $(f,g,h)$.
\eqref{spt.jacs}
demonstrates how
symplecticity becomes obscured in the interacting theory
when examined in terms of Poisson brackets.

\para{Covariant Symplectic Perturbations}

The above mathematical framework of symplectic perturbations
abstracts the Souriau method.
Based on our physical explorations in \Secs{NONCAN>YM}{NONCAN>BI},
we envision a further generalization of this formalization
as follows.

\enter
A \define{covariant symplectic perturbation}
on an almost-symplectic manifold $(\ps,\omega^\bullet)$
is a two-form $\omega' \inn \Omega^2(\ps)$
such that
$\omega = \omega^\bullet + \omega' \in \Omega^2(\ps)$
is nondegenerate.
A covariant symplectic perturbation is \define{proper} iff $d\omega = 0$.
A symplectic perturbation is \define{improper} iff $d\omega \neq 0$.

Proper covariant symplectic perturbations
deform almost-symplectic manifolds
to symplectic manifolds.
Improper covariant symplectic perturbations
deform almost-symplectic manifolds
to almost-symplectic manifolds.

Note that proper covariant symplectic perturbations are not closed on their own
unless $d\omega^\bullet = 0$.
If $d\omega^\bullet = 0$,
proper/improper covariant symplectic perturbations are respectively proper/improper symplectic perturbations.

In fact,
the name ``covariant symplectic perturbation''
will be fully justified
only when the almost-symplectic manifold $(\ps,\omega^\bullet)$
is equipped with a covariantizer $\varrho$,
defined below.

\enter

A \define{covariantizer}
on a smooth manifold $\ps$
is a unital graded algebra automorphism
$\varrho: \Omega^\blob(\ps) \to \Omega^\blob(\ps)$
of the $C^\infty(\ps)$-algebra $(\Omega^\blob(\ps),\wedge)$.
This means that $\varrho$ is
a $C^\infty(\ps)$-linear map on differential forms
that is
trivial on zero-forms,
degree-preserving,
and
invertible.

In particular, the action of $\varrho$
is \textit{completely determined 
by specifying its action on one-forms}.
Firstly, $\varrho$ is $C^\infty(\ps)$-linear and satisfies
$\varrho(f) \eqq f$ for all $f \inn C^\infty(\ps)$.
Secondly, its action is multiplicatively extended as 
$\varrho(\a^{(p)} \swedge \b^{(q)}) = \varrho(\a^{(p)}) \wedge \varrho(\b^{(q)})$
for all degrees $p$ and $q$,
where $\a^{(p)} \in \Omega^p(\ps)$ and $\b^{(q)} \in \Omega^q(\ps)$.

For each differential form $\a \in \Omega^p(\ps)$,
$\varrho(\a)$ and $\varrho^{-1}(\a)$
are referred to as the \define{covariant image} and \define{reverse covariant image} of $\a$,
respectively.

\enter

Generally speaking,
the action of a covariantizer $\varrho$ 
does not commute with the exterior derivative $d$:
$\varrho \hem\circ d \neq d \hhem\circ\nem \varrho$.
For instance, a covariantizer can be used to generate an almost-symplectic manifold
out of a symplectic manifold.
From this fact,
a concrete method arises for
constructing proper covariant symplectic perturbations.

Suppose a polarized symplectic manifold $(\ps,\omega^\circ)$
in which a symplectic potential (a one-form whose exterior derivative is the symplectic form)
$\theta^\circ$
is given in each patch.
Equip $(\ps,\omega^\circ)$ with a covariantizer $\varrho$ and 
let $\theta = \varrho(\theta^\circ)$.
Let $\omega = d\theta$
and define $\omega' = \omega - \omega^\bullet$
for $\omega^\bullet = \varrho(\omega^\circ)$.
From these manipulations,
a non-commutative diagram arises
between the actions of the covariantizer $\varrho$
and the exterior derivative $d$,
shown in \fref{cd-covspt}.

\newpage

The covariant image
$\omega^\bullet = \varrho(\omega^\circ)$,
is not necessarily closed.
Meanwhile, $d\omega = dd\theta = 0$ by construction.
Thus, $\omega'$ is a covariant symplectic perturbation
on the almost-symplectic manifold $(\ps,\omega^\bullet)$.

In particular,
we say that
$\omega'  \eqq [d,\varrho]\hem \theta^\circ$
is the \define{minimal covariant symplectic perturbation due to} 
the polarization and covariantizer data
$(\theta^\circ,\varrho)$
on $\ps$,
where
$[d,\varrho]$ denotes $ \varrho \hem\crc d - d \hhem\crc\nem \varrho$.
It will become self-evident soon that
this is the mathematical abstraction
of the \textit{minimal coupling} in physics.

\begin{figure}[t]
\begin{align*}
	\adjustbox{valign=c}{\begin{tikzpicture}
	    \node[empty] (O) at (0,0) {};
	    \node[empty] (X) at (4.1, 0) {};
	    \node[empty] (x) at (0.8, 0) {};
	    \node[empty] (Y) at (0, -1.9) {};
	    \node[b] (a) at ($(O)$) {$\theta^\circ$};
	    \node[b] (b) at ($(O)+(X)$) {$\theta$};
	    \node[b] (A) at ($(O)+(Y)$) {$\omega^\circ$};
	    \node[b] (B) at ($(O)+(X)+(Y)-(x)$) {$\omega^\bullet$};
	    \node[b] (B') at ($(O)+(X)+(Y)$) {};
	    \node[empty] (B'text) at ($(O)+(X)+(Y)+(x)+(0.1,0)$) {$\omega = \omega^\bullet + \omega'$};
	    \draw[->] (a)--(b) node[midway,above] {\scriptsize covariantize};
	    \draw[->] (A)--(B) node[midway,above] {\scriptsize covariantize};
	    \draw[->] (a)--(A) node[midway,left] {\scriptsize $d$};
	    \draw[->] (b)--(B') node[midway,right] {\scriptsize $d$};
	\end{tikzpicture}}
\end{align*}
	\caption{
		A non-commutative diagram
		arises from
		covariantization
		and
		exterior derivative.
	}
	\label{cd-covspt}
\end{figure}

\enter
Finally,
suppose a symplectic manifold $(\ps,\omega^\circ)$ equipped with
a covariantizer $\varrho$,
and let $\omega^\bullet = \varrho(\omega^\circ)$.
For any one-forms $\a,\b \in \Omega^1(\ps)$,
it holds that
\begin{align}
	\label{isom}
	\omega^\bullet{}^{-1}(\varrho(\a),\varrho(\b))
	\,=\,
		\omega^\circ{}^{-1}(\a,\b)
	\,.
\end{align}
\eqref{isom}
points out an isomorphic structure between
$\omega^\bullet{}^{-1}$ 
and
$\omega^\circ{}^{-1}$.

Let $\omega'$ be any covariant symplectic perturbation on $(\ps,\omega^\bullet)$,
so the pointwise inverse of $\omega = \omega^\bullet + \omega'$
is given in the bundle map notation as
\begin{align}
    \label{covspt}
    \omega^{-1} 
    \mem\,=\,\mem
	    \omega^\bullet{}^{-1}
	    \,-\,
            \omega^\bullet{}^{-1} \crc
            \omega' \crc
            \omega^\bullet{}^{-1}
	    \,+\,
            \omega^\bullet{}^{-1} \crc
            \omega' \crc
            \omega^\bullet{}^{-1} \crc
            \omega' \crc
            \omega^\bullet{}^{-1}
	    \,-\, \cdots
    \,\mem.
\end{align}
In local coordinates $\XQ^I$ on $\ps$,
\eqref{covspt} gives rise to the formula
\begin{align}
\label{covspt-elevator}
	&
	\omega^{-1}(\varrho(\a),\varrho(\b))
	\\
	&
	\,=\,
\begin{aligned}[t]
	&
		\omega^\circ{}^{-1}(\a,\b)
		- \omega^\circ{}^{-1}(\a,d\XQ^I) 
		\,(\varrho^{-1}\nem(\omega'))(\partial_I,\partial_J)\,
			\omega^\circ{}^{-1}(d\XQ^I,\b)
	\\
	&
		+ \omega^\circ{}^{-1}(\a,d\XQ^I) 
		\,(\varrho^{-1}\nem(\omega'))(\partial_I,\partial_J)\,
			\omega^\circ{}^{-1}(d\XQ^J,d\XQ^K)
		\,(\varrho^{-1}\nem(\omega'))(\partial_K,\partial_L)\,
			\omega^\circ{}^{-1}(d\XQ^L,\b)
		\\
		&
		- \cdots
	\,,
\end{aligned}
	\nonumber
\end{align}
which holds for all $\a,\b \in \Omega^1(\ps)$.
The isomorphism stated in \eqref{isom} plays a crucial role
in deriving \eqref{covspt}.
One considers the covariantized basis of one-forms,
$\varrho(d\XQ^I)$,
and uses the identity
$\omega' = \varrho(\varrho^{-1}(\omega')) = 
	\frac{1}{2}\, 
	(\varrho^{-1}(\omega'))(\partial_I,\partial_J)\,
	\varrho(d\XQ^I) \swedge \varrho(d\XQ^J)
$.

To provide a physical intuition,
the isomorphism in \eqref{isom} can be taken as 
a \textit{symplectic geometry implementation of Einstein's free-falling elevator},
mapping gauged dynamics to free dynamics locally.
In turn,
\eqref{covspt-elevator} emulates the curvature dynamics due to $\omega^{-1}$ in 
a flat background $\omega^\circ{}^{-1}$ via local frames.
In this context,
$(\varrho^{-1}\nem(\omega'))(\partial_I,\partial_J)$
in \eqref{covspt-elevator}
are the components of the covariant symplectic perturbation
in local frames.

In this setup,
a tentative definition of
\define{covariant Poisson bracket}
may be the bi-differential operator
$\pb{\blank}{\blank}_\cov : C^\infty(\ps) {\,\times\,} C^\infty(\ps) \to C^\infty(\ps)
: (f,g) \mapsto \pb{f}{g}_\cov
$, where
\begin{align}
	\label{covpb}
	\pb{f}{g}_\cov
	\,=\,
		\omega^{-1}(\varrho(df),\varrho(dg))
	\,.
\end{align}
We are obliged to clarify on
how the usual axioms of (almost-)Poisson bracket
are generalized.
It is easy that
$\R$-bilinearity and antisymmetry are retained.
The Leibniz property, 
$
\pb{fg}{h}_\cov
=
	\pb{f}{h}_\cov\, g
	+ f\, \pb{g}{h}_\cov
$,
is also straightforward from
the $C^\infty(\ps)$-linearity of $\varrho$.
Hence the discussion on the Jacobi identity only remains,
which we choose to perform in 
a revised setup
that will be introduced later.

%
%

%
%


Note that
\eqref{covspt-elevator} translates to
an expansion of
the covariant Poisson bracket:
\begin{align}
\label{covpb-expansion}
	\pb{f}{g}_\cov
	\,=\,
	&
		\pb{f}{g}^\circ
		- \pb{f}{\XQ^I}^\circ
		\,(\varrho^{-1}\nem(\omega'))(\partial_I,\partial_J)\,
			\pb{\XQ^J}{g}^\circ
	\\
	&
		+ \pb{f}{\XQ^I}^\circ
		\,(\varrho^{-1}\nem(\omega'))(\partial_I,\partial_J)\,
			\pb{\XQ^J}{\XQ^K}^\circ
		\,(\varrho^{-1}\nem(\omega'))(\partial_K,\partial_L)\,
			\pb{\XQ^L}{g}^\circ
		- \cdots
	\,.
	\nonumber
\end{align}
Physically speaking,
\eqref{covpb-expansion}
examines how interactions deform the Poisson structure
through the lens of Einstein's elevator.
It might be insightful to rewrite it as the following:
\begin{align}
	\label{cov-defl}
	\pb{f}{g}_\cov
	\,=\,
\begin{aligned}[t]
	&
	\BB{
		X_g^\circ
		- 
			\omega^\circ{}^{-1}
			\crc
			\varrho^{-1}\nem(\omega')
			\circ
			X_g^\circ
		+ \cdots
	}\act{
		f
	}
	\,.
\end{aligned}
\end{align}
The right-hand side describes an ordinary vector field acting on a smooth test function $f$.
Clearly, repeated insertions of the reverse image \smash{$\varrho^{-1}(\omega')$}
deflects the free-theory Hamiltonian vector field $X^\circ_g$
into a combination 
that yields the covariant Poisson bracket.
This describes the deflection of inertial trajectories
by tidal effects,
which is precisely the covariantized, Einstein-elevator version of
the interpretation given to \eqref{spt.X}.

It is the goal of \Sec{NONCOORD}
to substantiate these physical intuitions
through concrete examples,
which demonstrate and apply the abstract formalizations in this subsection.

\section{Covariant Yet Non-Coordinate Frames}
\label{NONCOORD}

\subsection{Gauge Theory}
\label{NONCOORD>YM}

\para{Covariantizer Construction}

Recall the free colored scalar particle,
defined on the phase space $\ps = T^*\mflat \moplus E$
via the symplectic potential $\theta^\circ$ in \eqref{colored.struct0}.
This $\theta^\circ$ is then promoted to $\theta$ in \eqref{YM.thetaA}
through the covariantization procedure  
familiar from standard physics textbooks,
which achieves
the coupling to the external gauge field $A \in \Omega^1(\mflat;\g)$.

This procedure
fits precisely into the universal grammar of minimal coupling
in \fref{cd-covspt}.
The promotion of $\theta^\circ$ to $\theta$ is mathematically formalized as the covariantizer
\begin{align}
	\label{YM.varrho}
	\varrho
	\,\,\,:\,\,\,
		\bigbig{
			dx^m
			,\mem
			dp_m
			,\mem
			d\vth^i
			,\mem
			d\bvth_i
		}
	\,\,\,\mapsto\,\,\,
		\bigbig{
			dx^m
			,\mem
			dp_m
			,\mem
			D\vth^i
			,\mem
			D\bvth_i
		}
	\,,
\end{align}
where 
$D$ denotes the covariant exterior derivative
so that
$D\vth^i = d\vth^i + A^i{}_j\mem \vth^j$
and $D\bvth_i = d\bvth_i - \bvth_j\mem A^j{}_i$.
Both sides in \eqref{YM.varrho}
define complete bases of one-forms on $\ps$,
ensuring that $\varrho$ is fully characterized as an invertible map.
Via \eqref{YM.varrho},
one finds
\begin{subequations}
\label{YM.cd}
\begin{align}
\label{YM.cd1}
	\theta^\circ
	\,&=\,
		p_m\mem dx^m
		+ i\mem \bvth_i\mem d\vth^i
	\,,\\
\label{YM.cd2}
	\theta
	\,=\,
		\varrho(\theta^\circ)
	\,&=\,
		p_m\mem dx^m
		+ i\mem \bvth_i\mem D\vth^i
	\,,\\
\label{YM.cd3}
	\omega^\bullet
	\,=\,
		\varrho(\omega^\circ)
	\,&=\,
		dp_m \wedge dx^m
		+ i\mem D\bvth_i \wedge D\vth^i
	\,,\\
\label{YM.cd4}
	\omega
	\,=\,
		d\theta
	\,&=\,
		dp_m \wedge dx^m
		+ i\mem D\bvth_i \wedge D\vth^i
		+ q_a\hem F^a
	\,,
\end{align}
\end{subequations}
which identifies the proper covariant symplectic perturbation
$\omega' = q_a\hem F^a$:
\begin{align}
	\label{YM.omega'}
	\varrho^{-1}(\omega')
	\,=\,
		\tfrac{1}{2}\,
			q_a\mem
			F^a{}_{mn}(x)\,
		dx^m \swedge dx^n
	\,.
\end{align}

\newpage

To elaborate,
the derivation of \eqref{YM.cd4}
is easy
if one notes the fact that
the actions of the ordinary and covariant exterior derivatives,
$d$ and $D$,
are identical on singlet-valued differential forms.
In particular,
\begin{align}
	\label{YM.mech}
    d\hem\BB{
        i\mem \bvth_i\mem D\vth^i
    }
    \,=\,
    D\hhem\BB{
        i\mem \bvth_i\mem D\vth^i
    }
    \,&=\,
        i\mem D\bvth_i \wedge D\vth^i
        + i\mem \bvth_i\mem D^2\vth^i
    \,.
\end{align}
Via $D^2\vth^i \eqq F^i{}_j\mem \vth^j$
and $q_a \eqq i\mem \bvth_i\mem (t_a)^i{}_j\mem \vth^j$,
one directly derives the covariant symplectic perturbation
$\omega' \eqq q_a\hem F^a$
in \eqref{YM.cd4}.
That is,
the very mechanism that generates the covariant symplectic perturbation
for gauge theory backgrounds
is the squaring $D^2$ of the covariant exterior derivative.

\para{Covariant Poisson Brackets}


Crucially,
\eqref{isom}
describes that
the free-theory Poisson brackets in \eqref{colored.pb0}
are \textit{directly recycled}
to the interacting theory as
\begin{align}
    \label{YM.pb00}
    \pib{dx^m}{dp_n}
    \,=\,
        \delta^m{}_n
    \,,\quad
    \pib{D\vth^i}{D\bvth_j}
    \,=\,
        -i\mem \delta^i{}_j
    \,.
\end{align}
Any other components of the almost-Poisson bivector $\omega^\bullet{}^{-1}$
vanish
when examined in the complete basis
$(dx^m,dp_m,D\vth^i,D\bvth_i)$ of one-forms.

With this understanding,
the covariant Poisson bracket in \eqref{covpb}
is readily obtained by
the expansion in \eqref{covpb-expansion}:
\begin{align}
\begin{split}
    \label{YM.covpb}
    \pb{x^m}{p_n}_\cov
    \,=\,
        \delta^m{}_n
    \,,\quad
    \pb{\vth^i}{\bvth_j}_\cov
    \,=\,
        -i\mem \delta^i{}_j
    \,,\quad
    \pb{p_m}{p_n}_\cov
    \,=\,
        q_a\hem F^a{}_{mn}(x)
    \,,
\end{split}
\end{align}
while other combinations, such as
$\pb{\vth^i}{p_r}_\cov$, all vanish.
Compare this with \eqref{YM.pb}.
Gauge covariance is manifest.
Bare connection terms do not occur.

To elaborate, the expansion in \eqref{covpb-expansion}
has described
\begin{align}
\begin{split}
	\label{YM.seesaw}
	\pb{p_m}{p_n}_\cov
	\,&=\,
		- \pb{p_m}{x^r}^\circ\hem
	    	{\hem\,q_a\hem F^a{}_{rs}(x)\,\hhem}
		\hem\pb{dx^s}{dp_n}^\circ
	\,=\,
		q_a\hem F^a{}_{mn}(x)
	\,,
\end{split}
\end{align}
precisely reincarnating 
the gymnastics in \eqref{EM.seesaw}
in the covariantized frame 
$(dx^m,dp_m,$ $D\vth^i,D\bvth_i)$
for non-abelian gauge theory.

\para{Dual Basis}

One may desire to find the bivectors $\omega^\bullet{}^{-1}$ and $\omega^{-1}$ explicitly.
To this end,
note that
$(\tpartial_m,\partial/\partial p_m,\partial/\partial \vth^i,\partial/\partial \bvth_i)$
serves as the \textit{dual basis} to 
$(dx^m,dp_m,D\vth^i,D\bvth_i)$,
where $\tpartial_m$ is defined by
\begin{align}
	\label{YM.hor}
	\tpartial_r
	\,=\,
		\frac{\partial}{\partial x^r}
		- \overleftarrow{
			\frac{\partial}{\partial \vth^i}
		}\,
			A^i{}_{jr}(x)\mem \vth^j
		+ \bvth_j\mem A^j{}_{ir}(x)\,
		\overrightarrow{
			\frac{\partial}{\partial \bvth_i}
		}
	\,.
\end{align}
This implies that \eqref{YM.pb00} is inverted as
\begin{align}
	\label{YM.ob}
	\omega^\bullet{}^{-1}
	\,=\,
		\tpartial_m \wedge \frac{\partial}{\partial p_m}
		-i\,
			\overleftarrow{\frac{\partial}{\partial \vth^i}}
		\wedge
			\overrightarrow{\frac{\partial}{\partial \bvth_i}}
	\,.
\end{align}
By \eqref{covspt},
it then follows that
\begin{align}
	\label{YM.om}
	\omega^{-1}
	\,=\,
		\tpartial_m \wedge \frac{\partial}{\partial p_m}
		-i\,
			\overleftarrow{\frac{\partial}{\partial \vth^i}}
		\wedge
			\overrightarrow{\frac{\partial}{\partial \bvth_i}}
		+ \frac{1}{2}\,
			q_a\hem F^a{}_{mn}(x)\,
		\frac{\partial}{\partial p_m}
			\wedge
		\frac{\partial}{\partial p_n}
	\,.
\end{align}
The basis $(\tpartial_m,\partial/\partial p_m,\partial/\partial \vth^i,\partial/\partial \bvth_i)$ of vector fields
must be gauge-covariant,
since the basis $(dx^m,dp_m,D\vth^i,D\bvth_i)$ of one-forms
is gauge-covariant.
Therefore, \eqref{YM.om}
is a manifestly gauge-covariant
representation of the interacting-theory Poisson bivector $\omega^{-1}$.

\newpage

\para{Ehresmann Connection}

In particular, it is left as an exercise to check that the vector field $\tpartial_r \inn \Gamma(T\ps)$ in \eqref{YM.hor}
is \textit{invariant} under the non-abelian gauge transformations.
In fact, the astute reader will point out that
it is the \textit{horizontal lift} of the spacetime vector field $\partial/\partial x^r \inn \Gamma(T\mflat)$
with respect to the
non-abelian gauge connection $A \in \Omega^1(\mflat;\g)$.

To elaborate,
the Ehresmann \cite{ehresmann1948connexions}
notion of a connection
on a fiber bundle $\pi : E \to \mflat$
is the direct sum decomposition
$TE = H \oplus V$
into horizontal and vertical bundles,
where $V = \ker(d\pi : TE \to T\mflat)$
is the vertical bundle.
For each $e \inn E$,
the horizontal subspace $H_e$ is a vector subspace of $T_e E$
so that $T_eE = H_e + V_e$.
If $\pi(e) = x$ for $e \inn E$,
the horizontal lift of a vector $v \in T_x\mflat$
is the unique element of $T_eE$
whose vertical components are zero.

In our case, the phase space $\ps = T^*E$ is also a fiber bundle over $\mflat$,
and the span of $\tpartial_r$ in \eqref{YM.hor}
defines the horizontal subspace at each point in $\ps$.
To see why this reincarnates the more familiar definitions of a connection,
consider the following computation:
\begin{align}
	\tpartial_r\act{
		q_a\hem F^a{}_{mn}(x)
	}
	\,=\,
		q_a\hem
		\BB{
			\partial_r F^a{}_{mn}(x)
			+ f^a{}_{cb}\mem A^c{}_r(x)\mem F^b{}_{mn}(x)
		}
	\,=\,
		q_a\hem D_r F^a{}_{mn}(x)
	\,.
\end{align}
In this manner, 
the horizontal derivative
\smash{$\tpartial_r$} 
implements
the familiar $D_r$.
Note also that
\begin{align}
	\comm{\tpartial_r}{\tpartial_s}
	\,=\,
		- \overleftarrow{
			\frac{\partial}{\partial \vth^i}
		}\,
			F^i{}_{jrs}(x)\mem \vth^j
		+ \bvth_j\mem F^j{}_{irs}(x)\,
		\overrightarrow{
			\frac{\partial}{\partial \bvth_i}
		}
	\quad\xleftrightarrow[]{\,\,\,\,\,\,}\quad
	\comm{D_r}{D_s}
	\,\sim\, F_{rs}
	\,.
\end{align}

\subsection{Gravity}
\label{NONCOORD>BI}

\para{Covariantizer Construction}

Next, we revisit the general-relativistic scalar particle in \Sec{NONCAN>BI},
formulated on the phase space $\ps = \Tstar\M$.
The issue with \eqref{BI.structA}
was that local Lorentz covariance is not manifest.
To resolve this problem,
we are free to employ \textit{any} Lorentz-valued (metric-preserving) connection $\tD$.
Since the symplectic potential $\theta = p_m\mem e^m$
is a differential one-form
valued in the singlet of local Lorentz representations,
its exterior derivative can be computed as
\begin{align}
	\label{BI.tDom}
	\omega
	\,=\,
		d\theta
	\,=\,
		\tD\theta
	\,=\,
		\tD p_m \wedge e^m
		+ p_m\mem \tD e^m
	\,.
\end{align}
This manifests invariances under
local Lorentz and spacetime coordinate transformations.

The Einstein equivalence principle asserts that
part of \eqref{BI.tDom} must be isomorphic to the free theory's $\omega^\circ$.
Indeed, the relevant covariantizer is
\begin{align}
	\label{BI.varrho}
	\varrho
	\,\,\,:\,\,\,
		\bigbig{
			dx^m
			,\mem
			dp_m
		}
	\,\,\,\mapsto\,\,\,
		\bigbig{
			e^m
			,\mem
			\tD p_m
		}
	\,,
\end{align}
which appeals to $\Tstar\M \cong T^*\mflat$ within local patches.
Both sides in \eqref{BI.varrho}
define complete bases of one-forms on $\ps \eqq \Tstar\M$,
while $x^m := \delta^m{}_\m\mem x^\m$.
Via \eqref{BI.varrho},
one finds
\begin{subequations}
\label{BI.cd}
\begin{align}
\label{BI.cd1}
	\theta^\circ
	\,&=\,
		p_m\mem dx^m
	\,,\\
\label{BI.cd2}
	\theta
	\,=\,
		\varrho(\theta^\circ)
	\,&=\,
		p_m\mem e^m
	\,,\\
\label{BI.cd3}
	\omega^\bullet
	\,=\,
		\varrho(\omega^\circ)
	\,&=\,
		\tD p_m \wedge e^m
	\,,\\
\label{BI.cd4}
	\omega
	\,=\,
		d\theta
	\,&=\,
		\tD p_m \wedge e^m
		+ p_m\hem \tT^m
	\,,
\end{align}
\end{subequations}
which identifies the proper covariant symplectic perturbation
$\omega' = p_m\hem \tT^m$.
\begin{align}
	\label{BI.omega'}
	\varrho^{-1}(\omega')
	\,=\,
		\tfrac{1}{2}\,
			p_k\mem
			\tT^k{}_{mn}(x)\,
		dx^m \swedge dx^n
	\,.
\end{align}
Here, we have identified the torsion two-form $\tT^m = \tD e^m$.
Now the mappings
$\varrho : \theta^\circ \mapsto \theta$
and
$\varrho : \omega^\circ \mapsto \omega$
literally implement Einstein's free-falling elevator
as local Lorentz frames.

\newpage

\para{Covariant Poisson Brackets}


Via the identity in
\eqref{isom},
the free-theory Poisson brackets in \eqref{scalar.pb0}
are reincarnated in local Lorentz frames as
\begin{align}
    \label{BI.pb00}
    \pib{e^m}{e^n}
    \,=\,
        0
    \,,\quad
    \pib{e^m}{\tD p_n}
    \,=\,
        \delta^m{}_n
    \,,\quad
    \pib{\tD p_m}{\tD p_n}
    \,=\,
        0
    \,.
\end{align}
Via the expansion in \eqref{covpb-expansion},
the covariant Poisson bracket in \eqref{covpb} describes
\begin{align}
\begin{split}
    \label{BI.covpb}
    \pb{x^m}{x^n}_\cov
    \,=\,
        0
    \,,\quad
    \pb{x^m}{p_n}_\cov
    \,=\,
    	\delta^m{}_n
    \,,\quad
    \pb{p_m}{p_n}_\cov
    \,=\,
    	p_k\mem \tT^k{}_{mn}(x)
    \,.
\end{split}
\end{align}
Compare this with \eqref{BI.pb}.
Unlike the anholonomy coefficients $\Omega^k{}_{mn}(x)$,
the torsion $\tT^k{}_{mn}(x)$ is a tensor.

To elaborate, the expansion in \eqref{covpb-expansion}
has described
\begin{align}
\begin{split}
	\label{BI.seesaw}
	\pb{p_m}{p_n}_\cov
	\,&=\,
		- \pb{p_m}{x^r}^\circ\mem
	    	{\hem\,p_k\hem \tT^k{}_{rs}(x)\,\hhem}
		\hem\pb{x^r}{p_n}^\circ
	\,=\,
		p_k\hem \tT^k{}_{mn}(x)
	\,,
\end{split}
\end{align}
precisely reincarnating 
the gymnastics in \eqref{EM.seesaw}
in the one-form basis $(e^m,\tD p_m)$.

\para{Dual Basis}

One may desire to find the bivectors $\omega^\bullet{}^{-1}$ and $\omega^{-1}$ explicitly.
To this end,
note that
$(\tE_m,\partial/\partial p_m)$
serves as the \textit{dual basis} to 
$(e^m,\tD p_m)$,
where $\tE_m$ is defined by
\begin{align}
	\label{BI.hor}
	\tE_r
	\,=\,
		E^\r{}_r(x)\mem
		\bb{
			\frac{\partial}{\partial x^\r}
			+ 
				p_n\mem \tgamma^n{}_{mr}(x)\,
				\frac{\partial}{\partial p_m}
		}
	\,.
\end{align}
Here, $\tgamma^n{}_{mr}(x)$ are the connection coefficients of $\tD$.
As a result, \eqref{BI.pb00} is inverted as
\begin{align}
	\label{BI.ob}
	\omega^\bullet{}^{-1}
	\,=\,
		\tE_m \wedge \frac{\partial}{\partial p_m}
	\,.
\end{align}
By \eqref{covspt},
it then follows that
\begin{align}
	\label{BI.om}
	\omega^{-1}
	\,=\,
		\tE_m \wedge \frac{\partial}{\partial p_m}
		+ \frac{1}{2}\,
			p_k\mem \tT^k{}_{mn}(x)\,
		\frac{\partial}{\partial p_m}
			\wedge
		\frac{\partial}{\partial p_n}
	\,.
\end{align}
The basis $(\tE_m,\partial/\partial p_m)$ of vector fields
must be covariant,
since the basis $(e^m,\tD p_m)$ of one-forms
is covariant.
Therefore, \eqref{BI.om}
is a manifestly gauge-covariant
representation of the interacting-theory Poisson bivector $\omega^{-1}$.

\para{Ehresmann Connection}

Again, it is left as an exercise to check that the vector field $\tE_r \inn \Gamma(T\ps)$ in \eqref{YM.hor}
is covariant under local Lorentz and spacetime coordinate transformations.
It is exactly the horizontal lift of the spacetime vielbein
$E_r \in \Gamma(T\M)$
with respect to the connection $\tD$.
Note the Lie bracket
\begin{align}
	\comm{\tE_r}{\tE_s}
	\,&=\,
		\Omega^k{}_{rs}(x)\,
			\tE_k
		+ p_m\mem \tR^m{}_{nrs}(x)\mem \frac{\partial}{\partial p_n}
	\,,
\end{align}
where $\tR^m{}_{nrs}(x)$ denotes the curvature tensor for $\tD$.

\para{Teleparallel Torsion as Gravitational Field Strength}

Amusingly, the parallel between the covariant symplectic perturbations in
\eqrefs{YM.omega'}{BI.omega'}
suggests that
the gravitational charge is the momentum $p_k$ ($\leftrightarrow q_a$)
while the gravitational field strength is the torsion $\tT^k{}_{mn}(x)$ ($\leftrightarrow F^a{}_{mn}(x)$).


Note that we have a complete freedom to choose the Lorentz-valued connection $\tD$
while not changing the physical content of the equations of motion.

\newpage

In standard Einstein gravity,
spacetime geometry
is dictated by the (dynamical) metric $g$.
As a result,
the only physically meaningful connection is often
the Levi-Civita connection, say $D$.
If we choose $\tD = D$, the torsion vanishes as $De^m = 0$,
so there is no covariant symplectic perturbation.
Hence there is no ``Lorentz force'' in the elevator.

If one is willing to accept
an auxiliary (background) connection $\tD$,
then one has employed a reference structure.
In this case, the covariant symplectic perturbation is non-vanishing
and gives rise to a spurious ``Lorentz force.''

A special case is a torsionful yet flat connection,
in particular $\tD = d$.
This is the teleparallel connection,
also known as Weitzenb\"ock connection
\cite{cartan1979letters}.
This is again a reference structure if one's goal is to describe Einstein gravity.

However, the interpretation changes if one's goal is to describe Born-Infeld theory \cite{born1934foundations}.
We point out that
classical Born-Infeld theory can be viewed as a field theory of a dynamical vielbein,
giving rise to a teleparallel geometry
where the Lorentz indices $m,n,r,s,\cdots$ are not gauged.
This is based on a new perspective on Born-Infeld theory 
due to color-kinematics duality \cite{cck}.
With the choice $\tD = d$,
the components of \eqref{BI.omega'} are
\begin{subequations}
\begin{align}
	\label{dc:BI.deriv1}
	p_k\mem
		\tT^k{}_{mn}
	\,=\,
	\pcan_\m\mem
		E^\m{}_k\mem
		\tT^k{}_{mn}
	\,=\,
	(-\pcan_\m)\mem
		\comm{E_m}{E_n}^\m
	\,,
\end{align}
where $\comm{E_m}{E_n}^\m = E^\r{}_m\mem \partial_\r E^\m{}_n - E^\r{}_n\mem \partial_\r E^\m{}_m$
computes the Lie bracket between the orthonormal frames.
Consider an expansion around a background,
$E^\m{}_m(x) = \delta^\m{}_m + A^\m{}_m(x)$.
Then \eqref{dc:BI.deriv1} boils down to
\begin{align}
	\label{dc:BI.deriv2}
	(-\pcan_\m)\mem
		\BB{
			\partial_m A^\m{}_n
			- \partial_n A^\m{}_m
			+ A^\r{}_m\mem \partial_\r A^\m{}_n
			- A^\r{}_n\mem \partial_\r A^\m{}_m
		}
	\,=\,
		(-\pcan_\m)\mem F^\m{}_{mn}
	\,.
\end{align}
\end{subequations}
Here, $F^\m{}_{mn}$ is literally the field strength
of a diffeomorphism-valued gauge potential $A^\m{}_m$,
which is known to reformulate Born-Infeld theory 
via the color-to-kinematics replacement
\cite{cck}.
Therefore, we find that
the framework of
covariant symplectic perturbations
provides a unique perspective that
identifies an exact double copy correspondence
\begin{align}
	\label{dc:BI}
	q_a\mem F^a{}_{mn}(x)
	\quad\xleftrightarrow[]{\,\,\,\,\,\,}\quad
	(-\pcan_\m)\hem F^\m{}_{mn}(x)
	\,,
\end{align}
mapping \textit{color generator} $q_a$ to \textit{diffeomorphism generator} $(-\pcan_\m)$
(the minus sign here implements the left action convention).
Recall the familiar formula
\begin{align}
	q_a\mem
		\BB{
			\partial_m A^a{}_n
			- \partial_n A^a{}_m
			+ f^a{}_{bc}\mem A^b{}_m\mem A^c{}_n
		}
	\,=\,
	q_a\mem
		F^a{}_{mn}
	\,.
\end{align}

Note that
an exact correspondence also arises between
the equations of motion of Yang-Mills theory and Born-Infeld theory
in terms of an equation $\pb{\pb{p_m}{p_n}}{p^n} = 0$
in the particle phase spaces.
Again, we had discovered this fact through the lens of symplectic perturbations;
we plan to elaborate more on in forthcoming work \cite{feynman}.
A brief outline was given in \rcite{dpb}.

Having witnessed this exact correspondence,
it is natural to ask the more ambitious question
on the double copy correspondence between Yang-Mills theory and Einstein gravity.
The above exploration
will identify the momentum as the gravitational charge,
which represents a diffeomorphism (local translation) charge.
It may be the case that we have to incorporate the other gravitational charge as well:
spin as a local Lorentz charge.

\newpage

\subsection{Gravity with Spin}
\label{NONCOORD>GR}

\para{Spinning Particle}

Consider the direct sum bundle
$\Aflat_2 = (T^* \moplus \Pi T^\C)\hem \mflat$
over flat spacetime
$\mflat = (\R^d,\eta)$.
This is a space with linear coordinates
$(x^m,p_m,\psi^m)$,
where $x^m$ and $p_m$ are real bosonic
while $\psi^m$ are complex fermionic
such that $[\psi^m]^* = \bpsi^m$.
It serves as the phase space for a massless spinning particle
in flat spacetime.
The symplectic structure is
\begin{align}
	\label{spinning.struct0}
	\theta^\circ
	\,=\,
		p_m\mem dx^m
		+ i\mem \bpsi_m\mem d\psi^m
	\qiq
	\omega^\circ
	\,=\,
		dp_m \wedge dx^m
		+ i\mem d\bpsi_m \wedge d\psi^m
	\,,
\end{align}
which gives rise to the canonical Poisson brackets
\begin{align}
    \label{spinning.pb0}
    \pb{x^m}{p_n}^\circ
    \,=\,
        \delta^m{}_n
    \,,\quad
    \pb{\psi^m}{\bpsi_n}^\circ
    \,=\,
        -i\mem \delta^m{}_n
    \,.
\end{align}
Two supersymmetry generators
act on this phase space,
from which the Hamiltonian arises.

\para{Covariant Coordinates}

To construct the general-relativistic theory of
this spinning particle,
take the direct sum bundle
$\A_2 = (T^* \moplus \Pi T^\C)\hem \M$
over $\M = (\R^d,g)$.
Local trivializations
due to an orthonormal frame
provides
local coordinates
$(x^\m,p_m,\psi^m)$
on $\A_2$.
The symplectic potential on $\A_2$ is
\begin{align}
    \label{GR.struct}
    \theta
    \,=\,
        p_m\mem e^m
        + i\mem \bpsi_m\mem D\psi^m
    \qiq
   	\omega
   	\,=\,
   		Dp_m \wedge e^m
   		+ i\mem D\bpsi_m \wedge D\psi^m
   		+ i\mem \bpsi_m\hem R^m{}_n\mem \psi^n
    \,,
\end{align}
where $e^m$ is the orthonormal coframe
and
$D$ is the covariant exterior derivative
with respect to the Levi-Civita connection.
For reference, the structure equations are
\begin{align}
	\label{ec}
	0 \,=\,
		De^m
	\,=\,
		de^m + \gamma^m{}_n \wedge e^n
	\,,\quad
	R^m{}_n
	\,=\,
		d\gamma^m{}_n + \gamma^m{}_k \swedge \gamma^k{}_n
	\,,
\end{align}
where $\gamma \in \Omega^1(\M,\so(1,d{\,-\mem}1))$ is the spin connection,
and $R \in \Omega^2(\M,\so(1,d{\,-\mem}1))$ is the Riemann curvature two-form.
For instance, $D\psi^m = d\psi^m + \gamma^m{}_n\mem \psi^n  = d\psi^m + \gamma^m{}_{n\r}(x)\mem \psi^n\mem dx^\r$.
Certainly, 
$(x^\m,p_m,\psi^m)$
serves as covariant coordinates on the phase space $\A_2$.

\para{Covariantizer Construction}

Again, 
$\theta$ in \eqref{GR.struct}
is isomorphic to
$\theta^\circ$ in \eqref{spinning.struct0}
by covariantization.
Similarly,
part of $\omega$ in \eqref{GR.struct}
is isomorphic to 
$\omega^\circ$ in \eqref{spinning.struct0}
by covariantization.
These observations are exactly the content of the Einstein equivalence principle
in general relativity.
They are mathematically formalized as the covariantizer
\begin{align}
	\label{GR.varrho}
	\varrho
	\,\,\,:\,\,\,
		\bigbig{
			dx^m
			,\mem
			dp_m
			,\mem
			d\psi^m
			,\mem
			d\bpsi_m
		}
	\,\,\,\mapsto\,\,\,
		\bigbig{
			e^m
			,\mem
			Dp_m
			,\mem
			D\psi^m
			,\mem
			D\bpsi_m
		}
	\,,
\end{align}
which appeals to $\A_2 \cong \Aflat_2$ within local patches.
Provided \eqref{GR.varrho},
one identifies
\begin{subequations}
\label{GR.cd}
\begin{align}
\label{GR.cd1}
	\theta^\circ
	\,&=\,
		p_m\mem dx^m
		+ i\mem \bpsi_m\mem d\psi^m
	\,,\\
\label{GR.cd2}
	\theta
	\,=\,
		\varrho(\theta^\circ)
	\,&=\,
		p_m\mem e^m
		+ i\mem \bpsi_m\mem D\psi^m
	\,,\\
\label{GR.cd3}
	\omega^\bullet
	\,=\,
		\varrho(\omega^\circ)
	\,&=\,
		Dp_m \wedge e^m
		+ i\mem D\bpsi_m \wedge D\psi^m
	\,,\\
\label{GR.cd4}
	\omega
	\,=\,
		d\theta
	\,&=\,
		Dp_m \wedge e^m
		+ i\mem D\bpsi_m \wedge D\psi^m
		- \tfrac{1}{2}\mem S_{mn}\hem R^{mn}
	\,,
\end{align}
\end{subequations}
so the proper covariant symplectic perturbation is
$\omega' = -\frac{1}{2}\mem S_{mn}\hem R^{mn}$:
\begin{align}
	\label{GR.omega'}
	\varrho^{-1}(\omega')
	\,=\,
		- \tfrac{1}{4}\,
			S_{mn}\mem
			R^{mn}{}_{rs}(x)\,
		dx^r \swedge dx^s
	\,.
\end{align}
Here, the particle's spin angular momentum is identified as
the Lorentz generator
(a composite variable valued in the coadjoint),
paralleling the color generator in \eqref{q}:
\begin{align}
	\label{S}
	S_\wrap{mn}
	\,=\,
		- 2 i\mem \bpsi_\wrap{[m} \psi_\wrap{n]}
	\,.
\end{align}

\newpage

Notice also the parallels between
\eqrefs{YM.varrho}{GR.varrho}
and also \eqrefs{YM.cd}{GR.cd}.
Our exposition above
views gravity as a gauge theory of the Lorentz group,
which lies within the spirit of covariant color-kinematics duality \cite{cck}.
The very mechanism that generates the covariant symplectic perturbation
$\omega' = -\frac{1}{2}\mem S_{mn}\hem R^{mn}$
is the squaring $D^2$ of the covariant exterior derivative,
so \eqref{YM.mech} is precisely paralleled as
\begin{align}
	\label{GR.mech}
    d\hem\BB{
        i\mem \bpsi_m\mem D\psi^m
    }
    \,=\,
    D\hhem\BB{
        i\mem \bpsi_m\mem D\psi^m
    }
    \,&=\,
        i\mem D\bpsi_m \wedge D\psi^m
        + i\mem \bpsi_m\mem D^2\psi^m
    \,.
\end{align}
Namely,
$
	i\mem \bvth_i\mem F^i{}_j\mem \vth^j
		= q_a\hem F^a 
	\,\leftrightarrow\,
	i\mem \bpsi_m\mem R^m{}_n\mem \psi^n
		= -\frac{1}{2}\mem S_{mn}\hem R^{mn}
$.

Note that we could also have employed the auxiliary Lorentz-valued connection $\tD$,
in which case the covariant symplectic perturbation develops a torsion term as well.

\para{Poisson Brackets}

On a related note,
computation shows that
the Poisson brackets between $(x^\m,p_m,\psi^m)$
are given by
\begin{align}
\begin{split}
	\label{GR.pb}
	\pb{x^\m}{p_r}
	\,&=\,
		E^\m{}_r(x)
	\,,\\
	\pb{p_r}{p_s}
	\,&=\,
		- p_k\mem \Omega^k{}_{rs}(x)
		- \tfrac{1}{2}\, S_{mn}\mem R^{mn}{}_{rs}(x)
	\,,\\
	\pb{\psi^m}{p_r}
	\,&=\,
    	- \gamma^m{}_{nr}(x)\mem \psi^n
    \,,
\end{split}
\end{align}
combining the features of
\eqrefs{YM.pb}{BI.pb}.
Crucially,
despite the covariance of the coordinates $(x^\m,p_m,\psi^m)$,
the Poisson brackets in \eqref{GR.pb}
do not manifest covariance.

\para{Covariant Poisson Brackets}

Meanwhile,
the isomorphism in \eqref{isom}
implies that
the free-theory Poisson brackets in \eqref{spinning.pb0}
are reincarnated in local Lorentz frames as
\begin{align}
    \label{GR.pb00}
    \pib{e^m}{\tD p_n}
    \,=\,
        \delta^m{}_n
    \,,\quad
    \pib{\tD \psi^m}{\tD \bpsi_n}
    \,=\,
        -i\mem \delta^m{}_n
    \,.
\end{align}
Via the expansion in \eqref{covpb-expansion},
the covariant Poisson bracket in \eqref{covpb} describes
\begin{align}
\begin{split}
    \label{GR.covpb}
    \pb{x^m}{p_n}_\cov
    \,&=\,
    	\delta^m{}_n
	\,,\\
	\pb{\psi^m}{\bpsi_n}_\cov
	\,&=\,
    	-i\mem \delta^m{}_n
    \,,\\
    \pb{p_m}{p_n}_\cov
    \,&=\,
    	- \tfrac{1}{2}\, S_{kl}\mem R^{kl}{}_{mn}(x)
    \,,
\end{split}
\end{align}
eliminating the non-tensorial objects in \eqref{GR.pb}.

To elaborate,
the expansion in \eqref{covpb-expansion}
has described
\begin{align}
\begin{split}
	\label{BI.seesaw}
	\pb{p_m}{p_n}_\cov
	\,&=\,
		- \pb{p_m}{x^r}^\circ\hem
	    \,\BB{
	    	- \tfrac{1}{2}\, S_{kl}\mem R^{kl}{}_{rs}(x)
	    }\,
		\hem\pb{x^s}{p_n}^\circ
	\,=\,
		- \tfrac{1}{2}\, S_{kl}\mem R^{kl}{}_{mn}(x)
	\,,
\end{split}
\end{align}
precisely reincarnating 
the gymnastics in \eqref{EM.seesaw}
in Einstein's elevator.

\para{Dual Basis}

One may desire to find the bivectors $\omega^\bullet{}^{-1}$ and $\omega^{-1}$ explicitly.
To this end,
note that
$(\tE_m,\partial/\partial p_m,\partial/\partial \psi^m,\partial/\partial \bpsi_m)$
serves as the \textit{dual basis} to 
$(e^m,Dp_m,D\psi^m,D\bpsi_m)$,
where $\tE_m$ is defined by
\begin{align}
	\label{GR.hor}
	\tE_r
	\,=\,
		E^\r{}_r(x)\mem
		\bb{
			\frac{\partial}{\partial x^\r \vphantom{\bpsi^0}}
			+ 
				p_n\mem \gamma^n{}_{mr}(x)\,
				\frac{\partial}{\partial p_m \vphantom{\bpsi^0}}
			- \overleftarrow{
				\frac{\partial}{\partial \psi^m \vphantom{\bpsi^0}}
			}\,
				\gamma^m{}_{nr}(x)\mem \psi^n
			+ \bpsi_n\mem \gamma^n{}_{mr}(x)\,
			\overrightarrow{
				\frac{\partial}{\partial \bpsi_m \vphantom{\bpsi^0}}
			}
		}
	\,,
\end{align}
combining the features of \eqrefs{YM.hor}{BI.hor}.
As a result, \eqref{GR.pb00} is inverted as
\begin{align}
	\label{GR.ob}
	\omega^\bullet{}^{-1}
	\,=\,
		\tE_m \wedge \frac{\partial}{\partial p_m\vphantom{\bpsi^0}}
		-i\,
			\overleftarrow{\frac{\partial}{\partial \psi^m}\vphantom{\bpsi^0}}
		\wedge
			\overrightarrow{\frac{\partial}{\partial \bpsi_m}\vphantom{\bpsi^0}}
	\,.
\end{align}
\newpage\noindent
By \eqref{covspt},
it then follows that
\begin{align}
	\label{GR.om}
	\omega^{-1}
	\,=\,
		\tE_m \wedge \frac{\partial}{\partial p_m\vphantom{\bpsi^0}}
		-i\,
			\overleftarrow{\frac{\partial}{\partial \psi^m}\vphantom{\bpsi^0}}
		\wedge
			\overrightarrow{\frac{\partial}{\partial \bpsi_m}\vphantom{\bpsi^0}}
		- \frac{1}{4}\,
			S_{kl}\mem R^{kl}{}_{mn}(x)\,
		\frac{\partial}{\partial p_m}
			\wedge
		\frac{\partial}{\partial p_n}
	\,.
\end{align}
The basis $(\tE_m,\partial/\partial p_m,\partial/\partial \psi^m,\partial/\partial \bpsi_m)$ of vector fields
must be covariant,
since the basis $(e^m,Dp_m,D\psi^m,D\bpsi_m)$ of one-forms
is covariant.
Therefore, \eqref{GR.om}
is a manifestly gauge-covariant
representation of the interacting-theory Poisson bivector $\omega^{-1}$.

\para{Ehresmann Connection}

Again, it is left as an exercise to check that the vector field $\tE_r \inn \Gamma(T\ps)$ in \eqref{YM.hor}
is covariant under local Lorentz and spacetime coordinate transformations.
It is exactly the horizontal lift of the spacetime vielbein
$E_r \in \Gamma(T\M)$
with respect to the Levi-Civita connection $D$.
The Lie bracket $\comm{\tE_r}{\tE_s}$ computes to
\begin{align}
\begin{split}
		\Omega^k{}_{rs}(x)\,
			\tE_k
		+ p_m\mem R^m{}_{nrs}(x)\mem \frac{\partial}{\partial p_n}
		- \overleftarrow{
			\frac{\partial}{\partial \psi^m \vphantom{\bpsi^0}}
		}\,
			R^m{}_{nrs}(x)\mem \psi^n
		+ \bpsi_n\mem R^n{}_{mrs}(x)\,
		\overrightarrow{
			\frac{\partial}{\partial \bpsi_m \vphantom{\bpsi^0}}
		}
	\,.
\end{split}
\end{align}

\para{Canonical Coordinates}


Since $(\A_2,\omega)$ is a symplectic manifold,
the Darboux theorem applies.
Explicitly, 
the canonical momentum can be identified as
\begin{align}
	\label{GR.pcan}
	\pcan_\m
	\,=\,
		p_m\mem e^m{}_\m(x)
		- \frac{1}{2}\, S_{mn}\mem \gamma^{mn}{}_\m(x)
	\qiq
	\theta
	\,=\,
		\pcan_\m\mem dx^\m
		+ i\mem \bpsi_m\mem d\psi^m
	\,.
\end{align}
Of course,
it is not a good idea to use canonical coordinates $(x^\m,\pcan_\m,\psi^m)$.
In particular, $\pcan_\m$ in \eqref{GR.pcan} is gauge-dependent
and is not a physical observable of the spinning particle:
one's choice of the orthonormal frame
changes its value.

\subsection{Ehresmann Frame}
\label{NONCOORD>FRAME}

As before, let us end with a coordinate-free characterization of the ideas
unearthed in \Secss{NONCOORD>YM}{NONCOORD>BI}{NONCOORD>GR}.

\para{Covariantizer Reimagined}

The astute reader may have noticed that
the covariantizer $\varrho$ is essentially a $(1,1)$-tensor on a phase space $\ps$,
mapping one-forms to one-forms:
to say,
``\hem$\a_I \mapsto \a_J\mem \varrho^J{}_I$''
in an index notation.
In particular,
\eqrefss{YM.varrho}{BI.varrho}{GR.varrho}
have mapped coordinate one-forms to non-coordinate one-forms,
like $\varrho(d\XQ^I) = \varrho^I{}_J\mem d\XQ^J$.

With this observation,
we introduce a conceptually cleaner framework
that replaces the covariantizer
with a \textit{non-coordinate coframe} $\bme^A = \bme^A{}_I(X)\mem dX^I$ on phase space.
Here,
$I,J,K,L,\cdots$ 
are coordinate indices of $\ps$,
while 
$A,B,C,D,\cdots$ 
are frame indices.
Both run through $(\dim \ps)$ integers.
Its dual gives a \textit{non-coordinate frame}
$\E_A = \E^I{}_A(\XQ)\mem \partial_I$.

The motivation for this refinement is to avoid some subtle or confusing aspects
of the covariantizer construction.
For instance,
our first version of
the covariant Poisson bracket in \eqref{covpb}
seems to yield uninsightful expressions when nested,
so a revision is desired.
Also, the notation $dx^m$ in \eqrefs{BI.omega'}{GR.omega'}
can be a source of unnecessary confusion.\footnote{
	Its heuristic idea traces back to
	an exploration on
	multi-valued coordinates
	\`a la \rcite{Kleinert:2008zzb},
	which may deviate from the standard differential geometry discourse.
}

\para{Frame}

A smooth manifold $\ps$
is \define{framed}
if it is equipped with
a set of vector fields $\E_A \in \Gamma(T\ps)$
for $A = 1,2,{\cdots},(\dim \ps)$
that is complete in the tangent space $T_p\ps$
for each point $p \inn \ps$.
In a framed smooth manifold $(\ps,\E)$,
the \define{coframe} refers to the set of one-forms
$\bme^A \in \Gamma(\Tstar\ps)$
that is dual to the frame:
$\langle \bme^A , \E_B \rangle = \delta^A{}_B$.

The \define{anholonomy coefficients} of a frame $\E$ on $\ps$
are $\Om^A{}_{BC} = \langle \bme^A , [\E_B,\E_C] \rangle$.
If any of the them are nonzero,
then the frame $\E$ said to be \define{non-coordinate}.

\para{Ehresmann Frame}

Let $\M$ be a framed smooth manifold.
Let $E_m$ be its frame,
where $m = 1,2,{\cdots},(\dim\M)$.
Suppose $\ps$ is a fiber bundle over $\M$
equipped with an Ehresmann connection.
Then there exists a natural frame on $\ps$,
which we call
the \define{Ehresmann frame}.

Concretely,
construct a local coordinate system $(x^\m,v^\a)$ on $\ps$
based on a local patch on $\M$.
Here, $x^\m$ coordinatize the base
while $v^\a$ coordinatize the fiber, so
$\m = 1,{\cdots},(\dim \M)$ and $\a = (1 \mplus \dim\M),{\cdots},(\dim \ps)$.
Let $\tE_m \inn \Gamma(T\ps)$ be the 
the horizontal lift of $E_m \in \Gamma(T\M)$
with respect tor the Ehresmann connection.
The Ehresmann frame on $\ps$
is $\E_A = (\tE_m,\partial/\partial v^\a)$.
From the fact that
$\tE_m$ span the horizontal subspaces
while $\partial/\partial v^\a$ span the vertical subspaces,
it should be clear that $\E_A$ is nondegenerate,
i.e., provides a complete basis in $T_p\ps$ for each $p \inn \ps$.
The dual $\bme^A$ is called the \define{Ehresmann coframe}.

\para{Einstein's Elevator via Ehresmann Frame}
A \define{free-theory structure} on 
an even-dimensional smooth manifold $\ps$
is a constant symplectic matrix
$\omega^\circ_{AB}$
of rank $\dim\ps$.
Let $(\omega^\circ{}^{-1})^{AB}$ be its inverse.
If $\ps$ is framed,
then it is an almost-symplectic manifold
via
\begin{align}
	\label{almoste}
	\omega^\bullet
	\,=\,
		\tfrac{1}{2}\:
		\omega^\circ_{AB}\,
			\bme^A \wedge \bme^B
	\,,\quad
	\omega^\bullet{}^{-1}
	\,=\,
		\tfrac{1}{2}\:
		(\omega^\circ{}^{-1})^{AB}\,
			\E_A \wedge \E_B
	\,.
\end{align}


We will be especially interested in the case when $\E$ is an Ehresmann frame on a phase space.
Physically, it will
specify a \textit{gauge-covariant yet non-coordinate basis} of vector fields.

In this geometrical language,
our explorations in 
\Secss{NONCOORD>YM}{NONCOORD>BI}{NONCOORD>GR}
describe
the process of
endowing a proper covariant symplectic perturbation $\omega'$
on this almost-symplectic manifold $(\ps,\omega^\bullet)$
to generate a symplectic manifold $(\ps,\omega = \omega^\bullet + \omega')$.

\para{Covariant Poisson Bracket Reimagined}

Consequently, we realize that ``covariant Poisson bracket''
means to examine the components of the Poisson bivector
in the non-coordinate basis.
That is, in a framed symplectic manifold $(\ps,\omega,\E)$ we compute
\begin{align}
	\label{cpb}
	\omega^{-1}(\bme^A,\bme^B)
	\,=\,
		(\omega^{-1})^{AB}
	\,.
\end{align}
Crucially, \eqref{almoste}
implies that the free-theory Poisson brackets are
recycled as
\begin{align}
	\label{isom*}
	\pib{\bme^A}{\bme^B}
	\,=\
		(\omega^\circ{}^{-1})^{AB}
	\,.
\end{align}
This is the symplectic geometry implementation of \textit{Einstein's elevator},
in which case
the covariant symplectic perturbation $\omega'$ encodes the \textit{tidal (curvature) effects}.
With \eqref{isom*}, \eqref{covspt} implies the expansion
\begin{align}
	\label{cpb-expansion}
	{}&{}
	\omega^{-1}(\bme^A,\bme^B)
	\\
	{}&{}
	\,=\,
		\pib{\bme^A}{\bme^B}
		- \pib{\bme^A}{\bme^C}
		\hhem\,\omega'(\E_A,\E_B)\,\mem
			\pib{\bme^D}{\bme^B}
		+ \cdots
	\nonumber
	\,,\\
	{}&{}
	\,=\,
		(\omega^\circ{}^{-1})^{AB}
		- (\omega^\circ{}^{-1})^{AC}
		\,\omega'_{AB}\,
			(\omega^\circ{}^{-1})^{DB}
		+ (\omega^\circ{}^{-1})^{AC}
		\,\omega'_{CD}\,
			(\omega^\circ{}^{-1})^{DE}
		\,\omega'_{EF}\,
			(\omega^\circ{}^{-1})^{FB}
		+ \cdots
	\,,
	\nonumber
\end{align}
where $\omega'_{AB} := \omega'(\E_A,\E_B)$.
Clearly, these are the refined versions of \eqrefs{isom}{covpb-expansion}.

\newpage

\eqrefss{cpb}{isom*}{cpb-expansion} will be utilized in actual calculations as
\begin{align}
\begin{split}
	&
	\omega^{-1}(\a,\b)
	\\
	&
	\,=\,
		\cont{\a}{\E_A}\,
			(\omega^{-1})^{AB}
		\,\cont{\E_B}{\b}
	\,,\\
	&
	\,=\,
		\cont{\a}{\E_A}\,
			(\omega^\circ{}^{-1})^{AB}
		\,\cont{\E_B}{\b}
		- 
		\cont{\a}{\E_A}\,
			(\omega^\circ{}^{-1})^{AC}
		\,\omega'_{CD}\,
			(\omega^\circ{}^{-1})^{DB}
		\,\cont{\E_B}{\b}
		+ \cdots
	\,,
\end{split}
\end{align}
where $\a, \b \in \Omega^1(\ps)$
will be covariant differentials.
Taking $\a \eqq df$ and $\b \eqq dg$ 
for singlet-valued phase space functions $f,g \in \Cinfty(\ps)$,
for instance,
will compute the Poisson bracket $\pb{f}{g} = \omega^{-1}(df,dg)$
while manifesting covariance at intermediate steps.

Once the covariant Poisson brackets $\omega^{-1}(\bme^A,\bme^B) = (\omega^{-1})^{AB}$
are fully known,
the ordinary Poisson brackets $\pb{\XQ^I}{\XQ^J} = (\omega^{-1})^{IJ}$
between phase space coordinates
are completely determined by a simple basis change:
$(\omega^{-1})^{IJ} = \E^I{}_A\mem \E^J{}_B\mem (\omega^{-1})^{AB}$.
In this way, one fully characterizes the Poisson geometry
of phase spaces in a manifestly covariant fashion.

\para{Covariant Jacobi Identity}

Finally,
``covariant Jacobi identity''
means to examine the Schouten-Nijenhuis bracket
$\comm{\omega^{-1}}{\omega^{-1}}$
in non-coordinate frames.

\App{GUISES}
explicates that
the vanishing of the Schouten-Nijenhuis bracket
$\comm{\omega^{-1}}{\omega^{-1}}$
is equivalent to the closure $d\omega = 0$,
yielding the Jacobi identity $\jac(f,g,h) = 0$
when fully contracted with the exact differentials $df,dg,dh$.
Specifically,
\begin{align}
	\label{jac}
	J \,:=\,
		- \tfrac{1}{2}\mem \comm{\omega^{-1}}{\omega^{-1}}
	\qiq
	\jac(f,g,h)
	\,=\,
		J(df,dg,dh)
	\,.
\end{align}


As a generalization,
we compute
the Schouten-Nijenhuis bracket
in the non-coordinate basis.
Recalling its definition 
\cite{schouten1940ueber,schouten1954differential,nijenhuis1955jacobi},
we find that
$J(\bme^A,\bme^B,\bme^C) = J^{ABC}$
computes to
\begin{align}
	\label{cjacobiator}
	J^{ABC}
	\,=\,
		3\mem\BB{
			\E_D\act{ (\omega^{-1})^{[AB|} }\mem (\omega^{-1})^{D|C]}
			- \Om^{[A|}{}_{EF}\mem (\omega^{-1})^{E|B|}\hem (\omega^{-1})^{F|C]}
		}
	\,,
\end{align}
which could be called the \define{covariant Jacobiator}.
Therefore,
an equivalent statement of
the closure $d\omega = 0$ is
\begin{align}
	\label{cjac}
	3\mem
	\omega^{-1}\BB{\mem
		d(\hhem{
			\omega^{-1}(\bme^{[A},\bme^B)
		})
		\mem,\mem
			\bme^{C]}
	\mem}
	\,=\, 
		3\mem
		\Om^{[A|}{}_{EF}\mem (\omega^{-1})^{E|B|}\hem (\omega^{-1})^{F|C]}
	\,.
\end{align}
\eqref{cjac} will be referred to as the \define{covariant Jacobi identity}.
Its left-hand side transcribes a typical combination,
while
its right-hand side represents apparent 
obstructions
due to the anholonomy.
We may refer to these as \textit{differential} and \textit{algebraic} parts, respectively.
The involved nature of \eqref{cjac}
is a vivid demonstration of
how symplecticity becomes obscured
for the sake of manifest gauge covariance.

Note that \eqref{cjac}
can also be derived by using the identity in \eqref{3id}:
\begin{align}
	\label{jdown}
    d\omega(\E_A^{\vphantom{\prime}},\E_B^{\vphantom{\prime}},\E_C^{\vphantom{\prime}})
    \,=\,
    	3\mem\BB{
	        \E^{\vphantom{\prime}}_{[A} \act{ \omega_{BC]}^{\vphantom{\prime}} }
	        + \omega_{[C|D}\mem \Om^D{}_{|AB]}^{\vphantom{\prime}}
	    }
	\,.
\end{align}

\para{Covariant Geometry of Phase Spaces}

In summary,
we conclude that
the covariant geometry of 
a particle phase space $\ps$,
fibering over spacetime to be equipped with an Ehresmann connection,
is defined by a triple of structures:
(a) free-theory structure $\omega^\circ$,
(b) Ehresmann frame $\E$,
and (c) covariant symplectic perturbation $\omega'$.
The goal of \Sec{CPB} is to 
revisit our examples
with this refined formulation:
gauge theory, gravity, and gravity with spin.

\newpage

\section{Covariant Poisson Brackets}
\label{CPB}

In this section, we provide a systematic analysis on the covariant symplectic geometry of particle phase spaces
and investigate how
covariant Poisson brackets obscure symplecticity.

\subsection{Gauge Theory}
\label{CPB>YM}

\para{Covariant Symplectic Geometry}

The phase space
$\ps = T^*\mflat \moplus E$
of the colored scalar particle
is characterized
by the covariant coordinates,
\begin{align}
	\label{YM.X}
	X^I
	\,=\,
		\bigbig{
			x^m
			,\mem
			p_m
			,\mem
			\vth^i
			,\mem
			\bvth_i
		}
	\,,
\end{align}
the Ehresmann coframe/frame,
\begin{align}
	\label{YM.EF}
	\bme^A
	\,=\,
		\bigbig{
			dx^m
			,\mem
			dp_m
			,\mem
			D\vth^i
			,\mem
			D\bvth_i
		}
	\,,\quad
	\E_A
	\,=\,
		\bb{
			\tpartial_m
			,\mem
			\frac{\partial}{\partial p_m}
			,\mem
			\frac{\partial}{\partial \vth^i}
			,\mem
			\frac{\partial}{\partial \bvth_i}
		}
	\,,
\end{align}
the canonical free-theory structure such that
\begin{align}
	\label{YM.o0}
	\omega^\bullet
	\,=\,
		\tfrac{1}{2}\,
		\omega^\circ_{AB}\,
			\bme^A \wedge \bme^B
	\,=\,
		Dp_m \wedge dx^m
		+ i\mem D\bvth_i \wedge D\vth^i
	\,,
\end{align}
and the proper covariant symplectic perturbation,
\begin{align}
	\label{YM.o1}
	\omega'
	\,=\,
		\tfrac{1}{2}\,
			q_a\mem F^a{}_{mn}(x)\,
				dx^m \swedge dx^n
	\,,
\end{align}
so $\omega = \omega^\bullet + \omega'$.
Here, it is convenient to
formally regard $\vth^i$ and $\bvth_i$ as independent variables.

\para{Anholonomy Coefficients}

The anholonomy coefficients are conveniently enumerated by
$-d\bme^A = \frac{1}{2}\mem \Om^A{}_{BC}\mem \bme^B \swedge \bme^C$:
\begin{align}
\begin{split}
	\label{YM.an}
	- d\bigbig{
		dx^m
	}
	\,&=\,
		0
	\,,\\
	- d\bigbig{
		dp_m
	}
	\,&=\,
		0
	\,,\\
	- d\bigbig{
		D\vth^i
	}
	\,&=\,
		A^i{}_{jm}(x)
		\,\BB{
			dx^m \swedge D\vth^j
		}
		- F^i{}_{jmn}(x)\mem \vth^j
		\,\BB{
			\tfrac{1}{2}\mem
				dx^m \swedge dx^n
		}
	\,,\\
	- d\bigbig{
		D\bvth_i
	}
	\,&=\,
		- A^j{}_{im}(x)
		\,\BB{
			dx^m \swedge D\bvth_j
		}
		+ \bvth_j\hem F^j{}_{imn}(x)
		\,\BB{
			\tfrac{1}{2}\mem
				dx^m \swedge dx^n
		}
	\,.
\end{split}
\end{align}
Otherwise, one can directly compute the Lie brackets $\comm{\E_A}{\E_B}$.

\para{Covariant Poisson Brackets}


\eqref{YM.o0} identifies an isomorphism to the free theory (pure-gauge dynamics),
while \eqref{YM.o1} encodes the curvature effect.
Hence, by straightforwardly recycling the free-theory Poisson brackets in \eqref{colored.pb0},
the covariant expansion in \eqref{cpb-expansion} derives that
the non-vanishing components of $\omega^{-1}$
in the Ehresmann basis are
\begin{align}
\begin{split}
	\label{YM.cpb}
	\pia{dx^m}{dp_n}
	\,&=\,
		\delta^m{}_n
	\,,\\
	\pia{D\vth^i}{D\bvth_j}
	\,&=\,
		-i\mem \delta^i{}_j
	\,,\\
	\pia{dp_m}{dp_n}
	\,&=\,
		q_a\hem F^a{}_{mn}(x)
	\,.
\end{split}
\end{align}
Evidently, gauge covariance is manifested.

\para{Covariant Jacobi Identity}

Suppose we want to diagnose symplecticity
directly from the covariant Poisson brackets in \eqref{YM.cpb}.
To this end, we check the covariant Jacobi identity
in \eqref{cjac}.
We compute its left-hand side and right-hand side separately.
It can be seen that
the nontrivial covariant Jacobiators are
$J(dp_m,dp_n,D\vth^i)$
and
$J(dp_m,dp_n,dp_r)$.

\newpage

First, we verify $J(dp_m,dp_n,D\vth^i) = 0$.
The differential part computes to
\begin{align}
	\label{YM.check1}
	\piA{
		D(
			q_a\hem F^a{}_{mn}(x)
		)
	}{
		D\vth^i
	}
	\,=\,
		F^a{}_{mn}(x)\,\mem
			\pia{Dq_a}{D\vth^i}
	\,=\,
		- F^i{}_{jmn}(x)\mem \vth^j
	\,,
\end{align}
where we have used the fact that
$d(q_a\hem F^a{}_{mn}(x)) = D(q_a\hem F^a{}_{mn}(x))$.
Indeed,
the third line in \eqref{YM.an} shows that
the algebraic part
exactly computes to
the same.

Second, we verify $J(dp_m,dp_n,dp_r) = 0$.
The differential part computes to
\begin{align}
	\label{YM.check2}
	\piA{
		D(
			q_a\hem F^a{}_{mn}(x)
		)
	}{
		Dp_r
	}
	\,=\,
		q_a\hem D_s F^a{}_{mn}(x)\,\mem
			\pia{dx^s}{Dp_r}
	\,=\,
		q_a\hem D_r F^a{}_{mn}(x)
	\,,
\end{align}
up to total anti-symmetrization on the free indices $m,n,r$.
This vanishes by the Bianchi identity of non-abelian gauge fields:
\begin{align}
	\label{YM.bianchi}
	D_\wrap{[r} F^a{}_\wrap{mn]} \,=\, 0
	\,.
\end{align}
Indeed, the second line in \eqref{YM.an} shows that
the algebraic part is exactly zero.


\subsection{Gravity}
\label{CPB>BI}

\para{Covariant Symplectic Geometry}

The phase space
$\ps = \Tstar\M$
of the general-relativistic scalar particle
is characterized
by the covariant coordinates,
\begin{align}
	\label{BI.X}
	X^I
	\,=\,
		\bigbig{
			x^\m
			,\mem
			p_m
		}
	\,,
\end{align}
the Ehresmann coframe/frame,
\begin{align}
	\label{BI.EF}
	\bme^A
	\,=\,
		\bigbig{
			e^m
			,\mem
			\tD p_m
		}
	\,,\quad
	\E_A
	\,=\,
		\bb{
			\tE_m
			,\mem
			\frac{\partial}{\partial p_m}
		}
	\,,
\end{align}
the canonical free-theory structure such that
\begin{align}
	\label{BI.o0}
	\omega^\bullet
	\,=\,
		\tfrac{1}{2}\,
		\omega^\circ_{AB}\,
			\bme^A \wedge \bme^B
	\,=\,
		\tD p_m \wedge e^m
	\,,
\end{align}
and the proper covariant symplectic perturbation,
\begin{align}
	\label{BI.o1}
	\omega'
	\,=\,
		\tfrac{1}{2}\,
			p_k\mem \tT^k{}_{mn}(x)\,
				e^m \swedge e^n
	\,,
\end{align}
so $\omega = \omega^\bullet + \omega'$.
For full generality,
we assume the arbitrary Lorentz-valued connection $\tD$.

\para{Anholonomy Coefficients}

The anholonomy coefficients are conveniently enumerated by
$-d\bme^A = \frac{1}{2}\mem \Om^A{}_{BC}\mem \bme^B \swedge \bme^C$:
\begin{align}
\begin{split}
	\label{BI.an}
	- d\bigbig{
		e^m
	}
	\,&=\,
		\Omega^m{}_{rs}(x)
		\,\BB{
			\tfrac{1}{2}\mem
				e^r \swedge e^s
		}
	\,,\\
	- d\bigbig{
		\tD p_m
	}
	\,&=\,
		- \tgamma^n{}_{mr}(x)
		\,\BB{
			e^r \swedge \tD p_n
		}
		+ p_n\hem \tR^n{}_{mrs}(x)
		\,\BB{
			\tfrac{1}{2}\mem
				e^r \swedge e^s
		}
	\,.
\end{split}
\end{align}

\para{Covariant Poisson Brackets}

\eqref{BI.o0} identifies an isomorphism to special relativity (pure-gauge dynamics)
in terms of the local orthonormal frame.
Hence, by straightforwardly recycling the free-theory Poisson brackets in \eqref{scalar.pb0},
the covariant expansion in \eqref{cpb-expansion} derives that
the non-vanishing components of $\omega^{-1}$
in the Ehresmann basis are
\begin{align}
\begin{split}
	\label{BI.cpb}
	\pia{e^m}{\tD p_n}
	\,&=\,
		\delta^m{}_n
	\,,\\
	\pia{\tD p_m}{\tD p_n}
	\,&=\,
		p_k\hem \tT^k{}_{mn}(x)
	\,.
\end{split}
\end{align}
Evidently, covariance is manifested.

\newpage

\para{Covariant Jacobi Identity}

To diagnose symplecticity
directly from the covariant Poisson brackets in \eqref{BI.cpb},
we check the covariant Jacobi identity
in \eqref{cjac}.
It can be seen that
the nontrivial covariant Jacobiators are
$J(\tD p_m,\tD p_n,e^r)$
and
$J(\tD p_m,\tD p_n,\tD p_r)$.

A preliminary calculation is
\begin{align}
\begin{split}
	d\bigbig{
		p_k\mem \tT^k{}_{mn}\,
		\xi^{mn}
	}
	\,=\,
	{}&{}
	\BB{
		\tD p_k\, \tT^k{}_{mn}
		+ p_k\mem \tD_r\tT^k{}_{mn}\, e^r
	}
			\,\xi^{mn}
	\\[-0.08\baselineskip]
	{}&{}
		+ p_k\mem \tT^k{}_{mn}
			\,\BB{
				d\xi^{mn}
				+ \tgamma^m{}_l\mem \xi^{ln}
				+ \tgamma^n{}_l\mem \xi^{ml}
			}
	\,,
\end{split}
\end{align}
where $\xi \in \Gamma(\wedge^2T\M)$
is an auxiliary bivector field on spacetime.
In turn,
\begin{align}
\begin{split}
	\label{BI.checkid}
	d\bigbig{
		p_k\mem \tT^k{}_{mn}
	}
	\,=\,
	{}&{}
		\tD p_k\mem \BB{
			\tT^k{}_{mn}
		}
		+ p_k\mem 
			\BB{
				\tD_r\tT^k{}_{mn}
				+ \tT^k{}_{ln}\mem \tgamma^l{}_{mr}
				+ \tT^k{}_{ml}\mem \tgamma^l{}_{nr}
			}\, e^r
	\,.
\end{split}
\end{align}
Namely, one needs to be careful about the fact that
\smash{$p_k\mem \tT^k{}_{mn}$} exhibits free indices $m,n$.
The occurrence of bare connection coefficients is unfortunate but seems unavoidable.

\enter
First, we verify $J(\tD p_m,\tD p_n,e^r) = 0$.
Using \eqref{BI.checkid},
the differential part gives
\begin{subequations}
\begin{align}
	\label{BI.check1a}
	\piA{
		d(
			p_k\hem \tT^k{}_{mn}
		)
	}{
		e^r
	}
	\,=\,
		\tT^k{}_{mn}\,\mem
			\pia{\tD p_k}{e^r}
	\,=\,
		- \tT^r{}_{mn}
	\,.
\end{align}
Meanwhile, 
The algebraic part computes to
\begin{align}
\begin{split}
	\label{BI.check1b}
	&
	\Om^{(e^r)}{}_\Wrap{(\tE_m)(\tE_n)}
	- \Om^{(\tD p_m)}{}_\Wrap{(\tE_n)(\psp p_r)}
	+ \Om^{(\tD p_n)}{}_\Wrap{(\tE_m)(\psp p_r)}
	=\,
		\Omega^r{}_{mn}
		+ \tgamma^r{}_{mn}
		- \tgamma^r{}_{nm}
	\,.
\end{split}
\end{align}
\end{subequations}
\eqrefs{BI.check1a}{BI.check1b}
do agree, since
$\tT^r = de^r + \tgamma^r{}_n \swedge e^n$
encodes the identity
$\tT^r{}_{mn} + \Omega^r{}_{mn} + 2\mem \tgamma^r{}_{[mn]} = 0$.

\enter
Second, we verify 
$J(\tD p_m,\tD p_n,\tD p_r) = 0$.
Using \eqref{BI.checkid},
the differential part gives
\begin{subequations}
\begin{align}
\begin{split}
	\label{BI.check2a}
	&
	\piA{
		d(
			p_k\hem \tT^k{}_{mn}
		)
	}{
		\tD p_r
	}
	\\[-0.07\baselineskip]
	&
	=\,
		p_k\mem 
			\BB{
				\tD_r\tT^k{}_{mn}
				+ \tT^k{}_{ln}\mem \tgamma^l{}_{mr}
				+ \tT^k{}_{ml}\mem \tgamma^l{}_{nr}
			}
		+ \BB{
			\tT^k{}_{mn}
		} \BB{
			p_l\mem \tT^l{}_{kr}
		}
	\\[-0.07\baselineskip]
	&
	=\,
		p_k\mem 
			\BB{
				\tD_r\tT^k{}_{mn}
				+ \tT^k{}_{ln}\mem \tgamma^l{}_{mr}
				- \tT^k{}_{lm}\mem \tgamma^l{}_{nr}
				+ \tT^k{}_{lr}\mem \tT^l{}_{mn}
			}
	\,,
\end{split}
\end{align}
up to the index anti-symmetrization.
Meanwhile, 
the algebraic part is
\begin{align}
	\label{BI.check2b}
	&
	\Om^{(\tD p_r)}{}_\Wrap{(\tE_m)(\tE_n)}
	+ \BB{
		\Om^{(\tD p_r)}{}_\Wrap{(\tE_m)(\psp p_l)}\,
			\pia{Dp_l}{Dp_n}
		- (m \mlra n)
	}
	\nonumber
	\\[-0.10\baselineskip]
	&
	=\,
		p_k\hem \tR^k{}_{rmn}
		- \tgamma^l{}_{rm}\,
			p_k\mem \tT^k{}_{ln}
		+ \tgamma^l{}_{rn}\,
			p_k\mem \tT^k{}_{lm}
	\nonumber
	\\[-0.07\baselineskip]
	&
	=\,
		p_k\mem
		\BB{
			 \tR^k{}_{rmn}
			- \tT^k{}_{ln}\mem \tgamma^l{}_{rm}
			+ \tT^k{}_{lm}\mem \tgamma^l{}_{rn}
		}
	\,,
\end{align}
\end{subequations}
up to the index anti-symmetrization.
Here, quadratic-in-$p$ terms
are suppressed since
$\Om^{(\tD p_r)}{}_\wrap{(\psp p_k)(\psp p_l)} = 0$.
Consequently,
\eqrefs{BI.check2a}{BI.check2b}
agree if
\begin{align}
	\label{BI.bianchi}
	\tR^k{}_\wrap{[rmn]}
	\,=\,
		\tD_\wrap{[r}\tT^k{}_\wrap{mn]}
		+ \tT^k{}_\wrap{l[r}\mem \tT^l{}_\wrap{mn]}
	\,,
\end{align}
as the bare connection terms cancel upon the total anti-symmetrization.
\eqref{BI.bianchi} is the first Bianchi identity in Riemann-Cartan geometry,
$R^k{}_m \wedge e^m = \tD \tD e^k = \tD T^k$.
Equivalently,
it can be shown that
\eqref{BI.bianchi}
arises from
$\comm{\comm{E_{[m}}{E_{n\vphantom{]}}}}{E_{r]}} = 0$,
which precisely is
\textit{the} Bianchi identity 
in \eqref{YM.bianchi}
for diffeomorphism gauge theory:
$\partial_{[r} F^\m{}_{mn]} + \comm{A_{[r}}{F{}_{mn]}}^\m = 0$.


Despite the simplicity of \eqref{BI.cpb},
the check of symplecticity is
substantially complicated.
Yet, adopting the Levi-Civita connection 
$\tD = D$
will simplify the process.

\newpage

\subsection{Gravity with Spin}
\label{CPB>GR}

\para{Covariant Symplectic Geometry}

The phase space
$\ps = \A_2 = (T^* \moplus \Pi T^\C)\hem \M$
of the general-relativistic spinning particle
is characterized
by the covariant coordinates,
\begin{align}
	\label{GR.X}
	X^I
	\,=\,
		\bigbig{
			x^\m
			,\mem
			p_m
			,\mem
			\psi^m
			,\mem
			\bpsi_m
		}
	\,,
\end{align}
the Ehresmann coframe/frame,
\begin{align}
	\label{GR.EF}
	\bme^A
	\,=\,
		\bigbig{
			e^m
			,\mem
			Dp_m
			,\mem
			D\psi^m
			,\mem
			D\bpsi_m
		}
	\,,\quad
	\E_A
	\,=\,
		\bb{
			\tE_m
			,\mem
			\frac{\partial}{\partial p_m\vphantom{\bpsi^0}}
			,\mem
			\frac{\partial}{\partial \psi^m\vphantom{\bpsi^0}}
			,\mem
			\frac{\partial}{\partial \bpsi_m\vphantom{\bpsi^0}}
		}
	\,,
\end{align}
the canonical free-theory structure such that
\begin{align}
	\label{GR.o0}
	\omega^\bullet
	\,=\,
		\tfrac{1}{2}\,
		\omega^\circ_{AB}\,
			\bme^A \wedge \bme^B
	\,=\,
		Dp_m \wedge e^m
		+ i\mem D\bpsi_m \wedge D\psi^m
	\,,
\end{align}
and the proper covariant symplectic perturbation,
\begin{align}
	\label{GR.o1}
	\omega'
	\,=\,
		-\tfrac{1}{4}\,
			S_{kl}\mem R^{kl}{}_{mn}(x)\,
				e^m \swedge e^n
	\,,
\end{align}
so $\omega = \omega^\bullet + \omega'$.
Here, it is convenient to
formally treat $\psi^m$ and $\bpsi_m$ as independent variables.

\para{Anholonomy Coefficients}

The anholonomy coefficients are conveniently enumerated by
$-d\bme^A = \frac{1}{2}\mem \Om^A{}_{BC}\mem \bme^B \swedge \bme^C$:
\begin{align}
\begin{split}
	\label{GR.an}
	- d\bigbig{
		e^m
	}
	\,&=\,
		\Omega^m{}_{rs}(x)
		\,\BB{
			\tfrac{1}{2}\mem
				e^r \swedge e^s
		}
	\,,\\
	- d\bigbig{
		\tD p_m
	}
	\,&=\,
		- \gamma^n{}_{mr}(x)
		\,\BB{
			e^r \swedge Dp_n
		}
		+ p_n\hem R^n{}_{mrs}(x)
		\,\BB{
			\tfrac{1}{2}\mem
				e^r \swedge e^s
		}
	\,,\\
	- d\bigbig{
		D\psi^m
	}
	\,&=\,
		\gamma^m{}_{nr}(x)
		\,\BB{
			e^r \swedge D\psi^n
		}
		- R^m{}_{nrs}(x)\mem \psi^n
		\,\BB{
			\tfrac{1}{2}\mem
				e^r \swedge e^s
		}
	\,,\\
	- d\bigbig{
		D\bpsi_m
	}
	\,&=\,
		- \gamma^n{}_{mr}(x)
		\,\BB{
			e^r \swedge D\bpsi_n
		}
		+ \bpsi_n\hem R^n{}_{mrs}(x)
		\,\BB{
			\tfrac{1}{2}\mem
				e^r \swedge e^s
		}
	\,.
\end{split}
\end{align}
\eqref{GR.an} combines the features of \eqrefs{YM.an}{BI.an}.

\para{Covariant Poisson Brackets}

\eqref{GR.o0} identifies an isomorphism to special relativity (pure-gauge dynamics),
while \eqref{GR.o1} encodes the tidal effect.
Hence, by straightforwardly recycling the free-theory Poisson brackets in \eqref{spinning.pb0},
the covariant expansion in \eqref{cpb-expansion} derives that
the non-vanishing components of $\omega^{-1}$
in the Ehresmann basis are
\begin{align}
\begin{split}
	\label{GR.cpb}
	\pia{e^m}{Dp_n}
	\,&=\,
		\delta^m{}_n
	\,,\\[-0.1\baselineskip]
	\pia{D\psi^m}{D\bpsi_n}
	\,&=\,
		-i\mem \delta^m{}_n
	\,,\\[-0.1\baselineskip]
	\pia{Dp_m}{Dp_n}
	\,&=\,
		- \tfrac{1}{2}\, S_{kl}\hem R^{kl}{}_{mn}(x)
	\,.
\end{split}
\end{align}
Evidently, covariance is manifested.

\para{Covariant Jacobi Identity}

To diagnose symplecticity
directly from the covariant Poisson brackets in \eqref{GR.cpb},
we check the covariant Jacobi identity
in \eqref{cjac}.
It can be seen that
the nontrivial covariant Jacobiators are
$J(Dp_m,Dp_n,e^r)$,
$J(Dp_m,Dp_n,D\psi^r)$,
and
$J(Dp_m,Dp_n,Dp_r)$.
The vanishing of the first one is easy
by $\Omega^k{}_{mn\vphantom{]}} + 2\gamma^k{}_{[mn]} = 0$.

A preliminary calculation is
\begin{align}
\begin{split}
	d\bigbig{
		S_{kl}\mem R^{kl}{}_{mn}\,
		\xi^{mn}
	}
	\,=\,
	{}&{}
	\smash{
		\BB{
			DS_{kl}\mem R^{kl}{}_{mn}
			+ S_{kl}\mem D_rR^{kl}{}_{mn}\, e^r
		}
			\,\xi^{mn}
	}
	\\[-0.08\baselineskip]
	{}&{}
		+ S_{kl}\mem R^{kl}{}_{mn}
			\,\BB{
				d\xi^{mn}
				+ \gamma^m{}_s\mem \xi^{sn}
				+ \gamma^n{}_s\mem \xi^{ms}
			}
	\,,
\end{split}
\end{align}
\newpage\noindent
where $\xi \in \Gamma(\wedge^2T\M)$
is an auxiliary bivector field on spacetime.
In turn,
\begin{align}
\begin{split}
	\label{GR.checkid}
	d\bigbig{
		S_{kl}\mem R^{kl}{}_{mn}
	}
	\,=\,
	{}&{}
		\tD S_{kl}\mem \BB{
			R^{kl}{}_{mn}
		}
		+ S_{kl}\mem 
			\BB{
				D_rR^{kl}{}_{mn}
				+ R^{kl}{}_{sn}\mem \gamma^s{}_{mr}
				+ R^{kl}{}_{ms}\mem \gamma^s{}_{nr}
			}\, e^r
	\,.
\end{split}
\end{align}
Namely, one needs to be careful about the fact that
\smash{$S_{kl}\mem R^{kl}{}_{mn}$} exhibits free indices $m,n$.
The occurrence of bare connection coefficients is unfortunate but seems unavoidable.

\enter
First, we verify $J(Dp_m,Dp_n,D\psi^r) = 0$.
Using \eqref{GR.checkid},
the differential part gives
\begin{subequations}
\begin{align}
	\label{GR.check1a}
	- \tfrac{1}{2}\:
	\piA{
		d(
			S_{kl}\mem R^{kl}{}_{mn}
		)
	}{
		D\psi^r
	}
	\,=\,
		- \tfrac{1}{2}\,
		R^{kl}{}_{mn}\,\mem
			\pia{DS_{kl}}{D\psi^r}
	\,=\,
		- R^r{}_{smn}\mem \psi^s
	\,.
\end{align}
This calculation combines the features of both \eqrefs{YM.check1}{BI.check1a}.
Indeed, 
the algebraic part computes to
\begin{align}
\begin{split}
	\label{GR.check1b}
	&
	\Om^{(D\psi^r)}{}_\Wrap{(\tE_m)(\tE_n)}
	=\,
		- R^r{}_{smn}\mem \psi^s
	\,,
\end{split}
\end{align}
\end{subequations}
reproducing \eqref{GR.check1a}.
This calculation is rather a mere revival of the gauge theory story.

\enter
Second, we verify 
$J(Dp_m,Dp_n,Dp_r) = 0$.
Using \eqref{GR.checkid},
the differential part gives
\begin{subequations}
\begin{align}
\begin{split}
	\label{GR.check2a}
	&
	-\tfrac{1}{2}\:
	\piA{
		d(
			S_{kl}\hem R^{kl}{}_{mn}(x)
		)
	}{
		Dp_r
	}
	\\[-0.07\baselineskip]
	&
	=\,
		-\tfrac{1}{2}\,
		S_{kl}\mem
			\BB{
				D_rR^{kl}{}_{mn}
				+ R^{kl}{}_{sn}\mem \gamma^s{}_{mr}
				- R^{kl}{}_{sm}\mem \gamma^s{}_{nr}
			}
	\,,
\end{split}
\end{align}
up to the index anti-symmetrization.
This calculation is simpler than \eqref{BI.check2a},
as we have set the torsion to zero
to focus on the curvature effects.
Meanwhile,
the algebraic part computes to
\begin{align}
	\label{GR.check2b}
	&
	\Om^{(\tD p_r)}{}_\Wrap{(\tE_m)(\tE_n)}
	+ \BB{
		\Om^{(\tD p_r)}{}_\Wrap{(\tE_m)(\psp p_s)}\,
			\pia{Dp_s}{Dp_n}
		- (m \mlra n)
	}
	\nonumber
	\\[-0.10\baselineskip]
	&
	=\,
		p_s\mem R^s{}_{rmn}
		- \tgamma^s{}_{rm}\,
			\BB{
				- \tfrac{1}{2}\mem S_{kl}\mem R^{kl}{}_{sn}
			}
		+ \tgamma^s{}_{rn}\,
			\BB{
				- \tfrac{1}{2}\mem S_{kl}\mem R^{kl}{}_{sm}
			}
	\nonumber
	\\[-0.07\baselineskip]
	&
	=\,
		p_s\mem R^s{}_{rmn}
		- \tfrac{1}{2}\mem S_{kl}\mem 
		\BB{
			- R^{kl}{}_{sn}\mem \tgamma^s{}_{rm}
			+ R^{kl}{}_{sm}\mem \tgamma^s{}_{rn}
		}
	\,,
\end{align}
\end{subequations}
up to the index anti-symmetrization.
Here, quadratic-in-$S$ terms
are suppressed since
$\Om^{(\tD p_r)}{}_\wrap{(\psp p_k)(\psp p_l)} = 0$.
This calculation is similar to \eqref{BI.check2b}.
By equating
\eqrefs{GR.check2a}{GR.check2b},
we find
\begin{align}
	\label{GR.bianchi}
	R^s{}_\wrap{[rmn]}
	\,=\,
		0
	\,,\quad
	D_\wrap{[r} R^{kl}{}_\wrap{mn]}
	\,=\, 
		0
	\,,
\end{align}
as the bare connection terms cancel upon the total anti-symmetrization of indices.
\eqref{GR.bianchi}
derives the first and second Bianchi identities in Riemannian geometry,
$R^s{}_r \wedge e^r = DDe^s = 0$
and
$DR^{kl} = 0$.
In light of our previous explorations in \Secs{CPB>YM}{CPB>BI},
the first Bianchi identity
will be attributed to the diffeomorphism charge (momentum),
tracing back to \eqref{BI.bianchi},
while
the second Bianchi identity
will be attributed to the local Lorentz charge (spin),
paralleling \eqref{YM.bianchi}.

Despite the simplicity of 
the covariant Poisson brackets in
\eqref{GR.cpb},
the check of symplecticity is
considerably complicated.

\newpage

\subsection{Symplectic Vielbein}
\label{CPB>SV}

Before moving on, let us introduce a useful and interesting idea:
\textit{symplectic vielbein formalism}.

\para{Local Flatness in Symplectic Geometry}

In general relativity,
the vielbein formalism
concerns a ``square root'' of the metric:
$g_{\m\n}(x) = \eta_{mn}\mem e^m{}_\m(x)\mem e^n{}_\n(x)$.
Here, $\eta_{mn}$ is a constant symmetric matrix.
As a result, the vielbein description of the curved geometry
fully embraces
the \textit{local flatness} in Riemannian geometry.

We propose to implement the vielbein formalism in symplectic geometry
by the formula
\begin{align}
	\label{sv-comp}
	\omega_{IJ}(X)
	\,=\,
		\omega^\circ_{AB}\,
			\ee^A{}_I(X)\,
			\ee^B{}_J(X)
	\,,
\end{align}
which concerns a ``square root'' of the symplectic form.
Here, $\omega^\circ_{AB}$ is a constant anti-symmetric matrix of full rank:
the free-theory structure.

The astute reader will sense that
\eqref{sv-comp} is precisely a compromise between manifest covariance and symplecticity.
In symplectic geometry,
the Darboux theorem \cite{darboux1882probleme}
states \textit{global flatness} in a sense,
meaning that
any symplectic form $\omega$
can be flawlessly ``ironed out''
to exhibit constant components ($\pm1$ and $0$)
within a finite region of nonzero extension.
This is by the use of canonical coordinates of course,
but the central lesson of \Sec{NONCAN}
has been that canonical coordinates are abandoned for manifest covariance.
The compromise is \textit{local flatness},
which means to achieve Darboux
only within the infinitesimal vicinity of each point.
In this way, manifest symplecticity is obstructed
yet to a slightly milder degree.

To be explicit,
recall the well-known result in Riemannian geometry
that the normal coordinates $y^m$ adapted to an orthonormal frame
describe
\begin{align}
	\label{normal.R}
	g
	\,=\,
	\bb{
		\eta_{mn}
		+ \frac{1}{3}\, R_{mrsn}(y\eqq0)\, y^r y^s
		+ \O(y^3)
	}\, dy^m \modot dy^n
	\,.
\end{align}
By \textit{local} flatness,
it means that the bracketed combination in \eqref{normal.R}
inevitably deviates from the flat metric $\eta_{mn}$
for $y \neq 0$,
in the presence of Riemannian curvature.
The vielbein describes the Jacobian factor
between generic and normal coordinates
at $y = 0$.

The Darboux theorem asserts that,
around any point in a symplectic manifold $(\ps,\omega)$,
there exist coordinates $\YQcan^I$
such that
\begin{align}
	\omega
	\,=\,
		\tfrac{1}{2}\:
			\omega^\circ_{IJ}\,
			d\YQcan^I \swedge d\YQcan^J
\end{align}
at points with any value of $\YQcan^I$.
It can be shown that a weaker statement holds:
there exist normal coordinates $Y^A$
around any point in a symplectic manifold $(\ps,\omega)$
such that
\begin{align}
	\label{normal.S}
	\omega
	\,=\,
		\tfrac{1}{2}\:
		\BB{
			\omega^\circ_{AB}
			+ \O(Y^2)
		}\,
		dY^A \swedge dY^B
	\,.
\end{align}
The symplectic vielbein describes the Jacobian factor
between generic and normal coordinates
at $Y = 0$.

Having described the significance of the idea
and sketched its intuition,
we provide a precise formalization below.

\para{Symplectic Vielbein}

Let $(\ps,\omega)$ be an almost-symplectic manifold.
A \define{locally canonical coframe}
is a set of one-forms
$\ee^A \in \Omega^1(\ps)$
such that
\begin{align}
	\label{sv}
	\omega
	\,=\,
		\tfrac{1}{2}\:
		\omega^\circ_{AB}\,
			\ee^A \wedge \ee^B
\end{align}
for a free-theory structure $\omega^\circ$.
This will be also referred to as a \define{symplectic vielbein}.

The dual of $\ee^A$
is the set of vector fields 
$\EE_A \in \Gamma(T\ps)$
such that
$\langle \ee^A , \EE_B \rangle = \delta^A{}_B$,
which will be called a \define{locally canonical frame}.
It should be clear that
\begin{align}
	\label{svi}
	\omega^{-1}
	\,=\,
		\tfrac{1}{2}\:
		(\omega^\circ{}^{-1})^{AB}\,
			\EE_A \wedge \EE_B
	\,.
\end{align}
The anholonomy coefficients of the locally canonical frame
are $\OOm^A{}_{BC} = \langle \ee^A , [\EE_B,\EE_C] \rangle$.

In the symplectic vielbein formalism,
the frame indices $A,B,C,D,\cdots$
may be raised and lowered by
the free-theory structure as
\begin{align}
	\label{raise-lower}
	\a^A \,=\, (\omega^\circ{}^{-1})^{AB}\mem \a_B
	\qfq
	\a_A \,=\, \omega^\circ_{AB}\mem \a^B
	\,.
\end{align}

\para{Covariant Jacobi Identity}
Recall the covariant Jacobiator in \eqref{cjacobiator},
which examined the Ehresmann frame components of the trivector
$J = -\frac{1}{2}\mem \comm{\omega^{-1}}{\omega^{-1}}$.
In the symplectic vielbein formalism,
one instead examines the components of this trivector
in the locally canonical frame.
Since $\omega^{-1}(\ee^A,\ee^B) = (\omega^\circ{}^{-1})^{AB}$
are just constants,
one finds
\begin{align}
	\label{cjacobiator0}
	J^{ABC}
	\,=\,
		- 3\mem \OOm^{[ABC]}
	\,.
\end{align}
In this way, the verification of symplecticity
becomes relatively simpler
as compared to the Ehresmann frame formalism in \Sec{NONCOORD>FRAME}.
This supports the intuition that
the symplectic vielbein is a compromise
between manifest covariance and symplecticity.
As a result, an equivalent statement of the closure $d\omega = 0$ is
\begin{align}
	\label{cjac0}
	\OOm^{ABC}
	+
	\OOm^{BCA}
	+
	\OOm^{CAB}
	\,=\, 
		0
	\,,
\end{align}
which is the covariant Jacobi identity in the symplectic vielbein formalism.

\para{Gauge Theory, Revisited}

Let us revisit \Sec{CPB>YM}
with the symplectic vielbein formalism.
The gauge-covariant \textit{and} locally canonical coframe and frame are
\begin{subequations}
\label{YM.LEF}
\begin{align}
	\label{YM.LE}
	\ee^A
	\,&=\,
		\BB{
			dx^m
			,\mem
			dp_m
			- \tfrac{1}{2}\,
				q_a\hem F^a{}_{mn}(x)\mem dx^n
			,\mem
			D\vth^i
			,\mem
			D\bvth_i
		}
	\,,\\
	\label{YM.LF}
	\EE_A
	\,&=\,
		\bb{
			\tpartial_m
			- \frac{1}{2}\,
				q_a\hem F^a{}_{mn}(x)\mem \frac{\partial}{\partial p_n}
			,\mem
			\frac{\partial}{\partial p_m}
			,\mem
			\frac{\partial}{\partial \vth^i}
			,\mem
			\frac{\partial}{\partial \bvth_i}
		}
	\,.
\end{align}
\end{subequations}
Together with the covariant coordinates $(x^m,p_m,\vth^i,\bvth_i)$,
\eqref{YM.LEF}
provides a succinct summary
of the exact phase space geometry
$({ \ps \eqq T^*\mflat \moplus E }\hem,\mem \omega)$.

The anholonomy coefficients 
are read off from
$-d\ee^A = \frac{1}{2}\mem \OOm^A{}_{BC}\mem \ee^B \swedge \ee^C$:
\begin{align}
\begin{split}
	\label{YM.Lan}
	- d\bigbig{
		dx^m
	}
	\,&=\,
		0
	\,,\\
	- d\bigbig{
		\pp_m
	}
	\,&=\,
		\tfrac{i}{2}\,
			D(\bvth_i\hem \vth^j)\mem \wedge F^i{}_{jmn}(x)\mem dx^n
		+
			\tfrac{1}{2}\mem
			q_a\hem D_mF^a{}_{rs}(x)
			\,\BB{
				\tfrac{1}{2}\mem
					dx^r \swedge dx^s
			}
	\,,\\
	- d\bigbig{
		D\vth^i
	}
	\,&=\,
		A^i{}_{jm}(x)
		\,\BB{
			dx^m \swedge D\vth^j
		}
		- F^i{}_{jmn}(x)\mem \vth^j
		\,\BB{
			\tfrac{1}{2}\mem
				dx^m \swedge dx^n
		}
	\,,\\
	- d\bigbig{
		D\bvth_i
	}
	\,&=\,
		- A^j{}_{im}(x)
		\,\BB{
			dx^m \swedge D\bvth_j
		}
		+ \bvth_j\hem F^j{}_{imn}(x)
		\,\BB{
			\tfrac{1}{2}\mem
				dx^m \swedge dx^n
		}
	\,.
\end{split}
\end{align}
Here,
$\pp_m := 
	dp_m
	- \tfrac{1}{2}\,
		q_a\hem F^a{}_{mn}\mem dx^n
$.
The Bianchi identity has simplified the second line.

\newpage

There are two sorts of 
nontrivial covariant Jacobiators.
The first kind vanishes easily:
\begin{align}
\begin{split}
	\label{YM.Lcheck1}
	\JJ{D\vth^i}{\pp_m}{\pp_n}
	\,&=\,
		- \OOm^{(D\vth^i)}{}_{(\XX_m)(\XX_n)}
		+ i\, \BB{
			\OOm^{(\pp_m)}{}_{(\psp\bvth_i)(\XX_n)}
			- (m \mlra n)
		}
	\,,\\
	\,&=\,
		\BB{
			F^i{}_{jmn}\mem \vth^j
		}
		+ i\, \BB{
			\tfrac{i}{2}\,
				F^i{}_{jmn}\mem \vth^j
			- (m \mlra n)
		}
	\,=\, 0
	\,.
\end{split}
\end{align}
Here,
$\XX_m := 
	\tpartial_m
	- \frac{1}{2}\,
		q_a\hem F^a{}_{mn}\mem \psp p_n
$.
The second kind vanishes by the Bianchi identity:
\begin{align}
\begin{split}
	\label{YM.Lcheck2}
	\JJ{\pp_m}{\pp_n}{\pp_r}
	\,&=\,
		- \OOm^{(\pp_r)}{}_{(\XX_m)(\XX_n)}
		- \OOm^{(\pp_m)}{}_{(\XX_n)(\XX_r)}
		- \OOm^{(\pp_n)}{}_{(\XX_r)(\XX_m)}
	\,,\\
	\,&=\,
		- \tfrac{3}{2}\mem
		q_a\hem D_\wrap{[m} F^a{}_\wrap{rs]}
	\,=\, 0
	\,.
\end{split}
\end{align}

\para{Gravity, Revisited}

Let us revisit \Sec{CPB>GR}
with the symplectic vielbein formalism.
The covariant \textit{and} locally canonical coframe and frame are
\begin{subequations}
\label{BI.LEF}
\begin{align}
	\label{BI.LE}
	\ee^A
	\,&=\,
		\BB{
			e^m
			,\mem
			\tD p_m
			- \tfrac{1}{2}\,
				p_k\hem \tT^k{}_{mn}(x)\mem e^n
		}
	\,,\\
	\label{BI.LF}
	\EE_A
	\,&=\,
		\bb{
			\tE_m
			- \frac{1}{2}\,
				p_k\hem \tT^k{}_{mn}(x)\mem \frac{\partial}{\partial p_n}
			,\mem
			\frac{\partial}{\partial p_m}
		}
	\,.
\end{align}
\end{subequations}
Together with the covariant coordinates $(x^\m,p_m)$,
\eqref{BI.LEF}
provides a succinct summary
of the exact phase space geometry
$({ \ps \eqq \Tstar\M }\hem,\mem \omega)$.

The anholonomy coefficients 
are read off from
$-d\ee^A = \frac{1}{2}\mem \OOm^A{}_{BC}\mem \ee^B \swedge \ee^C$:
\begin{align}
	\label{BI.Lan}
	- d\bigbig{
		e^m
	}
	\,&=\,
		\Omega^m{}_{rs}(x)
		\,\BB{
			\tfrac{1}{2}\mem
				e^r \swedge e^s
		}
	\,,\\
	- d\bigbig{
		\pp_m
	}
	\,&=\,
\begin{aligned}[t]
	&
		{-\BB{
			\tgamma^n{}_{mr}(x)
			+ \tfrac{1}{2}\,
						\tT^n{}_{mr}(x)
		}}
		\,\BB{
			e^r \swedge \pp_n
		}
		+ p_k\hem \BB{
			\tR^k{}_{mrs}(x)
			+ \Xi^k{}_{mrs}(x)
		}
		\,\BB{
			\tfrac{1}{2}\mem
				e^r \swedge e^s
		}
	\,.
\end{aligned}
\nonumber
\end{align}
Here,
$\pp_m := 
	\tD p_m
	- \tfrac{1}{2}\,
		p_k\hem \tT^k{}_{mn}\mem e^n
$.
Also, $\Xi^k{}_{mrs}(x) = -\Xi^k{}_{msr}(x)$
denotes a tensor field on $\M$.
It arises by
expanding
$\tD( \frac{1}{2}\, p_k\hem \tT^k{}_{mr}\mem e^r )$
and then replacing $\tD p_k$ with $\pp_k + \frac{1}{2}\, p_l\hem \tT^l{}_{ks}(x)\mem e^s$:
\begin{align}
\begin{split}
	\label{Xi}
	\Xi^k{}_{mrs}\mem
		e^r \swedge e^s
	\,&=\, 
		\tfrac{1}{2}\,
		\tT^k{}_{lr}\mem 
		\tT^l{}_{ms}\mem
			e^r \swedge e^s
		+ \tD\BB{ \tT^k{}_{mr}\mem e^r }
	\,,\\
	\,&=\,
		\BB{
			\tD_r \tT^k{}_{ms}
			- \tfrac{1}{2}\,
				\tT^k{}_{lm}\mem \tT^l{}_{rs}
			+ \tfrac{1}{2}\,
				\tT^k{}_{lr}\mem \tT^l{}_{ms}
		}\mem
			e^r \swedge e^s
	\,.
\end{split}
\end{align}


There are two sorts of 
nontrivial covariant Jacobiators.
The first kind vanishes by the identity 
$\tT^r{}_{mn} + \Omega^r{}_{mn} + 2\mem \tgamma^r{}_{[mn]} = 0$
used earlier in \Sec{CPB>BI}:
\begin{align}
\begin{split}
	\JJ{e^r}{\pp_m}{\pp_n}
	\,&=\,
		- \OOm^{(e^r)}{}_{(\XX_m)(\XX_n)}
		+ \BB{
			\OOm^{(\pp_n)}{}_{(\XX_m)(\psp p_r)}
			- (m \mlra n)
		}
	\,,\\
	\,&=\,
		\Omega^r{}_{mn}
		+ 2\mem \BB{
			\tgamma^r{}_\wrap{[mn]}(x)
			+ \tfrac{1}{2}\,
						\tT^r{}_\wrap{mn}
		}
	\,=\, 0
	\,.
\end{split}
\end{align}
Here,
$\XX_m := 
	\tE_m
	- \frac{1}{2}\,
		p_k\hem \tT^k{}_{mn}\mem \psp p_n
$.
The second kind vanishes by the Bianchi identity in \eqref{BI.bianchi}:
\begin{align}
\begin{split}
	\JJ{\pp_m}{\pp_n}{\pp_r}
	\,&=\,
		- \OOm^{(\pp_r)}{}_{(\XX_m)(\XX_n)}
		- \OOm^{(\pp_m)}{}_{(\XX_n)(\XX_r)}
		- \OOm^{(\pp_n)}{}_{(\XX_r)(\XX_m)}
	\,,\\
	\,&=\,
		- 3\mem p_k\hem \BB{
			\tR^k{}_{[rmn]}(x)
			+ \Xi^k{}_{[rmn]}(x)
		}
	\,=\, 0
	\,.
\end{split}
\end{align}
To see this, read off $\Xi^k{}_{mrs}$ from \eqref{Xi} to find
\begin{align}
\begin{split}
	\Xi^k{}_{[mrs]}
	\,=\,
		- \tD_{[m} \tT^k{}_{rs]}
		- 
			\tT^k{}_{l[m}\mem \tT^l{}_{rs]}
	\,=\,
		- \tR^k{}_{[mrs]}
	\,.
\end{split}
\end{align}

\newpage

\para{Gravity with Spin, Revisited}

Let us revisit \Sec{CPB>GR}
with the symplectic vielbein formalism.
The gauge-covariant \textit{and} locally canonical coframe and frame are
\begin{subequations}
\label{GR.LEF}
\begin{align}
	\label{GR.LE}
	\ee^A
	\,&=\,
		\BB{
			e^m
			,\mem
			Dp_m
			+ \tfrac{1}{4}\,
				S_{kl}\hem R^{kl}{}_{mn}(x)\mem e^n
			,\mem
			D\psi^m
			,\mem
			D\bpsi_m
		}
	\,,\\[-0.1\baselineskip]
	\label{GR.LF}
	\EE_A
	\,&=\,
		\bb{
			\tE_m
			+ \frac{1}{4}\,
				S_{kl}\hem R^{kl}{}_{mn}(x)\mem \frac{\partial}{\partial p_n}
			,\mem
			\frac{\partial}{\partial p_m\vphantom{\bpsi^0}}
			,\mem
			\frac{\partial}{\partial \psi^m\vphantom{\bpsi^0}}
			,\mem
			\frac{\partial}{\partial \bpsi_m\vphantom{\bpsi^0}}
		}
	\,.
\end{align}
\end{subequations}
Together with the covariant coordinates $(x^m,p_m,\psi^m,\bpsi_m)$,
\eqref{GR.LEF}
provides a succinct summary
of the exact phase space geometry
$({ \ps \eqq \A_2 = (T^* \moplus \Pi T^\C)\M }\hem,\mem \omega)$.

The anholonomy coefficients 
are read off from
$-d\ee^A = \frac{1}{2}\mem \OOm^A{}_{BC}\mem \ee^B \swedge \ee^C$:
\begin{align}
	\label{GR.Lan}
	- d\bigbig{
		e^m
	}
	\,&=\,
		\Omega^m{}_{rs}(x)
		\,\BB{
			\tfrac{1}{2}\mem
				e^r \swedge e^s
		}
	\,,\\[-0.1\baselineskip]
	\nonumber
	- d\bigbig{
		\pp_m
	}
	\,&=\,
		\tfrac{i}{2}\,
			D(\bpsi_k\hem \psi^l)\mem \wedge R^k{}_{lmn}(x)\mem e^n
		+\BB{
			p_k\hem \tR^k{}_{mrs}(x)
			-
				\tfrac{1}{4}\mem
				S_{kl}\hem D_mR^{kl}{}_{rs}(x)
		}
			\BB{
				\tfrac{1}{2}\mem
					e^r \swedge e^s
			}
	\,,\\[-0.1\baselineskip]
	\nonumber
	- d\bigbig{
		D\psi^m
	}
	\,&=\,
		\gamma^m{}_{nr}(x)
		\,\BB{
			e^r \swedge D\psi^n
		}
		- R^m{}_{nrs}(x)\mem \psi^n
		\,\BB{
			\tfrac{1}{2}\mem
				e^r \swedge e^s
		}
	\,,\\[-0.1\baselineskip]
	\nonumber
	- d\bigbig{
		D\bpsi_m
	}
	\,&=\,
		- \gamma^n{}_{mr}(x)
		\,\BB{
			e^r \swedge D\bpsi_n
		}
		+ \bpsi_n\hem R^n{}_{mrs}(x)
		\,\BB{
			\tfrac{1}{2}\mem
				e^r \swedge e^s
		}
	\,.
\end{align}
Here,
$\pp_m := 
	Dp_m
	+ \tfrac{1}{4}\,
		S_{kl}\hem R^{kl}{}_{mn}\mem e^n
$.
The second Bianchi identity has been used to simplify the second line a bit.

There are three sorts of 
nontrivial covariant Jacobiators.
The first kind vanishes easily:
$J(Dp_m,Dp_n,e^r) = 0$.
The second kind
vanishes by mimicking \eqref{YM.Lcheck1}:
\begin{align}
\begin{split}
	\label{GR.Lcheck1}
	\JJ{D\psi^r}{\pp_m}{\pp_n}
	\,&=\,
		- \OOm^{(D\psi^r)}{}_{(\XX_m)(\XX_n)}
		+ i\, \BB{
			\OOm^{(\pp_m)}{}_{(\psp\psi^r)(\XX_n)}
			- (m \mlra n)
		}
	\,,\\[-0.1\baselineskip]
	\,&=\,
		\BB{
			R^r{}_{smn}\mem \psi^s
		}
		+ i\, \BB{
			\tfrac{i}{2}\,
				R^r{}_{smn}\mem \psi^s
			- (m \mlra n)
		}
	\,=\, 0
	\,.
\end{split}
\end{align}
Here,
$\XX_m := 
	\tE_m
	+ \frac{1}{4}\,
		S_{kl}\hem R^{kl}{}_{mn}\mem \psp p_n
$.
The third kind vanishes by
the first and second Bianchi identities of Riemannian geometry,
shown in \eqref{GR.bianchi}:
\begin{align}
\begin{split}
	\JJ{\pp_m}{\pp_n}{\pp_r}
	\,&=\,
		- \OOm^{(\pp_r)}{}_{(\XX_m)(\XX_n)}
		- \OOm^{(\pp_m)}{}_{(\XX_n)(\XX_r)}
		- \OOm^{(\pp_n)}{}_{(\XX_r)(\XX_m)}
	\,,\\[-0.1\baselineskip]
	\,&=\,
		- 3\mem 
			\BB{
				p_k\hem \tR^k{}_{[mrs]}(x)
				-
					\tfrac{1}{4}\mem
					S_{kl}\hem D_{[m}R^{kl}{}_{rs]}(x)
			}
	\,=\, 0
	\,.
\end{split}
\end{align}

\para{Remarks}
Arguably,
the process of checking covariant Jacobi identities
is simpler
in the symplectic vielbein formalism,
as the 
differential part
is made to vanish by construction.
Especially,
the preliminary calculations
in \Secs{CPB>BI}{CPB>GR},
which employed the auxiliary bivector field $\xi$
and produced extra bare connection terms,
could be entirely omitted.

On a related note,
it is interesting to compare between
the (covariant) Jacobi identities in
(a) the ordinary coordinate frame formalism,
(b) the Ehresmann frame formalism,
and
(c) the symplectic vielbein formalism.
In case (a), information is stored solely on the 
differential part.
In case (b), both 
the differential and algebraic parts
are used.
In case (c), information is stored solely on the 
algebraic part.

Lastly, the simplest test of symplecticity is of course
the direct examination of $d\omega = 0$.
In the symplectic vielbein formalism,
it is not difficult to see that
\begin{align}
	\label{down0}
	d\omega
	\,=\,
		d\ee^A \wedge \ee_A
	\,=\,
		\tfrac{1}{2}\:
			\OOm_{ABC}\,
				\ee^A \swedge \ee^B \swedge \ee^C
	\,=\,
		- \tfrac{1}{3!}\:
			J_{ABC}\,
				\ee^A \swedge \ee^B \swedge \ee^C
	\,.
\end{align}
It is also left as an easy exercise to compute $d\omega$ for our example systems,
in which case one generates the Bianchi identities in \eqrefss{YM.bianchi}{BI.bianchi}{GR.bianchi}
immediately.

\newpage

\section{Covariant Equations of Motion}
\label{CEOM}

Finally, we present the most important application of the covariant Poisson bracket:
manifestly covariant derivations of classical equations of motion.

The geometrical statement of the Hamiltonian equations of motion
is
\begin{align}
	\label{heom}
	\frac{d}{d\t}
	\,=\,
		\Xvec_{\ham(\t)}
	\,.
\end{align}
The \define{time-evolution vector field} $d/d\t \inn \Gamma(T\ps)$ 
defines the time evolution on $\ps$.
The \define{time-evolution generator}
$\ham(\t)$ is a one-parameter family of functions on $\ps$,
where $\t$ is the time parameter.
The coordinate-based approach to the Hamiltonian equations of motion
contracts the each side of \eqref{heom}
with coordinate differentials:
\begin{align}
	\label{heom.1}
	\dot{X}^I
	\,=\,
		\i_{d/d\t} \mem dX^I
	\,=\,
		\cont{dX^I}{\Xvec_{\ham(\t)}}
	\,=\,
		\pb{X^I}{\ham(\t)}
	\,=\,
		\pia{dX^I}{d\ham(\t)}
	\,.
\end{align}
Even if the coordinates $X^I$ are gauge-covariant and non-canonical,
however, \eqref{heom.1} does not manifest covariance
because the bare time derivative $\dot{X}^I$ will not be covariant.

Instead of following this textbook path,
we propose to contract each side of \eqref{heom}
with the gauge-covariant yet non-coordinate coframe,
i.e., the Ehresmann coframe:
\begin{align}
	\label{ceom}
		\i_{d/d\t} \mem \bme^A
	\,=\,
		\pia{\bme^A}{d\ham(\t)}
	\,=\,
		\pia{\bme^A}{\bme^B}\,
		\E_B\act{
			\ham(\t)
		}
	\,.
\end{align}
Since $\ham(\t)$ must be a gauge singlet,
\eqref{ceom} determines the classical equations of motion
in a manifestly covariant way
via the covariant Poisson brackets $\pia{\bme^A}{\bme^B}$.
We refer to \eqref{ceom} as 
the \define{covariant Hamiltonian equations of motion}.

\subsection{Gauge Theory}
\label{CEOM>YM}

\para{Covariant Poisson Geometry}

The colored scalar particle
is a Hamiltonian system
whose phase space $\ps = T^*\mflat \moplus E$
describes physical variables 
$x^m$\:(position),
$p_m$\:(momentum),
and
$q_a$\:(color charge).
In the presence of a background non-abelian gauge field,
their covariant Poisson brackets are
\begin{align}
\begin{split}
	\label{YM.cpb.re}
	\pia{dx^m}{dp_n}
	\,&=\,
		\delta^m{}_n
	\,,\\
	\pia{Dq_a}{Dq_b}
	\,&=\,
		q_c\mem f^c{}_{ab}
	\,,\\
	\pia{dp_m}{dp_n}
	\,&=\,
		q_a\hem F^a{}_{mn}(x)
	\,.
\end{split}
\end{align}

\para{Covariant Equations of Motion}

Contracting
the Hamiltonian vector field of $\frac{1}{2}\mem p^2$
with the covariant one-forms $dx^m$, $dp_m$, and $Dq_a$,
it is easily found that
\begin{align}
\begin{split}
	\label{YM.ceom}
	\dot{x}^m
	\,=\,
		\i_{d/d\t}\mem dx^m
	\,&=\,
		\pia{dx^m}{dp_n}\, p^n
	\,=\,
		p^m
	\,,\\
	\dot{p}_m
	\,=\,
		\i_{d/d\t}\mem dp_m
	\,&=\,
		\pia{dp_m}{dp_n}\, p^n
	\,=\,
		q_a\hem F^a{}_{mn}(x)\, p^n
	\,,\\
	\frac{Dq_a}{d\t}
	\,=\,
		\i_{d/d\t}\mem Dq_a
	\,&=\,
		\pia{Dq_a}{dp_n}\, p^n
	\,=\,
		0
	\,,
\end{split}
\end{align}
where we have used $d(\frac{1}{2}\mem p^2) = dp_m\mem p^m$.
\eqref{YM.ceom} directly reproduces the Wong's equations \cite{wong1970field}
in a manifestly covariant fashion.
Compare this with the textbook derivation in \eqref{YM.eom}
which used the ordinary Poisson bracket.

To elaborate, the time-evolution generator for the colored scalar particle is
\begin{align}
	\ham(\t)
	\,=\,
		\frac{\k^0(\t)}{2}\,
			p^2
		+ \k^1(\t)\mem \bigbig{
			\bvth_i\hem\vth^i - w
		}
	\,,
\end{align}
where
the Lagrange multipliers
$\k^0(\t)$ and $\k^1(\t)$ are sole functions of $\t$,
and $w$ is a constant;
recall the brief discussion in \Sec{NONCAN>YM}.
By worldline parametrization, we are free to choose $\k^0(\t) = 1$.
On the other hand, $\k^1(\t)$ can be simply left agnostic in deriving \eqref{YM.ceom}.
From the complete covariant Poisson brackets in \eqref{YM.cpb},
it is easy to see that
the $\U(1)$ constraint $\bvth_i\hem\vth^i - w$ commutes with
the color charge in the covariant sense:\footnote{
	In the covariantizer mode of thinking,
	we recycle the known Poisson bracket in the free theory
	to immediately recognize
	$\omega^{-1}(Dq_a,D(\bvth_i\hem\vth^i))
		= \omega^\bullet{}^{-1}(Dq_a,D(\bvth_i\hem\vth^i))
		= \omega^\circ{}^{-1}(dq_a,d(\bvth_i\hem\vth^i))
		= \pb{q_a}{\bvth_i\hem\vth^i}^\circ
	$.
}
\begin{align}
\begin{split}
	\pia{Dq_a}{
		d(\bvth_i\hem\vth^i)
	}
	\,&=\,
	\pia{Dq_a}{
		D(\bvth_i\hem\vth^i)
	}
	\,,\\
	\,&=\,
		\pia{Dq_a}{
			D\bvth_i
		}\mem \vth^i
		+
		\bvth_i\,\hem
		\pia{Dq_a}{
			D\vth^i
		}
	\,,\\
	\,&=\,
		\bvth_j\mem (t_a)^j{}_i\mem \vth^i
		- \bvth_i\mem (t_a)^i{}_j\mem \vth^j
	\,=\,
		0
	\,.
\end{split}
\end{align}
If desired, one can also easily derive the covariant equations of motion for $\vth^i$:
\begin{align}
	\label{YM.fermion}
	\frac{D\vth^i}{d\t}
	\,=\,
		\i_{d/d\t}\mem D\vth^i
	\,&=\,
		\pia{D\vth^i}{dp_n}\, p^n
		+ \k^1\,
			\pia{D\vth^i}{D\bvth_j}\, \vth^j
	\,=\,
		- i\mem \k^1\, \vth^i
	\,.
\end{align}

\enter
\subsection{Gravity}
\label{CEOM>BI}

\para{Covariant Poisson Geometry}

The general-relativistic scalar particle
is a Hamiltonian system
whose phase space $\ps = \Tstar\M$
describes physical variables 
$x^\m$\:(position)
and
$p_m$\:(momentum).
In the gravitational background,
their covariant Poisson brackets are
\begin{align}
\begin{split}
	\label{BI.cpb.re(D)}
	\pia{e^m}{e^n}
	\,=\,
		0
	\,,\quad
	\pia{e^m}{Dp_n}
	\,=\,
		\delta^m{}_n
	\,,\quad
	\pia{Dp_m}{Dp_n}
	\,=\,
		0
	\,,
\end{split}
\end{align}
where we have employed the Levi-Civita connection $D$.

\para{Covariant Equations of Motion}

Contracting
the Hamiltonian vector field of $\frac{1}{2}\mem p^2$
with the covariant one-forms $e^m$ and $Dp_m$,
it is easily found that
\begin{align}
\begin{split}
	\label{BI.ceom(D)}
	e^m{}_\m(x)\, \dot{x}^\m
	\,=\,
		\i_{d/d\t}\mem e^m
	\,&=\,
		\pia{e^m}{D p_n}\, p^n
	\,=\,
		p^m
	\,,\\
	\frac{D p_m}{d\t}
	\,=\,
		\i_{d/d\t}\mem D p_m
	\,&=\,
		\pia{D p_m}{D p_n}\, p^n
	\,=\,
		0
	\,,
\end{split}
\end{align}
where we have used $d(\frac{1}{2}\mem p^2) = D(\frac{1}{2}\mem p^2) = D p_m\mem p^m$.
Obviously,
\eqref{BI.ceom(D)} states the geodesic equation
in the first-order form.
Switching to the coordinate frame immediately yields
\begin{align}
	\frac{D}{d\t}\, \dot{x}^\m
	\,=\,
		0
	\qfq
	\ddot{x}^\m + \Gamma^\m{}_{\r\s}(x)\, \dot{x}^\r\mem \dot{x}^\s
	\,=\,
		0
	\,,
\end{align}
where $\Gamma^\m{}_{\r\s}(x)$ are the Christoffel symbols.
Compare this with the textbook derivation 
sketched around \eqref{BI.pb-can},
which uses the ordinary Poisson bracket
and generates partial derivatives of the inverse metric.
The physical meaning is also clear---%
\eqref{BI.ceom(D)} is a direct lift of the free-theory equations of motion
via Einstein's elevator (local inertial frame):
\begin{align}
\begin{split}
	\label{BI.ceom(D)0}
	\dot{x}^\m
	\,=\,
		\i_{d/d\t}\mem dx^m
	\,&=\,
		\pif{x^m}{p_n}\, p^n
	\,=\,
		p^m
	\,,\\
	\dot{p}_m
	\,=\,
		\i_{d/d\t}\mem dp_m
	\,&=\,
		\pif{p_m}{p_n}\, p^n
	\,=\,
		0
	\,.
\end{split}
\end{align}

\newpage

To elaborate, the time-evolution generator for the general-relativistic scalar particle is
\begin{align}
	\ham(\t)
	\,=\,
		\frac{\k^0(\t)}{2}\,
			p^2
	\,,
\end{align}
where $p^2 = \eta^{mn}\hem p_m\hhem p_n$.
Without loss of generality,
we adopt
an affine parameter: $\k^0 = 1$.

\para{With Generic Connection}

More generally, we can also repeat the above derivation with an arbitrary Lorentz-valued connection $\tD$.
The covariant Poisson brackets are
\begin{align}
\begin{split}
	\label{BI.cpb.re}
	\pia{e^m}{e^n}
	\,=\,
		0
	\,,\quad
	\pia{e^m}{\tD p_n}
	\,=\,
		\delta^m{}_n
	\,,\quad
	\pia{\tD p_m}{\tD p_n}
	\,=\,
		p_k\mem \tT^k{}_{mn}(x)
	\,.
\end{split}
\end{align}
Contracting
the Hamiltonian vector field of $\frac{1}{2}\mem p^2$
with the covariant one-forms $e^m$ and $\tD p_m$,
it is easily found that
\begin{align}
\begin{split}
	\label{BI.ceom}
	e^m{}_\m(x)\, \dot{x}^\m
	\,=\,
		\i_{d/d\t}\mem e^m
	\,&=\,
		\pia{e^m}{\tD p_n}\, p^n
	\,=\,
		p^m
	\,,\\
	\frac{\tD p_m}{d\t}
	\,=\,
		\i_{d/d\t}\mem \tD p_m
	\,&=\,
		\pia{\tD p_m}{\tD p_n}\, p^n
	\,=\,
		p_k\mem \tT^k{}_{mn}(x)\, p^n
	\,,
\end{split}
\end{align}
where we have used $d(\frac{1}{2}\mem p^2) = \tD(\frac{1}{2}\mem p^2) = \tD p_m\mem p^m$.
\eqref{BI.ceom} still describes the geodesic equation,
regardless of one's choice of $\tD$.

To see this, 
let $\delta\gamma^m{}_{nr} := \tgamma^m{}_{nr} - \gamma^m{}_{nr}$
describe the difference between the arbitrary connection $\tD$
and the Levi-Civita connection $D$,
which is a \textit{tensor}.
Then the second equation in \eqref{BI.ceom} boils down to
\begin{align}
\begin{split}
	0
	\,&=\,
		\frac{\tD p_m}{d\t}
		- p_k\mem \tT^k{}_{mn}\, p^n
	\,,\\
	\,&=\,
		\frac{Dp_m}{d\t}
		- p_k\mem \BB{
			\delta\gamma^k{}_{mn} + \tT^k{}_{mn}
		}\, p^n
	\,,\\
	\,&=\,
		\frac{Dp_m}{d\t}
		- p_k\mem \BB{
			2\mem \delta\gamma^k{}_{[mn]} + \tT^k{}_{mn}
		}\, p^n
	\,.
\end{split}
\end{align}
Meanwhile, from the torsion-free condition of the Levi-Civita connection,
one finds
\begin{align}
	0
	\,=\,
		De^k
	\,=\,
		\tD e^k
		- \bigbig{\delta\gamma^k{}_{nm}\mem e^m} \wedge e^n
	\,=\,
		\tT^k
		+ \delta\gamma^k{}_{[mn]}\mem e^m \wedge e^n
	\,,
\end{align}
which implies
$\tT^k{}_{mn} + 2\mem \delta\gamma^k{}_{[mn]} = 0$.
Therefore, the actual contents of
\eqrefs{BI.ceom(D)}{BI.ceom}
are the same.

\eqref{BI.ceom}
provides an exotic yet intriguing reformulation of the geodesic equation
in the form of the Lorentz force law,
where $p_k\mem \tT^k{}_{mn}$ serves as the ``field strength.''
Namely,
$p_k\mem \tT^k{}_{mn}(x)\, p^n$ in \eqref{BI.ceom}
is the \textit{gravitational force}
represented as a Lorentz-type force.
It will be interesting to assess this fact
from the angle of gravito-electromagnetism \cite{wald2010general}.
In the case of Born-Infeld theory as a diffeomorphism-valued gauge theory,
this Lorentz force interpretation is not a mere analogy
but an exact, literal statement,
due to the discussion in \Sec{NONCOORD>BI}.

The reason why such Lorentz force formulation arises
should be clear:
we have fixed the Hamiltonian constraint to the free-theory form
as a full implementation of the Souriau philosophy,
so time evolution must describe successive Lorentz transformations
on the orthonormal-frame momentum $p_m$:
the momentum-squared $p^2$ is dynamically conserved.
Any force should describe a Lorentz-type force
if formulated as a sole effect of symplectic perturbations.

\newpage

\subsection{Gravity with Spin}
\label{CEOM>GR}

\para{Covariant Poisson Geometry}

The general-relativistic spinning particle
is a Hamiltonian system
whose phase space $\ps = \A_2 \eqq (T^* \moplus \Pi T^\C)\M$
describes
real bosonic variables $x^\m$, $p_m$
and a complex fermionic variable $\psi^m$.
While $x^\m$ and $p_m$ describe
position and momentum,
the worldline fermion $\psi^m$ implements
spin $S_{mn} \eqq {-2i\mem \bpsi_{[m} \psi_{n]}}$
as a composite object.
In the gravitational background,
their covariant Poisson brackets are
\begin{align}
\begin{split}
	\label{GR.cpb.re}
	\pia{e^m}{Dp_n}
	\,&=\,
		\delta^m{}_n
	\,,\\
	\pia{D\psi^m}{D\bpsi_n}
	\,&=\,
		-i\mem \delta^m{}_n
	\,,\\
	\pia{Dp_m}{Dp_n}
	\,&=\,
		i\mem \bpsi_k\mem R^k{}_{lmn}(x)\mem \psi^l
	\,.
\end{split}
\end{align}

\para{Supersymmetry and Constraints}

This particle is a constrained system
characterized by $\mathcal{N} \eqq\, 2$ supersymmetry.
The supercharges and the R-symmetry generator are
\begin{align}
	\label{susy2}
	Q \,=\, p_m\mem \psi^m
	\,,\quad
	\bQ \,=\, p_m\mem \bpsi^m
	\,,\quad
	N \,=\, \bpsi_m\hem \psi^m - w
	\,,
\end{align}
where $w$ is a constant.
The definitions in \eqref{susy2}
are kept the same
regardless of the gravitational interactions.
The mass-shell constraint $H$, however,
is taken as a derived object
defined as
\begin{align}
	\label{H2}
	H
	\,=\,
		\frac{i}{2}\,
		\pb{Q}{\bQ}
	\,=\,
		\frac{1}{2}\,
			p^2
		+ \frac{1}{8}\,
			S^{kl}\hem R_{klmn}(x)\mem S^{mn}
	\,.
\end{align}
Again, 
the covariant Poisson bracket
serves as an excellent tool for
deriving \eqref{H2} in a manifestly covariant fashion:
\begin{align}
\begin{split}
	\label{QQ2.pb}
	\pb{Q}{\bQ}
	\,&=\,
		\pia{dQ}{d\bQ}
	\,=\,
		\pia{DQ}{D\bQ}
	\,,\\
	\,&=\,
		p_m\,
		\pia{D\psi^m}{D\bpsi^n}
		\,p_n
		+
		\psi^m\,
		\pia{Dp_m}{Dp_n}
		\,\bpsi^n
	\,,\\
	\,&=\,
		-i\mem p^2
		+
		\psi^m\,
		\BB{
			i\mem \bpsi_k\mem R^k{}_{lmn}(x)\mem \psi^l
		}
		\,\bpsi^n
	\,.
\end{split}
\end{align}
Unlike in our non-spinning examples,
the mass-shell constraint now exhibits a deformation in the presence of curvature.
Physically,
the Hamiltonian perturbation
$
	\frac{1}{8}\,
		S^{kl}\hem R_{klmn}(x)\mem S^{mn}
$
in \eqref{H2}
incorporates a quadrupolar spin coupling
(known as gravimagnetic ratio \cite{Khriplovich:1997ni}).

\para{Hamiltonian Flows}

From $dQ = DQ = Dp_m\mem \psi^m + p_m\mem D\psi^m$,
the supersymmetry action on the phase space
is covariantly characterized by
\begin{align}
\begin{split}
	\label{GR.Qaction}
	\cont{e^m}{\Xvec_Q}
	\,&=\,
		\pia{e^m}{DQ}
	\,=\,
		\psi^m
	\,,\\
	\cont{Dp_m}{\Xvec_Q}
	\,&=\,
		\pia{Dp_m}{DQ}
	\,=\,
		i\mem \bpsi_k\mem R^k{}_{lmn}\mem \psi^l\mem \psi^n
	\,,\\
	\cont{D\psi^m}{\Xvec_Q}
	\,&=\,
		\pia{D\psi^m}{DQ}
	\,=\,
		0
	\,,\\
	\cont{D\bpsi_m}{\Xvec_Q}
	\,&=\,
		\pia{D\bpsi_m}{DQ}
	\,=\,
		-i\mem p_m
	\,.
\end{split}
\end{align}
In the same way,
the action of the mass-shell constraint
is covariantly characterized by
\begin{align}
	\label{GR.Haction}
	\cont{e^m}{\Xvec_H}
	\,&=\,
		\pia{e^m}{DH}
	\,=\,
		p^m
	\,,\\
\nonumber
	\cont{Dp_m}{\Xvec_H}
	\,&=\,
		\pia{Dp_m}{DH}
	\,=\,
		i\mem \bpsi_k\mem R^k{}_{lmn}\mem \psi^l
			\mem p^n
		+ \tfrac{1}{2}\,
			D_mR_{klrs}\mem 
				\bpsi^k \psi^l\hem \bpsi^r \psi^s
	\,,\\
\nonumber
	\cont{D\psi^m}{\Xvec_H}
	\,&=\,
		\pia{D\psi^m}{DH}
	\,=\,
		i\mem
			R^m{}_{nrs}\mem 
				\psi^n\hem \bpsi^r \psi^s
	\,,\\
\nonumber
	\cont{D\bpsi_m}{\Xvec_H}
	\,&=\,
		\pia{D\bpsi_m}{DH}
	\,=\,
		i\mem
			\bpsi_n\mem
			R^n{}_{mrs}\mem 
				\bpsi^r \psi^s
	\,.
\end{align}
The action of the R-symmetry is easy.

\newpage

\para{Constraint Algebra}

The constraints $Q$, $\bQ$, $H$, and $N$ are of the first class:
they form a closed algebra under the Poisson bracket.
The nontrivial computations are
\begin{subequations}
\label{calg2}
\begin{align}
	\label{calg2.QQ}
	\pb{Q}{Q} 
	\,&=\, 
		i\mem \bpsi_k\mem R^k{}_{lmn}\mem \psi^l\,
			\psi^m \psi^n
	\,=\,
		0
	\,,\\
	\label{calg2.QH}
	\pb{Q}{H}
	\,&=\,
		\tfrac{1}{2}\,
			\psi^m\mem
			D_mR_{klrs}\mem 
				\bpsi^k \psi^l\hem \bpsi^r \psi^s
		+ i\mem p^m\mem
		\BB{
			R_{mnrs}
			- R_{rsmn}
		}\mem \psi^n\hem \bpsi^r \psi^s
	\,=\,
		0
	\,,
\end{align}
\end{subequations}
which all vanish by the Bianchi identities and the index symmetry
of the Riemann tensor.
It may be interesting to examine the Lie brackets between
$\Xvec_Q$, $\Xvec_\bQ$, $\Xvec_H$, $\Xvec_N$
via \eqref{homomorphism.nc}.

%

\para{Covariant Equations of Motion}

The time-evolution generator
that summarizes all the constraints is
\begin{align}
	\label{GR.ham}
	\ham(\t)
	\,=\,
		\k^0(\t)\mem H
		\mem+\mem \chi^1(\t)\mem Q
		\mem+\mem \bQ\mem \bchi^1(\t)
		\mem+\mem \k^1(\t)\mem N
	\,,
\end{align}
where $\k^0(\t)$ and $\k^1(\t)$ are
real bosonic
Lagrange multipliers
while $\chi^1(\t)$ is complex fermionic.

%

From \eqrefss{GR.Qaction}{GR.Haction}{GR.ham},
the covariant Hamiltonian equations of motion are readily found.
Due to the first-class property,
$\dot{Q}$, $\dot{\bQ}$, $\dot{H}$, and $\dot{N}$ all identically vanish on the constraint surface
while
demanding no condition on the Lagrange multipliers.
Hence,
without loss of generality,
we fix $\k^0 = 1$, $\chi^1 = 0$, and $\k^1 = 0$
to find
\begin{subequations}
\label{GR.ceom.simp}
\begin{align}
\label{GR.ceom.simp.x}
	&e^m{}_\m(x)\, \dot{x}^\m
	\,=\,
		p^m
	\,,\\
\label{GR.ceom.simp.p}
	&\frac{Dp_m}{d\t}
	\,=\,
		\BB{
			- \tfrac{1}{2}\mem
			R_{mnrs}\mem S^{rs}
		}\mem p^n
		+ \tfrac{1}{8}\mem
			D_mR_{klrs}\mem
				S^{kl} S^{rs}
	\,,\\
\label{GR.ceom.simp.psi}
	&\frac{D\psi^m}{d\t}
	\,=\,
		\BB{
			- \tfrac{1}{2}\mem
			R^m{}_{nrs}\mem S^{rs}
		}\hem \psi^n
	\,.
\end{align}
\end{subequations}
Physically, \eqref{GR.ceom.simp.p} describes the spin-induced gravitational Lorentz force
plus a derivative interaction 
$\tfrac{1}{8}\mem
			D_mR_{klrs}\mem
				S^{kl} S^{rs}
$
as a Stern-Gerlach type force.

In the regime where the derivatives of the Riemann curvature are negligible,
\eqref{GR.ceom.simp} describes that
the momentum $p^m$ and the worldline fermion $\psi^m$
exhibit the same precession frequency.
This can be identified as the ideal spin precession behavior,
which pinpoints a specific value of the gravimagnetic ratio
(cf. \rcite{sst-asym}).


\subsection{Gauge Theory with Spin}

\label{CEOM>YMS}

Finally, we derive the covariant Hamiltonian equations of motion for a new example
for the sake of completeness:
colored spinning particle.

\para{Covariant Poisson Geometry}

The colored spinning particle
is a straightforward extension
of the colored scalar particle
by a fermionic sector.
The phase space is $\ps = \A_1 := (T^* \moplus \Pi T)\mflat \oplus E$,
which describes
real bosonic coordinates $x^m$, $p_m$,
real fermionic coordinates $\psi^m$,
and
complex fermionic coordinates $\vth^i$.
%
The spin angular momentum
is a composite,
$S_{mn} \eqq {-2i\mem \psi_{[m} \psi_{n]}}$.
The symplectic form for the $\psi$ sector is
$\tfrac{i}{2}\, d\psi_m \swedge d\psi^m$.
As a result, the following covariant Poisson brackets arise
in the gauge theory background:
\begin{align}
\begin{split}
	\label{YMS.cpb}
	\pia{dx^m}{dp_n}
	\,&=\,
		\delta^m{}_n
	\,,\\
	\pia{dp_m}{dp_n}
	\,&=\,
		q_a\hem F^a{}_{mn}(x)
	\,,\\
	\pia{d\psi^m}{d\psi^n}
	\,&=\,
		-i\mem \eta^{mn}
	\,,\\
	\pia{Dq_a}{Dq_b}
	\,&=\,
		q_c\mem f^c{}_{ab}
	\,.
\end{split}
\end{align}
\newpage

\para{Supersymmetry and Constraints}

This particle is a constrained system
characterized by $\mathcal{N} \eqq\, 1$ supersymmetry.
The supercharge and the mass-shell constraint are
\begin{align}
	\label{susy1}
	Q \,=\, p_m\mem \psi^m
	\,,\quad
	H
	\,=\,
		\frac{i}{2}\,
		\pb{Q}{Q}
	\,=\,
		\frac{1}{2}\,
			p^2
		- \frac{1}{4}\,
			q_a\hem F^a{}_{mn}(x)\mem S^{mn}
	\,.
\end{align}
Again, a Hamiltonian perturbation
$
	- \frac{1}{4}\,
	q_a\hem F^a{}_{mn}(x)\mem S^{mn}
$
arises by fixing the definition of the supercharge.
This incorporates the dipolar spin coupling, i.e., the gyromagnetic ratio $g$.
Lastly, we recall
the $\U(1)$ constraint in the color sector
from \Sec{NONCAN>YM}:
$N = \bvth_i\hem \vth^i - w$.

$Q$, $H$, and $N$ are first-class constraints.
In particular, $\pb{Q}{H} = 0$ holds by the Bianchi identity:
$D_r F^a{}_{mn}\mem \psi^r \psi^m \psi^n = 0$.

\para{Hamiltonian Flows}

From $dQ = DQ = Dp_m\mem \psi^m + p_m\mem D\psi^m$,
the supersymmetry action on the phase space
is covariantly characterized by
\begin{align}
\begin{split}
	\label{YMS.Qaction}
	\cont{dx^m}{\Xvec_Q}
	\,&=\,
		\pia{dx^m}{DQ}
	\,=\,
		\psi^m
	\,,\\
	\cont{dp_m}{\Xvec_Q}
	\,&=\,
		\pia{dp_m}{DQ}
	\,=\,
		q_a F^a{}_{mn}\mem \psi^n
	\,,\\
	\cont{d\psi^m}{\Xvec_Q}
	\,&=\,
		\pia{d\psi^m}{DQ}
	\,=\,
		-i\mem p^m
	\,,\\
	\cont{Dq_a}{\Xvec_Q}
	\,&=\,
		\pia{Dq_a}{DQ}
	\,=\,
		0
	\,.
\end{split}
\end{align}
In the same way,
the action of the mass-shell constraint
is covariantly characterized by
\begin{align}
	\label{YMS.Haction}
	\cont{dx^m}{\Xvec_H}
	\,&=\,
		\pia{dx^m}{DH}
	\,=\,
		p^m
	\,,\\
\nonumber
	\cont{dp_m}{\Xvec_H}
	\,&=\,
		\pia{dp_m}{DH}
	\,=\,
		q_a F^a{}_{mn}
			\mem p^n
		+ \tfrac{1}{4}\,
			q_a\hem 
			D_m F^a{}_{rs}\mem 
				S^{rs}
	\,,\\
\nonumber
	\cont{d\psi^m}{\Xvec_H}
	\,&=\,
		\pia{d\psi^m}{DH}
	\,=\,
		q_a F^a{}_{mn}\mem \psi^n
	\,,\\
\nonumber
	\cont{Dq_a}{\Xvec_H}
	\,&=\,
		\pia{Dq_a}{DH}
	\,=\,
		\tfrac{1}{4}\,
			q_b\mem f^b{}_{ca}\,
			F^c{}_{mn}(x)\mem S^{mn}
	\,.
\end{align}
The action of the $\U(1)$ constraint $N$ is easy.

\para{Covariant Equations of Motion}

The time-evolution generator
that summarizes all the constraints is
\begin{align}
	\label{YMS.ham}
	\ham(\t)
	\,=\,
		\k^0(\t)\mem H
		\mem+\mem Q\mem \chi^1(\t)
		\mem+\mem \k^1(\t)\mem N
	\,,
\end{align}
where $\k^0(\t)$ and $\k^1(\t)$ are
real bosonic
Lagrange multipliers
while $\chi^1(\t)$ is complex fermionic.

From \eqrefss{YMS.Qaction}{YMS.Haction}{YMS.ham},
the covariant Hamiltonian equations of motion are readily found.
Again, we can set $\chi^1 = 0$ without loss of generality
to have
\begin{align}
\begin{split}
	\label{YMS.ceom.simp}
	\dot{x}^m
	\,&=\,
		p^m
	\,,\\
	\dot{p}_m
	\,&=\,
		q_a F^a{}_{mn}\mem p^n
		+ \tfrac{1}{4}\,
			q_a\mem D_mF^a{}_{rs}\mem S^{rs}
	\,,\\
	\dot{\psi}^m
	\,&=\,
		q_a F^a{}^m{}_n\mem \psi^n
	\,,\\
	\frac{Dq_a}{d\t}
	\,&=\,
		\frac{1}{4}\,
			q_b\mem f^b{}_{ca}\,
			F^c{}_{mn}\mem S^{mn}
\end{split}
\end{align}
Physically, \eqref{YMS.ceom.simp} describes the non-abelian Lorentz force
plus a derivative interaction 
$\frac{1}{4}\,
	q_a\mem D_mF^a{}_{rs}\mem S^{rs}
$
as a Stern-Gerlach type force.

Again,
the orbital and spin precessions are synchronized 
in the small-derivative regime.
On account of the Bargmann-Michel-Telegdi equations \cite{Bargmann:1959gz},
this spin precession behavior
describes the gyromagnetic ratio $g = 2$: the Dirac value.
See \cite{Holstein:2006wi}
for a discussion on
the ideal nature of $g = 2$.

\newpage

\section{Variational Principle and Path Integral Origins}
\label{PI}

Lastly, we explicate how
the covariant Hamiltonian equations of motion in \Sec{CEOM}
are reproduced from the variational principle.
We also identify a path integral origin of the covariant Poisson bracket,
based on a tree-level computation.

\para{First-Order Variations}

Firstly, let us recall how the Hamiltonian equations of motion
are derived from the variational principle.
A generic phase-space action takes the form
\begin{align}
	\label{action.coord}
	S[X]
	\,=\,
		\int d\t\,\,
			\BB{
				\theta_I(X(\t))\mem \dot{X}^I\hnem(\t)
				- \ham(X(\t),\t)
			}
	\,,
\end{align}
which defines a one-dimensional sigma model
to a polarized symplectic manifold $(\ps,\mem \omega \eqq d\theta)$
with a time-dependent Hamiltonian.

An infinitesimal variation is a worldline field
$\t \mapsto \dX^I\hnem(\t)$
that describes vectors attached to the worldline.
Based on the Cartan formula
$\pounds_{{\d}X} \theta = \i_{{\d}X} \omega + d(\i_{{\d}X}\theta)$,
we discard a boundary term
to identify
the first-order bulk variation of the action in \eqref{action.coord}
as
\begin{align}
	\label{vaction.coord}
	{\d}^{(1)}\nem S[X,\dX]
	\,=\,
		\int d\t\,\,
			\dX^I\hnem(\t)\, 
			\BB{
				\omega_{IJ}(X)\mem \dot{X}^I\hnem(\t)
				- \ham_{,I}(X(\t),\t)
			}
	\,.
\end{align}
Imposing ${\d}^{(1)}\nem S[X,\dX] = 0$ for all $\dX^I$,
we derive the Hamiltonian equations of motion:
\begin{align}
	\dot{X}^I(\t)
	\,=\,
		(\omega^{-1}\hnem(X(\t)))^{IJ}\mem \ham_{,J}(X(\t),\t)
	\,.
\end{align}

\para{Second-Order Variations}

Secondly,
let us
review the fact \cite{deformation-quantification}
that
the Poisson bracket
arises from computing tree-level correlation functions
with a topological model.

Consider the action
\begin{align}
	\label{top.coord}
	S_\top[X]
	\,=\,
		\int d\t\,\,
			\theta_I(X(\t))\mem \dot{X}^I\hnem(\t)
	\,,
\end{align}
which defines a one-dimensional sigma model
to $(\ps,\mem \omega \eqq d\theta)$.
This is a topological theory
in the sense of reparametrization invariance \cite{deformation-quantification}:
it describes \textit{zero Hamiltonian}.

The saddle of \eqref{top.coord} describes a point at rest:
$\dot{X}^I\hnem(\t) = 0$.
We perform a background field theory computation around this constant-point saddle.
By again using the Cartan formula
to compute ${\pounds_{\dX}}^2 \theta$,
the second-order variation of \eqref{top.coord},
discarding boundary terms,
and imposing $\dot{X}^I\hnem(\t) = 0$ on the background,
can be found as
\begin{align}
	\label{2ndS}
	S_\top^{(2)}[X,\dX]
	\,&=\,
		\int d\t\,\,
			\omega_{IJ}(X)\, \dX^I \d\dot{X}^J
	\,.
\end{align}
To clarify, $X$ in \eqref{2ndS} is the constant point background while $\dX$ is dynamical.
The free propagator due to \eqref{2ndS} is
\begin{align}
	\expval{
		\dX^I\hnem(\t_1)\,
		\dX^J\hnem(\t_2)
	}
	\,=\,
		\frac{i\hbar}{2}\:
		(\omega^{-1}\hnem(X))^{IJ}
		\: \Theta(\t_1,\t_2)
	\,,
\end{align}
where $\Theta(\t_1,\t_2)$ is a symmetric first-order Green's function:
\begin{align}
	\label{prop0}
	\Theta(\t_1,\t_2)
	\,=\,
		\left\{\:\,
		\begin{aligned}[c]
			+ 1 
			&\quad
				(\t_1{\:>\:}\t_2)
			\\
			- 1 
			&\quad
				(\t_1{\:<\:}\t_2)
		\end{aligned}
		\:\right\}
	\,.
\end{align}
Note that \eqref{prop0} is only sensitive to the topological information between the two times $\t_1,\t_2 \in \R$,
i.e., their relative ordering.
This reflects that the theory is topological \cite{deformation-quantification}.

\newpage

\begin{figure}
	\begin{align*}
		\Big\langle\,\,
		\adjustbox{valign=c}{\begin{tikzpicture}[]
				\node (F) at (0,0) {};
				\node (G) at (0.75,0) {};
				\node (xshift) at (0.4,0) {};
				\node[empty] (Fe) at ($(F)-(xshift)$) {};
				\node[empty] (Ge) at ($(G)+(xshift)$) {};
				\draw[dprop] (Fe)--(F)--(G)--(Ge);
				\node[dot] (f) at ($(F)$) {};
				\node[dot] (g) at ($(G)$) {};
		\end{tikzpicture}}
		\,\,\Big\rangle
		\,\,\,\,=\,\,\,\,
		\adjustbox{valign=c}{\begin{tikzpicture}[]
				\node[empty] (Fe) at ($(F)-(xshift)$) {};
				\node[empty] (Ge) at ($(G)+(xshift)$) {};
				\draw[dprop] (Fe)--(F)--(G)--(Ge);
				\node[dot] (f) at ($(F)$) {};
				\node[dot] (g) at ($(G)$) {};
		\end{tikzpicture}}
		\,\,+\,\,
		\adjustbox{valign=c}{\begin{tikzpicture}[]
				\node[empty] (Fe) at ($(F)-(xshift)$) {};
				\node[empty] (Ge) at ($(G)+(xshift)$) {};
				\draw[dprop] (Fe)--(F)--(G)--(Ge);
				\node[dot] (f) at ($(F)$) {};
				\node[dot] (g) at ($(G)$) {};
				\draw[l] (f) -- (g);
		\end{tikzpicture}}
		\,\,+\,\,
		\adjustbox{valign=c}{\begin{tikzpicture}[]
				\node[empty] (Fe) at ($(F)-(xshift)$) {};
				\node[empty] (Ge) at ($(G)+(xshift)$) {};
				\draw[dprop] (Fe)--(F)--(G)--(Ge);
				\node[dot] (f) at ($(F)$) {};
				\node[dot] (g) at ($(G)$) {};
				\draw[l] (f) to[out=+60, in=+120, looseness=1.1] (g);
				\draw[l] (f) to[out=-60, in=-120, looseness=1.1] (g);
		\end{tikzpicture}}
		\,\,+\,\,
			\cdots
	\end{align*}
	\caption{
		The operator product from path integral.
	}
	\label{moyal}
\end{figure}

If desired,
higher-order variations
can also be systematically identified by
constructing a normal coordinate system around the constant point
\cite{gde,CSG-qu}.
In this paper,
it suffices to work out the lowest nontrivial orders
since we restrict our attention to tree-level diagrams.

Suppose two functions $f,g\in\Cinfty(\ps)$ on the phase space.
We compute their correlation function
by inserting them at two separate times
$\t_1 = \e>0$ and $\t_2 = 0$:
$\langle f(X(\e))\mem g(X(0)) \rangle$.
After the background-fluctuation split, this is
\begin{align}
	\label{expval.fg}
	\Big\langle{\:
		\BB{
			f(X) + \dX^I\hnem(\e)\mem f_{,I}(X) + \cdots
		}
		\BB{
			g(X) + \dX^J\hnem(0)\mem g_{,J}(X) + \cdots
		}
	\:}\Big\rangle
	\,.
\end{align}
Here, the expectation value describes
the path integral
\begin{align}
	\label{pi.coord}
	\int \mathcal{D}\dX\,\,
		\exp\bb{
			-\frac{1}{i\hbar}
			\int d\t\,\,
			\BB{
				\omega_{IJ}(X)\, \dX^I\hnem(\t)\mem \d\dot{X}^J\hnem(\t)
				+ \cdots
			}
		}
	\,.
\end{align}
At tree level,
\eqref{expval.fg} evaluates to
\begin{align}
	\label{tree.fg}
		f(X)\mem g(X)
		+ \frac{i\hbar}{2}\:
			f_{,I}(X)
				\,
				(\omega^{-1}\hnem(X))^{IJ}
				\hem
			g_{,J}(X)
		+ \O(\hbar^2)
	\,.
\end{align}
In principle, one could proceed to higher loop orders as illustrated in \fref{moyal}.

In this exploration,
the time ordering $\t_1 > \t_2$ on the worldline
is used as a trick for
implementing the operator ordering.
By taking the limit $\e \too 0$ with $\e > 0$,
\rcite{deformation-quantification}
identifies \eqref{tree.fg} as encoding the quantum operator product
$\hat{f}\hem \hat{g}$ realized at the constant point $X$.
By the correspondence principle,
the $\O(\hbar^0)$ part of
$(\hat{f}\hem \hat{g} - \hat{g}\hem \hat{f})/i\hbar$
must be the Poisson bracket.
This derives
\begin{align}
	\label{tree-res.coord}
	\pb{f}{g}
	\,=\,
		f_{,I}
			\,
			(\omega^{-1})^{IJ}
			\hem
		g_{,J}
	\,.
\end{align}

\para{Covariant Variations}

The idea of covariant variation
is to perform a \textit{worldline field redefinition}
for the fluctuation field,
based on an Ehresmann frame given to $\ps$:
\begin{align}
	\label{wl-fr}
	\dX^A\hnem(\t)
	\,:=\,
		\bme^A{}_I(X(\t))\, \dX^I\hnem(\t)
	\,.
\end{align}
In the field basis $(X^I,\dX^A)$,
\eqref{vaction.coord} appears as
\begin{align}
	\label{vaction.nc}
	{\d}^{(1)}\nem S[X,\dX]
	\,=\,
		\int d\t\,\,
			\dX^A\hnem(\t)\, 
			\BB{
				\omega_{AB}(X)\mem \bme^A{}_I(X(\t))\mem \dot{X}^I\hnem(\t)
				- \ham_{|A}(X(\t),\t)
			}
	\,.
\end{align}
Imposing ${\d}^{(1)}\nem S[X,\dX] = 0$ for all $\dX^I$,
we obtain
\begin{align}
	\bme^A{}_I(X(\t))\mem \dot{X}^I\hnem(\t)
	\,=\,
		(\omega^{-1}\hnem(X(\t)))^{AB}\, \ham_{|B}(X(\t),\t)
	\,.
\end{align}
This is precisely the covariant Hamiltonian equations of motion in \eqref{ceom}.
Here, we have denoted $f_{|A} := \E_A\act{ f }$.
Therefore, the covariant Hamiltonian equations of motion
are directly derived by varying the phase space action
by defining the worldline fluctuations in the Ehresmann frame.

\newpage

The field redefinition in \eqref{wl-fr}
can be also applied to
the topological action in \eqref{2ndS}.
In this case,
the perturbation theory around the constant-point saddle
describes the path integral
\begin{align}
	\label{pi.nc}
	\int \mathcal{D}\dX\,\,
		\exp\bb{
			-\frac{1}{i\hbar}
			\int d\t\,\,
			\BB{
				\omega_{AB}(X)\, \dX^A\hnem(\t)\mem {\d}\dot{X}^B\hnem(\t)
				+ \cdots
			}
		}
	\,.
\end{align}
The bare propagator due to \eqref{pi.nc} is
\begin{align}
	\label{prop.nc}
	\expval{
		\dX^A\hnem(\t_1)\,
		\dX^B\hnem(\t_2)
	}
	\,=\,
		\frac{i\hbar}{2}\:
		(\omega^{-1}\hnem(X))^{AB}
		\: \Theta(\t_1,\t_2)
	\,.
\end{align}
By using the same trick of implementing the operator ordering
via time ordering,
\eqref{prop.nc} implies that, in the operator notation,
\begin{align}
	\frac{1}{i\hbar}\,
	\BB{
		\hat{\dX}{}^A\mem \hat{\dX}{}^B
		- \hat{\dX}{}^B\mem \hat{\dX}{}^A
	}
	\,=\,
		(\omega^{-1}\hnem(X))^{AB}
		+ \O(\hbar^1)
	\,.
\end{align}
The correspondence principle thus asserts that
the Poisson bracket ``$\pb{\dX^A}{\dX^B}$'' 
is exactly
$(\omega^{-1}\hnem(X))^{AB}$.
This deciphers the path integral origin of
the covariant Poisson bracket
in \eqref{cpb}:
the tree-level correlator between
worldline fluctuations
in the gauge-covariant field basis.

One can also compute the correlator
$\langle f(X(\e))\mem g(X(0)) \rangle$.
In the gauge-covariant perturbation scheme, this describes
\begin{align}
	\label{expval.fg.nc}
	\Big\langle{\:
		\BB{
			f(X) + \dX^A\hnem(\e)\mem f_{|A}(X) + \cdots
		}
		\BB{
			g(X) + \dX^B\hnem(0)\mem g_{\hhem|B}(X) + \cdots
		}
	\:}\Big\rangle
	\,.
\end{align}
At tree level, \eqref{expval.fg.nc} evaluates to
\begin{align}
	\label{tree.fg.nc}
		f(X)\mem g(X)
		+ \frac{i\hbar}{2}\:
			f_{|I}(X)
				\,
				(\omega^{-1}\hnem(X))^{IJ}
				\hem
			g_{|J}(X)
		+ \O(\hbar^2)
	\,,
\end{align}
which, through the correspondence principle, derives
\begin{align}
	\label{tree-res.nc}
	\pb{f}{g}
	\,=\,	
		\omega^{-1}(df,dg)
	\,=\,
		f_{|A}
			\,
			(\omega^{-1})^{AB}
			\hem
		g_{|B}
	\,=\,
		\E_A\act{ f }
			\,
			(\omega^{-1})^{AB}
			\,
		\E_B\act{ g }
	\,.
\end{align}

Overall, the lesson is that
the covariant frameworks
arise by defining and quantizing the fluctuation fields
in the gauge-covariant yet non-coordinate fashion.
It remains to elaborate on why such non-coordinate variations
are physically demanded
in each of our examples.

\subsection{Gauge Theory}

\para{Covariant Variation}

The first-order action of the colored scalar particle is
\begin{align}
	\label{action.YM}
		\int d\t\,\,
		\bb{
			p_m\mem \dot{x}^m
			+ i\mem \bvth_i\mem 
				\BB{
					\dot{\vth}^i
					+ A^i{}_{jr}(x)\mem \vth^j
						\,\dot{x}^r
				}
			- \ham(x,p,\vth,\bvth,\t)
		}
	\,.
\end{align}
In the spacetime picture,
\eqref{action.YM} describes a worldline $\t \mapsto x^m(\t)$
attached with
covectors $\t \mapsto p_m(\t)$
and color variables $\t \mapsto \vth^i(\t)$, $\t \mapsto \bvth_i(\t)$.

Now suppose a variation of this worldline:
\begin{align}
	\label{action.YM.prime}
		\int d\t\,\,
		\bb{
			p'_m\mem \dot{x}'^m
			+ i\mem \bvth'_i\mem 
				\BB{
					\dot{\vth}'^i
					+ A^i{}_{jr}(x')\mem \vth'^j
						\,\dot{x}'^r
				}
			- \ham(x',p',\vth',\bvth',\t)
		}
	\,.
\end{align}
The worldline changes its shape in spacetime.
The covector and color variables are now attached to this new worldline.
In particular, the new color variable $\vth'^i$
belongs to the point $x'$,
while the original color variable $\vth^i$
belongs to the point $x$.
Their behaviors
under gauge transformations
are different:
the former transforms by the gauge parameter at $x'$,
while the latter transforms by the gauge parameter at $x$.

For this reason,
one should not define the variation as
${\d}\vth^i =$ ``\hem$\vth'^i - \vth^i$'',
since it is utterly nonsensical to compare between vectors at different points as such:
they belong to different fibers of the vector bundle $E$.
To resolve this issue,
one must first \textit{parallel transport}
the new color variable $\vth'^i$
to the original point $x$
by employing a connection.
In particular, the parallel transportation with respect to the gauge connection $A \inn \Omega^1(\mflat;\g)$ gives
\begin{align}
	\label{propervar.vth}
	{\d}\vth^i
	\,=\,
		\BB{
			\delta^i{}_j
			+ A^i{}_{jr}(x)\mem {\d}x^r
		}\mem
		\vth'^j
		- \vth^i
	\,=\,
		\vth'^i - \vth^i + A^i{}_{jr}(x)\mem \vth^j\, {\d}x^r
		+ \O(\d^2)
	\,,
\end{align}
where we have set $x'^m = x^m + {\d}x^m$ for infinitesimal ${\d}x^m$.
The combination
$\delta^i{}_j + A^i{}_{jr}(x)\mem {\d}x^r$ in \eqref{propervar.vth}
describes an infinitesimal Wilson line
from $x' = x + {\d}x$ to $x$.

With this understanding,
the physically sensible variation is
\begin{align}
\begin{split}
	\label{variation.YM}
	x'^m
	\,=\,
		x^m + {\d}x^m
	&\,,\quad
	\vth'^i
	\,=\,
		\vth^i + {\d}\vth^i - A^i{}_{jr}(x)\mem \vth^j\, {\d}x^r
	\,,\\
	p'_m
	\,=\,
		p_m + {\d}p_m
	&\,,\quad
	\bvth'_i
	\,=\,
		\bvth_i + {\d}\bvth_i + \bvth_j\mem A^j{}_{ir}(x)\, {\d}x^r
	\,.
\end{split}
\end{align}
\eqref{variation.YM} describes a vector
\begin{align}
	{\d}x^m
		\,\bb{
			\frac{\partial}{\partial x^m}
			- A^i{}_{jr}(x)\mem \vth^j
			\,\frac{\partial}{\partial \vth^i}
			+ \bvth_j\mem A^j{}_{ir}(x)
			\,\frac{\partial}{\partial \bvth_i}
		}
	+ {\d}p_m
		\,\frac{\partial}{\partial p_m}
	+ \overleftarrow{\frac{\partial}{\partial \vth^i}}
		{\d}\vth^i
	+ {\d}\bvth_i
		\,\overrightarrow{\frac{\partial}{\partial \bvth_i}}
	\,,
\end{align}
which is precisely $\dX^A\mem \E_A$
with $\E_A$ in \eqref{YM.EF}
and $\dX^A = ({\d}x^m,{\d}p_m,{\d}\vth^i,{\d}\bvth_i)$.

\para{Covariant Equations of Motion}

By plugging in \eqref{variation.YM} to \eqref{action.YM.prime},
extracting the part linear in the fluctuation fields,
and using the definition of the non-abelian field strength,
one finds that
the first-order variation is
\begin{align}
	\label{action.YM.v1}
		\int d\t\,\,
		\lrp{
		\begin{aligned}[c]
		&
			{\d}p_m\mem \dot{x}^m
			+ p_m\mem {\d}\dot{x}^m
			+ i\mem {\d}\bvth_i\mem \frac{D\vth^i}{d\t}
			+ i\mem \bvth_i\mem \frac{D{\d}\vth^i}{d\t}
			+ q_a F^a{}_{mn}(x)\mem {\d}x^m\hem \dot{x}^n
		\\
		&
			- {\d}x^m\mem \tpartial_m \ham(\t)
			- {\d}p_m\mem \frac{\partial\ham(\t)}{\partial p_m}
			- \frac{\overleftarrow{\partial}\ham(\t)}{\partial \vth^i}\mem {\d}\vth^i
			- {\d}\bvth_i\mem \frac{\overrightarrow{\partial}\ham(\t)}{\partial \bvth_i}
		\end{aligned}
		}
	\,.
\end{align}
Therefore, the classical equations of motion are
\begin{align}
\begin{split}
	\dot{x}^m
	\,=\,
		\frac{\partial\ham(\t)}{\partial p_m}
	&\,,\quad
	\dot{p}_m
	\,=\,
		- \frac{\partial\ham(\t)}{\partial x^m}
		+ q_a F^a{}_{mn}(x)\mem \dot{x}^n
	\,,\\
	\frac{D\vth^i}{d\t}
	\,=\,
		-i\mem \frac{\overrightarrow{\partial}\ham(\t)}{\partial \bvth_i}
	&\,,\quad
	\frac{D\bvth_i}{d\t}
	\,=\,
		i\mem \frac{\overleftarrow{\partial}\ham(\t)}{\partial \vth^i}
	\,.
\end{split}
\end{align}
These reproduce the covariant equations of motion in \eqrefs{YM.ceom}{YM.fermion}.

\para{Covariant Path Integral}

Similarly, it is left as an exercise to check that
the second-order covariant variation
the topological action
around the constant point
is given by
\begin{align}
	\label{action.YM.v2}
		\int d\t\,\,
		\bb{
			\frac{1}{2}\mem\BB{
				{\d}p_m\mem {\d}\dot{x}^m
				- {\d}\dot{p}_m\mem {\d}x^m
			}
			+ \frac{i}{2}\mem\BB{
				{\d}\bvth_i\mem {\d}\dot{\vth}^i
				- {\d}\dot{\bvth}_i\mem {\d}\vth^i
			}
			+ q_a F^a{}_{mn}(x)\mem {\d}x^m\hem {\d}\dot{x}^n
		}
	\,,
\end{align}
\newpage\noindent
where $D{\d}\vth^i \nem/d\t = {\d}\dot{\vth}^i$
via $\dot{x}^r = 0$.
The bare propagators due to \eqref{action.YM.v2} are
\begin{align}
\begin{split}
	\label{YM.prop}
	\expval{
		{\d}x^m(\t_1)\,
		{\d}p_n(\t_2)
	}
	\,&=\,
		\frac{i\hbar}{2}\:
		\delta^m{}_n\, \Theta(\t_1,\t_2)
	\,,\\
	\expval{
		{\d}\vth^i(\t_1)\,
		{\d}\bvth_j(\t_2)
	}
	\,&=\,
		\frac{\hbar}{2}\:
		\delta^i{}_j\, \Theta(\t_1,\t_2)
	\,,\\
	\expval{
		{\d}p_m(\t_1)\,
		{\d}p_n(\t_2)
	}
	\,&=\,
		\frac{i\hbar}{2}\:
		q_a F^a{}_{mn}(x)\, \Theta(\t_1,\t_2)
	\,.
\end{split}
\end{align}
Certainly, \eqref{YM.prop} reproduces the covariant Poisson brackets in \eqref{YM.cpb}.

\subsection{Gravity}

\para{Covariant Variation}

The first-order action of the general-relativistic scalar particle is
\begin{align}
	\label{action.BI}
		\int d\t\,\,
		\bb{
			p_m\mem e^m{}_\m(x)\mem \dot{x}^\m
			- \ham(x,p,\t)
		}
	\,.
\end{align}
This describes a worldline in a curved spacetime,
attached with
covectors $\t \mapsto p_m(\t)$.

Suppose a variation of this worldline:
\begin{align}
	\label{action.BI.prime}
		\int d\t\,\,
		\bb{
			p'_m\mem e^m{}_\m(x')\mem \dot{x}'^\m
			- \ham(x',p',\t)
		}
	\,.
\end{align}
The covector variable $p'_m$ now dwells in the cotangent space at $x'$.
Obviously, it is nonsensical to define
${\d}p_m =$ ``\mem$p'_m {\mem-\,} p_m$'',
in which case one is
subtracting
an element of $T^*_x\M$
from an element of $T^*_{x'}\M$.

To this end, one needs an isomorphism between 
$T^*_x$ and $T^*_{x'}$,
which can be provided by
any connection on $\Tstar\M$
in terms of a Wilson line.
A generic connection $\tgamma$
facilitates
a legitimate definition of the variation ${\d}p_m$ as
\begin{align}
	{\d}p_m
	\,=\,
		p'_n\mem
		\BB{
			\delta^n{}_m
			- \tgamma^n{}_{mr}(x)\mem {\d}x^r
		}\mem
		- p_m
	\,=\,
		p'_m - p_m - p_n\mem \tgamma^n{}_{mr}(x)\mem {\d}x^r
		+ \O(\d^2)
	\,,
\end{align}
where we have set
$x'^\m = x^\m + E^\m{}_m(x)\mem {\d}x^m$
for an infinitesimal ${\d}x^m$.
In sum, a sensible variation is
\begin{align}
\begin{split}
	\label{variation.BI}
	x'^\m
	\,=\,
		x^\m + E^\m{}_m(x)\mem {\d}x^m
	\,,\quad
	p'_m
	\,=\,
		p_m + {\d}p_m + p_n\mem \tgamma^n{}_{mr}(x)\mem {\d}x^r
	\,.
\end{split}
\end{align}
\eqref{variation.BI} describes a vector
\begin{align}
	{\d}x^r
		\,\bb{
			\frac{\partial}{\partial x^r}
			+ p_n\mem \tgamma^n{}_{mr}(x)
			\,\frac{\partial}{\partial p_m}
		}
	+ {\d}p_m
		\,\frac{\partial}{\partial p_m}
	\,,
\end{align}
which is precisely $\dX^A\mem \E_A$
with $\E_A$ in \eqref{BI.EF}
and $\dX^A = ({\d}x^m,{\d}p_m)$.

\para{Covariant Equations of Motion}

By plugging in \eqref{variation.BI} to \eqref{action.BI.prime},
extracting the part linear in the fluctuation fields,
and using the definition of the torsion two-form,
one finds that
the first-order variation is
\begin{align}
	\label{action.BI.v1}
		\int d\t\,\,
		\lrp{
		\begin{aligned}[c]
		&
			{\d}p_m\mem e^m{}_\m(x)\mem \dot{x}^\m
			+ p_m\mem \frac{\tD {\d}x^m}{d\t}
			+ p_k\hem \tT^k{}_{mn}(x)\mem {\d}x^m\hem \dot{x}^n
		\\
		&
			- {\d}x^m\mem \tE_m\act{\ham(\t)}
			- {\d}p_m\mem \frac{\partial\ham(\t)}{\partial p_m}
		\end{aligned}
		}
	\,.
\end{align}
\newpage\noindent
Note that ${\d}x^m$ is a vector variable:
$\tD {\d}x^m \nem/ d\t = {\d}\dot{x}^m \mplus \tgamma^m{}_{n\r}(x)\mem {\d}x^n\mem \dot{x}^\r$.
The classical equations of motion are
\begin{align}
\begin{split}
	e^m{}_\m(x)\mem \dot{x}^\m
	\,=\,
		\frac{\partial\ham(\t)}{\partial p_m}
	&\,,\quad
	\frac{\tD p_m}{d\t}
	\,=\,
		- \tE_m\act{ \ham(\t) }
		+ p_k\mem \tT^k{}_{mn}(x)\mem \dot{x}^n
	\,.
\end{split}
\end{align}
These reproduce the covariant equations of motion in \eqref{BI.ceom}.

Note that the metric-preserving condition on the auxiliary condition $\tgamma$
is a choice that simplifies calculations for Hamiltonians that involve
the metric, such as $\eta^{-1}{}^{mn}\hem p_m\hhem p_n$.
Also, the choice relevant for general relativity would be 
the Levi-Civita connection $\tgamma = \gamma$,
in which case the torsion term in \eqref{action.BI.v1} does not occur.
Lastly, it should be also clear that
the covariant variations are also directly viable with spacetime indices:
$x'^\m = x^\m + {\d}x^\m$,
$p'_\m = p_\m + {\d}p_\m + p_\n\mem \bar{\Gamma}^\n{}_{\m\r}(x)\mem {\d}x^\r$.

\para{Covariant Path Integral}

Similarly, it is left as an exercise to check that
the second-order covariant variation
the topological action
around the constant-point saddle
is
\begin{align}
	\label{action.BI.v2}
		\int d\t\,\,
		\bb{
			\frac{1}{2}\mem\BB{
				{\d}p_m\mem {\d}\dot{x}^m
				- {\d}\dot{p}_m\mem {\d}x^m
			}
			+ p_k\mem \tT^k{}_{mn}(x)\mem {\d}x^m\hem {\d}\dot{x}^n
		}
	\,,
\end{align}
where $\tD {\d}x^m \nem/d\t = {\d}\dot{x}^m$ via $\dot{x}^\r = 0$.
Hence, the bare propagators in the topological theory are
\begin{align}
\begin{split}
	\label{BI.prop}
	\expval{
		{\d}x^m(\t_1)\,
		{\d}p_n(\t_2)
	}
	\,&=\,
		\frac{i\hbar}{2}\:
		\delta^m{}_n\, \Theta(\t_1,\t_2)
	\,,\\
	\expval{
		{\d}p_m(\t_1)\,
		{\d}p_n(\t_2)
	}
	\,&=\,
		\frac{i\hbar}{2}\:
		p_k\mem \tT^k{}_{mn}(x)\, \Theta(\t_1,\t_2)
	\,.
\end{split}
\end{align}
Certainly, \eqref{BI.prop} reproduces the covariant Poisson brackets in \eqref{BI.cpb}.

\subsection{Gravity with Spin}

\para{Covariant Variation}

The first-order action of the general-relativistic spinning particle is
\begin{align}
	\label{action.GR}
		\int d\t\,\,
		\bb{
			p_m\mem e^m{}_\m(x)\mem \dot{x}^\m
			+ i\mem \bpsi_m\mem 
				\BB{
					\dot{\psi}^m
					+ \gamma^m{}_{n\r}(x)\mem \psi^n
						\,\dot{x}^\r
				}
			- \ham(x,p,\psi,\bpsi,\t)
		}
	\,.
\end{align}
This describes a worldline in curved spacetime,
attached with covectors
$\t \mapsto p_m(\t)$
and fermionic variables $\t \mapsto \psi^m(\t)$, $\t \mapsto \bpsi_m(\t)$.

Now suppose a variation of this worldline:
\begin{align}
	\label{action.GR.prime}
		\int d\t\,\,
		\bb{
			p'_m\mem e^m{}_\m(x')\mem \dot{x}'^\m
			+ i\mem \bpsi'_m\mem 
				\BB{
					\dot{\psi}'^m
					+ \gamma^m{}_{n\r}(x')\mem \psi'^n
						\,\dot{x}'^\r
				}
			- \ham(x,p,\psi,\bpsi,\t)
		}
	\,.
\end{align}
Evidently, it is a mathematical nonsense
to define 
${\d}\psi^m =$ ``\mem$\psi'^m {\mem-\,} \psi^m$'',
in which case one is
advocating to subtract
an element of $\Pi T^\C_x\M$
from an element of $\Pi T^\C_{x'}\M$.

To this end, one can employ any connection on $T^\C\hnem\M$.
For simplicity, we take the (complexified) Levi-Civita connection.
For $x'^\m = x^\m + E^\m{}_m(x)\mem {\d}x^m$
with an infinitesimal ${\d}x^m$,
the Wilson line from $x\mplus{\d}x$ to $x$
describes
\begin{align}
	{\d}\psi^m
	\,=\,
		\BB{
			\delta^m{}_n
			+ \gamma^m{}_{nr}(x)\mem {\d}x^r
		}\mem
		\psi'^n
		- \psi^m
	\,=\,
		\psi'^m - \psi^m + \gamma^m{}_{nr}(x)\mem \psi^n\, {\d}x^r
		+ \O(\d^2)
	\,.
\end{align}
As a result, the physically sensible variation for the general-relativistic spinning particle is
\begin{align}
\begin{split}
	\label{variation.GR}
	x'^\m
	\,=\,
		x^\m + E^\m{}_m(x)\mem {\d}x^m
	&\,,\quad
	\psi'^m
	\,=\,
		\psi^m + {\d}\psi^m - \gamma^m{}_{nr}(x)\mem \psi^n\, {\d}x^r
	\,,\\
	p'_m
	\,=\,
		p_m + {\d}p_m
	&\,,\quad
	\bpsi'_m
	\,=\,
		\bpsi_m + {\d}\bpsi_m + \bpsi_n\mem \gamma^n{}_{mr}(x)\, {\d}x^r
	\,.
\end{split}
\end{align}
\eqref{variation.GR} describes a vector
\begin{align}
\begin{split}
	&
	{\d}x^r
		\,\bb{
			E^\r{}_r(x)\mem
				\frac{\partial}{\partial x^\r\vphantom{\bpsi^0}}
			- \gamma^m{}_{nr}(x)\mem \psi^n
			\,\frac{\partial}{\partial \psi^m\vphantom{\bpsi^0}}
			+ \bpsi_n\mem \gamma^n{}_{mr}(x)
			\,\frac{\partial}{\partial \bpsi_m\vphantom{\bpsi^0}}
		}
	\\
	&
	+ {\d}p_m
		\,\frac{\partial}{\partial p_m}
	+ \overleftarrow{\frac{\partial}{\partial \psi^m\vphantom{\bpsi^0}}}
		{\d}\psi^m
	+ {\d}\bpsi_m
		\,\overrightarrow{\frac{\partial}{\partial \bpsi_m\vphantom{\bpsi^0}}}
	\,,
\end{split}
\end{align}
which is precisely $\dX^A\mem \E_A$
with $\E_A$ in \eqref{YM.EF}
and $\dX^A = ({\d}x^m,{\d}p_m,{\d}\psi^m,{\d}\bpsi_m)$.

\para{Covariant Equations of Motion}

By plugging in \eqref{variation.GR} to \eqref{action.GR.prime},
extracting the part linear in the fluctuation fields,
and using the definition of the non-abelian field strength,
one finds that
the first-order variation is
(see also \rcite{Gibbons:1993ap})
\begin{align}
	\label{action.GR.v1}
		\int d\t\,\,
		\lrp{
		\begin{aligned}[c]
		&
			{\d}p_m\mem e^m{}_\m(x)\mem \dot{x}^\m
			+ p_m\mem \frac{D{\d}x^m}{d\t}
			+ i\mem {\d}\bpsi_m\mem \frac{D\psi^m}{d\t}
			+ i\mem \bpsi_m\mem \frac{D{\d}\psi^m}{d\t}
		\\
		&
			+ i\mem \bpsi_k\hem R^k{}_{lmn}(x)\mem \psi^l
				\, {\d}x^m\hem e^n{}_\n(x)\mem \dot{x}^\n
		\\
		&
			- {\d}x^m\mem \tE_m\act{ \ham(\t) }
			- {\d}p_m\mem \frac{\partial\ham(\t)}{\partial p_m}
			- \frac{\overleftarrow{\partial}\ham(\t)}{\partial \vth^i}\mem {\d}\vth^i
			- {\d}\bvth_i\mem \frac{\overrightarrow{\partial}\ham(\t)}{\partial \bvth_i}
		\end{aligned}
		}
	\,.
\end{align}
Therefore, the classical equations of motion are
\begin{align}
\begin{split}
	e^m{}_\m(x)\mem \dot{x}^\m
	\,=\,
		\frac{\partial\ham(\t)}{\partial p_m}
	&\,,\quad
	\frac{Dp_m}{d\t}
	\,=\,
		- \tE_m\act{ \ham(\t) }
		+ i\mem \bpsi_k\hem R^k{}_{lmn}(x)\mem \psi^l\mem e^n{}_\n(x)\mem \dot{x}^\n
	\,,\\
	\frac{D\psi^m}{d\t}
	\,=\,
		-i\mem \frac{\overrightarrow{\partial}\ham(\t)}{\partial \bpsi_m\vphantom{\bpsi^0}}
	&\,,\quad
	\frac{D\bpsi_m}{d\t}
	\,=\,
		i\mem \frac{\overleftarrow{\partial}\ham(\t)}{\partial \psi^m\vphantom{\bpsi^0}}
	\,.
\end{split}
\end{align}
These reproduce the covariant equations of motion in \eqref{GR.ceom.simp}.

\para{Covariant Path Integral}

Similarly, it is left as an exercise to check that
the second-order covariant variation
the topological action
around the constant-point background
is
\begin{align}
	\label{action.GR.v2}
		\int d\t\,\,
		\lrp{\,
		\begin{aligned}[c]
		&
			\tfrac{1}{2}\mem\BB{
				{\d}p_m\mem {\d}\dot{x}^m
				- {\d}\dot{p}_m\mem {\d}x^m
			}
			+ \tfrac{i}{2}\mem\BB{
				{\d}\bpsi_m\mem {\d}\dot{\psi}^m
				- {\d}\dot{\bar\psi}_m\mem {\d}\psi^m
			}
		\\
		&
			+ i\mem \bpsi_k\hem R^k{}_{lmn}(x)\mem \psi^l\,
				{\d}x^m\hhem {\d}\dot{x}^n
		\end{aligned}
		\,}
	\,.
\end{align}
Hence, the bare propagators in the topological theory are
\begin{align}
\begin{split}
	\label{GR.prop}
	\expval{
		{\d}x^m(\t_1)\,
		{\d}p_n(\t_2)
	}
	\,&=\,
		\frac{i\hbar}{2}\:
		\delta^m{}_n\, \Theta(\t_1,\t_2)
	\,,\\
	\expval{
		{\d}\psi^m(\t_1)\,
		{\d}\bpsi_n(\t_2)
	}
	\,&=\,
		\frac{\hbar}{2}\:
		\delta^m{}_n\, \Theta(\t_1,\t_2)
	\,,\\
	\expval{
		{\d}p_m(\t_1)\,
		{\d}p_n(\t_2)
	}
	\,&=\,
		\frac{i\hbar}{2}\:
		i\mem \bpsi_k\hem R^k{}_{lmn}(x)\mem \psi^l\, \Theta(\t_1,\t_2)
	\,.
\end{split}
\end{align}
Certainly, \eqref{GR.prop} reproduces the covariant Poisson brackets in \eqref{GR.cpb}.

\newpage
\section{Summary}

This paper establishes
a manifestly gauge-covariant formulation of Hamilton mechanics for classical particles
at the expense of manifest symplecticity.
\vspace{1.5pt}
\begin{enumerate}[label=\arabic*.,itemsep=-0.45pt]
	\item
		Symplecticity is a principle in classical mechanics
		that encodes the conservation of classical probability
		while ensuring consistency of Poisson bracket.
		It is trivialized in canonical coordinates
		and is obscured for the sake of manifest gauge covariance.
	\item
		The obstruction to canonical coordinates is formalized as
		symplectic perturbations.
	\item
		For non-abelian interactions,
		the non-coordinate nature of covariant differentials
		necessitates the framework of
		covariant symplectic perturbations.
	\item
		A proper covariant symplectic perturbation is
		a deformation of an almost-symplectic structure $\omega^\bullet$
		to a symplectic structure $\omega = \omega^\bullet + \omega'$.
		Physically, $\omega^\bullet$ is the covariantization of a free-theory symplectic structure $\omega^\circ$
		while $\omega'$ encodes curvature/tidal effects.
		This precisely implements the idea of Einstein's equivalence principle
		in symplectic geometry,
		for both gauge theory and gravity.
	\item
		The covariantizer $\varrho$ is a mathematical formalization
		of covariantization.
		It was used as an intermediate, temporary construct
		from \Sec{NONCAN>SPT} to \Sec{NONCOORD>GR}
		but then was deprecated,
		together with the tentative definition envisioned in \eqref{covpb}.
	\item
		A classical particle's phase space
		is typically a fiber bundle over spacetime.
		Its covariant symplectic geometry
		is defined by
			a free-theory structure $\omega^\circ_{AB}$,
			an Ehresmann frame $\E_A$,
		and 
			a covariant symplectic perturbation
			$\omega = \omega^\bullet + \omega'$.
		The almost-symplectic form $\omega^\bullet = \frac{1}{2}\, \omega^\circ_{AB}\, \bme^A \swedge \bme^B$
		arises by the dual $\bme^A$ of the Ehresmann frame.
		The Ehresmann frame is specified by an Ehresmann connection on the phase space.
		The Ehresmann frame is gauge covariant;
		$\omega^\bullet$ and $\omega'$ are separately gauge invariant.
	\item
		The covariant Poisson brackets
		are the components of the Poisson bivector $\omega^{-1}$
		in the Ehresmann frame:
		$(\omega^{-1})^{AB} \eqq \omega^{-1}(\bme^A,\bme^B)$.
		They admit a geometric series expansion
		due to 
		$\omega^{-1} = \omega^\bullet{}^{-1} - \omega^\bullet{}^{-1} \crc \omega' \crc \omega^\bullet{}^{-1} + \cdots$
		and the ``elevator identity''
		$\omega^\bullet{}^{-1}(\bme^A,\bme^B) = (\omega^\circ{}^{-1})^{AB}$.
		Physically, this characterizes
		the curvature/tidal deformation of the Poisson structure
		in the manifestly covariant fashion.
	\item
		The covariant Hamiltonian equations of motion
		evaluate the components of the time-evolution vector field
		in the Ehresmann frame:
		$\i_{d/d\t}\hem \bme^A = \langle \bme^A , D\ham(\t) \rangle$.
		They are directly derived from the variational principle
		if the variations are properly defined.
	\item
		In background perturbation theory for the topological phase space action,
		the covariant Poisson brackets
		encode tree-level correlation functions 
		between worldline fluctuation fields
		in the covariant field basis.
	\item
		The symplectic vielbein formalism
		is a compromise between symplecticity and gauge covariance,
		which describes locally canonical frames such that
		$\omega = \frac{1}{2}\, \omega^\circ_{AB}\, \bme^A \swedge \bme^B$.
		It can be associated with
		a weaker version of Darboux theorem.
	\item
		(Covariant) symplectic perturbations
		provide a unified formulation of gauge and gravitational interactions,
		pointing out several interesting correspondences that are sometimes exact
		in the context of color-kinematics duality.
\end{enumerate}

\appendix
\section{Equivalent Statements of Symplecticity}
\label{GUISES}

In this appendix,
we familiarize ourselves with the notion of symplecticity
by reviewing its equivalent mathematical definitions
and their physical significance.

\para{Basic Definitions, Notations, and Conventions}

On a smooth manifold $\CX$,
$\Omega^k(\CX)$ denotes the space of smooth differential $k$-forms,
while
$Z^k(\CX)$ denotes the space of closed smooth differential $k$-forms.
The pairing between a $p$-form $\a \inn \Omega^p(\CX)$ and a $p$-vector $v \inn \Gamma(\wedge^pT\CX)$
is $\langle\a,v\rangle$.
The action of a vector field $v \inn \Gamma(T\CX)$
on a scalar field $f \inn \Cinfty(\CX)$
is denoted as $v\act{f} = \langle df , v \rangle = \i_v\hem df$.
The interior product with respect to a vector field $v$
is denoted as $\i_v$,
while
the Lie derivative is denoted as $\pounds_v$.
The Cartan magic formula
reads $\pounds_v = d\mem \i_v + \i_v\hem d$.
The Schouten-Nijenhuis bracket \cite{schouten1940ueber,schouten1954differential,nijenhuis1955jacobi}
between two multivector fields
$v \inn \Gamma(\wedge^pT\CX)$
and
$w \inn \Gamma(\wedge^qT\CX)$
is denoted as $[v,w]$.
For $p \eqq q \eqq 1$, this is the usual Lie bracket between vector fields.

We record some useful identities.
For any $\a \inn \Omega^2(\CX)$ and $u,v,w \inn \Gamma(T\CX)$,
it holds that
\begin{align}
\begin{split}
\label{3id}
    d\a(u,v,w)
    \,=\,
    \begin{aligned}[t]
    &
          u\act{ \a(v,w) }
        + v\act{ \a(w,u) }
        + w\act{ \a(u,v) }
    \\
    &
        - \a(\comm{u}{v},w)
        - \a(\comm{v}{w},u)
        - \a(\comm{w}{u},v)
    \,.
    \end{aligned}
\end{split}
\end{align}
Also,
for any $u_1,u_2,v_1,v_2 \inn \Gamma(T\CX)$,
the Schouten-Nijenhuis bracket between
$(u_1 \swedge u_2) \in \Gamma(\wedge^2T\CX)$
and
$(v_1 \swedge v_2) \inn \Gamma(\wedge^2T\CX)$
is defined \cite{schouten1940ueber,schouten1954differential,nijenhuis1955jacobi} as
\begin{align}
\begin{split}
	\label{SN2}
	\comm{
		u_1 \swedge u_2
	}{
		v_1 \swedge v_2
	}
	\,=\,
	\begin{aligned}[t]
		&
			\comm{u_1}{v_1} \wedge u_2 \wedge v_2
			- \comm{u_1}{v_2} \wedge u_2 \wedge v_1
		\\
		&
			- \comm{u_2}{v_1} \wedge u_1 \wedge v_2
			+ \comm{u_2}{v_2} \wedge u_1 \wedge v_1
		\,.
	\end{aligned}
\end{split}
\end{align}

We also import the basic definitions and conventions
about almost-symplectic and symplectic geometries
in \Sec{NONCAN>SPT}.
In particular, recall that
an almost-symplectic manifold
is a smooth manifold $\ps$ equipped with
a non-degenerate two-form
$\omega \inn \Omega^2(\ps)$
\cite{gelfand1997fedosov}.
The \define{Liouville measure}
of an almost-symplectic manifold $(\ps,\omega)$
is the top form
$\mu = \omega^{\wedge 2n} \hnem/ n! \in \Omega^{2n}(\ps)$,
where $2n = \dim \ps$.

\para{Statements of Symplecticity}

In an almost-symplectic manifold $(\ps,\omega)$,
the following are equivalent statements:
\enter
\hrule
\begin{enumerate}
    \item[(A\textsubscript{1})]
        $d\omega = 0$,
    \item[(A\textsubscript{2})]
        Local existence of $\theta \in \Omega^1(\ps)$
        such that
        $\omega = d\theta$,
    \item[(A\textsubscript{3})]
        Local existence of 
        coordinates $\XQcan^I$
        such that
        $\omega = \frac{1}{2}\: \omega^\circ_{IJ}\mem d\XQcan^I \swedge d\XQcan^J$
        with $\omega^\circ_{IJ}$ a constant symplectic matrix,
    \item[(B\textsubscript{1})]
        $\pounds_{\Xvec_f} \omega = 0$
        \,
        $\forall f \in \Cinfty(\ps)$,
    \item[(B\textsubscript{2})]
        $\pounds_{\Xvec_f} \mu = 0$
        \,
        $\forall f \in \Cinfty(\ps)$,
    \item[(C\textsubscript{1})]
        $\comm{\Xvec_f}{\Xvec_g} = - \Xvec_{\pb{f}{g}}$
        \,
        $\forall f,g \in \Cinfty(\ps)$,
    \item[(C\textsubscript{2})]
        $\comm{\omega^{-1}}{\omega^{-1}} = 0$,
    \item[(C\textsubscript{3})]
        $
            \Jac{f}{g}{h}
            = 0
        $
        \,
        $\forall f,g,h \in \Cinfty(\ps)$.
\end{enumerate}
\hrule
\enter
To prove this claim, we show that 
statement (A\textsubscript{1}) is necessary and sufficient for
the others.

\newpage

\hrule
\enter
\begin{itemize}[label=\adjustbox{valign=c,scale=0.7,raise=1.5pt}{$\blacktriangleright$}]
\item
Firstly,
	(A\textsubscript{2}) is by Poincar\'e lemma.
	(A\textsubscript{3}) is by 
	the celebrated Darboux theorem \cite{darboux1882probleme}.

\enter
\item
Secondly,
	(B\textsubscript{1}) is easy by the Cartan magic formula.
	Similarly, (B\textsubscript{2}) is by the Cartan magic formula
	and the non-degeneracy of $\omega$.

\enter
\item
Thirdly,
	(C\textsubscript{1}) is by
	computing $\i_{[\Xvec_f,\Xvec_g]} \omega$ as
	$
	    \pounds_{\Xvec_f} \i_{\Xvec_g} \omega
	    - \i_{\Xvec_g} \pounds_{\Xvec_f} \omega
	$.

\enter
\item
Lastly, it remains to investigate 
	(C\textsubscript{2}) and (C\textsubscript{3}).
To this end,
we perform component computations 
in local coordinates $X^I$
while denoting $\Pi := \omega^{-1}$
to avoid clutter.

\begin{itemize}
\item
	First,
	we establish that
	statements (C\textsubscript{2}) and (A\textsubscript{1}) are equivalent.
	By using the defining identity in \eqref{SN2},
	it follows that
	\begin{align}
	\label{computeJ.coord}
		&
		\comm{\Pi}{\Pi}
		\\
	\nonumber
		\,&=\,
		\frac{1}{4}\lrp{\,
		\begin{aligned}[c]
			&
				\comm{\partial_{I_1}}{\partial_{J_1}} \wedge \Pi^{I_1I_2}\mem \partial_{I_2} \wedge \Pi^{J_1J_2}\mem \partial_{J_2}
				- \comm{\partial_{I_1}}{\Pi^{J_1J_2}\mem \partial_{J_2}} \wedge \Pi^{I_1I_2}\mem \partial_{I_2} \wedge \partial_{J_1}
			\\&
				- \comm{\Pi^{I_1I_2}\mem \partial_{I_2}}{\partial_{J_1}} \wedge \partial_{I_1} \wedge \Pi^{J_1J_2}\mem \partial_{J_2}
				+ \comm{\Pi^{I_1I_2}\mem \partial_{I_2}}{\Pi^{J_1J_2}\mem \partial_{J_2}} \wedge \partial_{I_1} \wedge \partial_{J_1}
		\end{aligned}
		\,}
		\,,\\
	\nonumber
		\,&=\,
		\frac{1}{4}\lrp{\,
		\begin{aligned}[c]
			&
				0
				- \partial_{I_1}\Pi^{J_1J_2}\mem \partial_{J_2} \wedge \Pi^{I_1I_2}\mem \partial_{I_2} \wedge \partial_{J_1}
				+ \partial_{J_1}\Pi^{I_1I_2}\mem \partial_{I_2} \wedge \partial_{I_1} \wedge \Pi^{J_1J_2}\mem \partial_{J_2}
			\\&
			+ \BB{
				{\Pi^{I_1I_2}\mem \partial_{I_2}}{\Pi^{J_1J_2}\mem \partial_{J_2}} 
					\wedge \partial_{I_1} \wedge \partial_{J_1}
				- {\Pi^{J_1J_2}\mem \partial_{J_2}}{\Pi^{I_1I_2}\mem \partial_{I_2}} 
					\wedge \partial_{I_1} \wedge \partial_{J_1}
			}
		\end{aligned}
		\,}
		\,,\\
	\nonumber
		\,&=\,
		\frac{1}{4}\lrp{\,
		\begin{aligned}[c]
			&
				- \Pi^{L_1L_2}{}_{,K}\mem \Pi^{KL_3}
				- \Pi^{L_1L_2}{}_{,K}\mem \Pi^{KL_3}
			\\&
				+ \Pi^{L_1K}\mem \Pi^{L_2L_3}{}_{,K}
				+ \Pi^{L_1K}\mem \Pi^{L_2L_3}{}_{,K}
		\end{aligned}
		\,}\,
			\partial_{L_1} \wedge \partial_{L_2} \wedge \partial_{L_3}
		\,,\\
	\nonumber
		\,&=\,
		- \Pi^{L_1L_2}{}_{,K}\mem \Pi^{KL_3}\,
			\partial_{L_1} \wedge \partial_{L_2} \wedge \partial_{L_3}
		\,.
	\end{align}
	This means that the components of the trivector $- \tfrac{1}{2}\: \comm{\Pi}{\Pi}$ are
	\begin{align}
		\label{SNB.coord}
		- \tfrac{1}{2}\:
			\comm{\Pi}{\Pi}^{I_1I_2I_3}
		\,=\,
			3\mem \Pi^{[I_1I_2|}{}_{,K}\mem \Pi^{K|I_3]}
		\,.
	\end{align}
	Clearly,
	$[\Pi,\Pi] = 0$ iff
	$\Pi^{[I_1I_2|}{}_{,K}\mem \Pi^{K|I_3]} = 0$.
	Since 
	$
		\Pi^{I_1I_2}{}_{,K}
		= -\Pi^{I_1J_1}\, \omega_{J_1J_2,K}\,\hhnem \Pi^{J_2I_2}
	$
	and $\Pi$ is non-degenerate,
	this is again equivalent to
	$\omega_{[J_1J_2,J_3]} = 0$,
	which is nothing but statement (A\textsubscript{1}).
	
\enter
\item
	Second, we establish that
	statement (C\textsubscript{2})
	holds iff 
	statement (C\textsubscript{3})
	holds.
	By using \eqref{SNB.coord} and $\comm{\partial_I}{\partial_J} = 0$,
	one finds
	\begin{align}
	\begin{split}
		\label{plug.coord}
		- \tfrac{1}{2}\:
			\comm{\Pi}{\Pi}(df,dg,dh)
		\,=\,
			3\mem \Pi^{[I_1I_2|}{}_{,K}\mem \Pi^{K|I_3]}
			\,
			f_{,I_1}\mem g_{,I_2}\mem h_{,I_3}
		\,=\,
			\Jac{f}{g}{h}
		\,.
	\end{split}
	\end{align}
\end{itemize}

	In conclusion,
	(A\textsubscript{1}),
	(C\textsubscript{2}),
	and
	(C\textsubscript{3})
	are equivalent statements.
	
\end{itemize}
\enter
\hrule

\enter
\enter
Note that various pathways can exist
for proving the equivalence between these statements.
The following is a summary of several identities in almost-symplectic geometry:
\begin{align}
\begin{split}
	\label{Jac3}
	\Jac{f}{g}{h}
	\,&=\,
		d\omega(\Xvec_f,\Xvec_g,\Xvec_h)
	\,,\\
	\,&=\,
		-\tfrac{1}{2}\:
			\comm{\omega^{-1}}{\omega^{-1}}(df,dg,dh)
	\,,\\
	\,&=\,
       	- \BB{
       		\comm{\Xvec_f}{\Xvec_g} + \Xvec_{\pb{f}{g}}
       	}\act{
       		h
       	}
 \,.
\end{split}
\end{align}
Note that this paper adopts the right-action convention for Hamiltonian vector fields,
$\Xvec_f = \omega^{-1}(\blank,df)$.
The left-action convention will simplify the sign factors in \eqref{Jac3}.

\para{Physical Significance of Symplecticity}

In a general setup,
a \define{Hamiltonian system} is a pair $(\ps,\omega,\ham)$,
where $(\ps,\omega)$ is an almost-symplectic manifold
while $\ham$ represents
a one-parameter family of smooth functions
$\ham(\t) \inn \Cinfty(\ps)$
for $\t \inn \R$,
called the \define{time-evolution generator}.
A Hamiltonian system is \define{proper} iff $\omega$ is symplectic
and \define{improper} iff $\omega$ is not symplectic.

We use the term
\define{symplecticity}
for referring to any of the above eight properties 
from (A\textsubscript{1}) to (C\textsubscript{3})
in proper Hamiltonian systems.

\enter

Let us explicate the significance of each statement.
\enter
\begin{itemize}
\item
	Physically, (A\textsubscript{1}) or equivalently (A\textsubscript{2}) means
	the \textit{existence of an action principle}.
	That is, symplecticity means the \emph{theory is Lagrangian}.
	
	A symplectic potential $\theta$ facilitates a (patchwise defined) first-order action
	of the form $\int \theta - \ham(\t)\mem d\t$.
	Conversely, any Lagrangian can be converted to a phase space action via the standard procedures of Legendre transform,
	from which a symplectic potential $\theta$ is read off
	to define a symplectic form $\omega = d\theta$
	so that $d\omega = dd\theta = 0$.

\enter
\item
	(A\textsubscript{3}) means that
	symplecticity is made manifest in \emph{canonical coordinates}.
	The closure 
	$
		d\bigbig{
			\frac{1}{2}\: \omega^\circ_{IJ}\mem d\XQcan^I \swedge d\XQcan^J
		} = 0
	$
	trivially holds,
	as $d\XQcan^I$ are coordinate (exact) differentials.
	Said in another way,
	the role of canonical coordinates
	is to trivialize the components of the symplectic form
	to constants: $\pm1$ or $0$.
	In the same fashion, the canonical Poisson brackets describe $\pm1$ or $0$
	so that the Jacobi identity becomes trivial.

	It is a widespread customary practice in
	standard textbooks
	to exclusively employ canonical coordinates
	and restrict the coordinate transformations on phase spaces
	to canonical transformations.
	This is usually implemented without detailed justifications.
	However, we point out that
	the rationale could be manifest symplecticity.

\enter
\item
	(B\textsubscript{2}) is referred to as the Liouville theorem.
	Physically, this encodes 
	the \emph{conservation} \emph{of classical probability}
	for ensembles of proper Hamiltonian systems.
	In classical statistical mechanics,
	classical probability distributions
	are defined 
	on the phase space
	with respect to the Liouville measure $\mu$.
	The fact that time evolution preserves the Liouville measure
	is vital for the consistency of the entire discipline.
	
	In this light,
	one may regard symplecticity as the classical avatar of unitarity in quantum mechanics.
	There are several senses in which this argument can be viewed precise.
	For example,
	take the two-sphere $S^2$ as a compact phase space,
	in which case the quantum theory describes the finite-dimensional Hilbert space $\mathcal{H}$ of a spin-$j$ system
	of dimension $\dim \mathcal{H} = \int_\ps\mem (\omega/2\pi\hbar) = 2j+1$
	\cite{Varilly:1989sv}.
	Hence the preservation of $\omega$ due to
	statement (B\textsubscript{1})
	translates to the preservation of Hilbert space dimension,
	which expresses conservation of quantum probability.

\enter
\item
	(C\textsubscript{1}) implies that
	$f {\:\mapsto\:} {-\Xvec_f}$ is a \emph{Lie algebra homomorphism}.
	This property is important when discussing symmetries.
	Symmetry actions are represented as Hamiltonian flows (momentum map),
	and symmetry algebras are represented by the Poisson algebra.
	This homomorphism property
	is also a crucial element in the Magnus expansion approach to time evolution
	that identifies an effective Hamiltonian
	\cite{Kim:2025sey,Kim:2025ebl}.
	

	
	The quantum-mechanical origin of this homomorphism
	is the \textit{preservation of the operator algebra}
	by the adjoint action
	\cite{Kim:2025ebl}.
	Namely, the adjoint action (arising from the operator commutator)
	is a derivation on the operator algebra.
	
\enter
\item
	(C\textsubscript{3}) is the \emph{Jacobi identity} for the Poisson bracket.
	(C\textsubscript{2}) is essentially the coordinate-free statement of the Jacobi identity.
	The Jacobi identity can be interpreted as the preservation of the Poisson bracket.
	This
	ensures \emph{consistency of the time derivative},
	for instance:
	$\frac{d}{d\t}\mem \pb{f}{g} = \pb{\dot{f}}{g} + \pb{f}{\dot{g}}$.
	
	The quantum-mechanical origin of the Jacobi identity
	is the \textit{associativity of the operator algebra} \cite{kontsevich,Kim:2025ebl}.

\enter
\end{itemize}
This discussion should justify
taking symplecticity as an essential principle in classical mechanics.

\para{Symplectic Geometry in Non-Coordinate Frames}

Finally, let us revisit the eight statements of symplecticity
in non-coordinate frames.
This analysis represents how symplecticity becomes obscured
if there is a physically motivated obstruction 
to Darboux charts or coordinate frames,
e.g., manifest gauge covariance.

Recall from \Secs{NONCOORD>FRAME}{CPB>SV}
that a generic frame $\E$ and its dual $\bme$ on an almost-symplectic manifold $(\ps,\omega)$
describe the formulae
\begin{align}
	\label{mobile}
	\omega
	\,=\,
		\tfrac{1}{2}\:
			\omega_{AB}(X)\,
				\bme^A \wedge \bme^B
	\,,\quad
	\omega^{-1}
	\,=\,
		\tfrac{1}{2}\:
			(\omega^{-1}\hnem(X))^{AB}\,
				\E_A \wedge \E_B
	\,.
\end{align}
Similarly,
a symplectic vielbein $\EE$
and its dual $\ee$ on
$(\ps,\omega)$
describe the formulae
\begin{align}
	\label{mobile0}
	\omega
	\,=\,
		\tfrac{1}{2}\:
			\omega^\circ_{AB}\,
				\ee^A \wedge \ee^B
	\,=\,
		\tfrac{1}{2}\:
			\ee^A \wedge \ee_A
	\,,\quad
	\omega^{-1}
	\,=\,
		\tfrac{1}{2}\:
			(\omega^\circ{}^{-1})^{AB}\,
				\EE_A \wedge \EE_B
	\,=\,
		\tfrac{1}{2}\:
			\EE_A \wedge \EE^A
	\,.
\end{align}
In the symplectic vielbein formalism, we raise and lower frame indices
with the constant matrices $\omega^\circ_{AB}$ and $(\omega^\circ{}^{-1})^{AB}$
as in \eqref{raise-lower}.
The anholonomy coefficients are defined as
$\Om^A{}_{BC} = \langle \bme^A , \comm{\E_B}{\E_C} \rangle$
and
$\OOm^A{}_{BC} = \langle \ee^A , \comm{\EE_B}{\EE_C} \rangle$.

\enter
\enter
\hrule
\enter
\begin{itemize}[label=\adjustbox{valign=c,scale=0.7,raise=1.5pt}{$\blacktriangleright$}]
\item
	(A\textsubscript{3})
	can be approached in terms of the symplectic vielbein.
	The formulae in \eqref{mobile0} represent the \textit{obstruction to canonical coordinates},
	directly generalizing (A\textsubscript{3}).
	In non-canonical charts,
	the constant components $\omega^\circ_{AB}$ or $(\omega^\circ{}^{-1})^{AB}$
	cannot be achieved with the coordinate bases.
	Thus one retreats to the \textit{locally Darboux} realization of the symplectic structure in \eqref{mobile0}.
	
	Due to the anholonomy of the symplectic vielbein,
	this Darboux-esque realization 
	does not trivialize the closure statement in (A\textsubscript{1}):
	\begin{align}
		\label{domega.sv}
		d\omega
		\,=\,
			d\ee^A \wedge \ee_A
		\,=\,
			\tfrac{1}{2}\:
				\OOm_{ABC}\, \ee^A \wedge \ee^B \wedge \ee^C
		\,.
	\end{align}
	Compare this computation with 
	$
		d\bigbig{
			\frac{1}{2}\: \omega^\circ_{IJ}\mem d\XQcan^I \swedge d\XQcan^J
		} = 0
	$.
	We conclude that
	the statement of symplecticity in the symplectic vielbein formalism is
	\begin{align}
		\label{asw}
		\OOm_{[ABC]} = 0
		\,.
	\end{align}
	Regarding (A\textsubscript{2}),
	we remark that the straightforward construction of a symplectic potential 
	$
		\frac{1}{2}\: \omega^\circ_{IJ}\mem d\XQcan^I \swedge d\XQcan^J
		= d\bigbig{
			\frac{1}{2}\: \omega^\circ_{IJ}\mem \XQcan^I d\XQcan^J
		}
	$
	is not available in the symplectic vielbein description
	since $\ee^A$ is non-coordinate.
	Namely, there is no $\XQcan^A$ such that ``\mem$\ee^A = d\XQcan^A$.''
	
	\newpage
	In generic frames,
	(A\textsubscript{1}) states that
	\begin{align}
		\label{A1.frame}
		0 \,=\,
			d\omega(\E_A,\E_B,\E_C)
		\,=\,
			3\mem \E_{[A}\act{
				\omega_{BC]}
			}
			+ 3\mem \Om^D{}_{[AB}\mem \omega_{C]D}
		\,,
	\end{align}
	which follows from using the identity in \eqref{3id}.

\enter
\item
	Since $ddf = 0$, 
	the statement of symplecticity in
	(B\textsubscript{1})
	is represented as the following:
	\begin{align}
		\Xvec_f\act{\omega_{AB}}
		\,=\,
			\Xvec_f^C\,
			\E_{C}\act{
				\omega_{AB}
			} 
		\,=\,
			-2\mem 
				\Xvec_f^C\,
				\E_{[A}\act{
					\omega_{B]C}
				}
			- 3\mem \Om^D{}_{[AB}\mem \omega_{C]D}\mem \Xvec_f^C
		\,.
	\end{align}
	Here, the Liouville property is obscured by
	anholonomy terms on the right-hand side.
	
\enter
\item
	(C\textsubscript{1}) is approached as
	\begin{align}
		\label{homomorphism.nc}
		\Xvec_f^D\,\mem \E_D\act{
			\Xvec_g^C
		}
		-
		\Xvec_g^D\,\mem \E_D\act{
			\Xvec_f^C
		}
		\,=\,
			-\Xvec^C_{\pb{f}{g}}
			- \Om^C{}_{AB}\, \Xvec_f^A\, \Xvec_g^B 
		\,,
	\end{align}
	which evaluates
	$
		\cont{
			\bme^C
		}{
			\comm{\Xvec_f}{\Xvec_g}
		}
		= 
		\cont{
			\bme^C
		}{
			-\Xvec_{\pb{f}{g}}
		}
	$.
	Here, the homomorphism property is spuriously obstructed
	by anholonomy terms on the right-hand side.

\enter
\item
For
	(C\textsubscript{2}) and (C\textsubscript{3}),
we perform component computations 
while denoting $\Pi := \omega^{-1}$
to avoid clutter.

\begin{itemize}
\item
	For (C\textsubscript{2}), we compute
	\begin{align}
	\begin{split}
	\label{computeJ.noncoord}
		\comm{\Pi}{\Pi}
		\,&=\,
		\frac{1}{4}\lrp{\,
		\begin{aligned}[c]
			&
				\comm{\E_{A_1}}{\E_{B_1}} \wedge \Pi^{A_1A_2}\mem \E_{A_2} \wedge \Pi^{B_1B_2}\mem \E_{B_2}
			\\&
				- \comm{\E_{A_1}}{\Pi^{B_1B_2}\mem \E_{B_2}} \wedge \Pi^{A_1A_2}\mem \E_{A_2} \wedge \E_{B_1}
			\\&
				- \comm{\Pi^{A_1A_2}\mem \E_{A_2}}{\E_{B_1}} \wedge \E_{A_1} \wedge \Pi^{B_1B_2}\mem \E_{B_2}
			\\&
				+ \comm{\Pi^{A_1A_2}\mem \E_{A_2}}{\Pi^{B_1B_2}\mem \E_{B_2}} \wedge \E_{A_1} \wedge \E_{B_1}
		\end{aligned}
		\,}
		\,,\\
		\,&=\,
		\frac{1}{4}\lrp{\,
		\begin{aligned}[c]
			&
				\Om^C{}_{A_1B_1}\, \E_C \wedge \Pi^{A_1A_2}\mem \E_{A_2} \wedge \Pi^{B_1B_2}\mem \E_{B_2}
			\\&
				- \E_{A_1}\act{\Pi^{B_1B_2}}\,
					\E_{B_2} \wedge \Pi^{A_1A_2}\mem \E_{A_2} \wedge \E_{B_1}
			\\&
				- \Pi^{B_1B_2}\mem \Om^C{}_{A_1B_2}\mem \E_C
					\wedge \Pi^{A_1A_2}\mem \E_{A_2} \wedge \E_{B_1}
			\\&
				+ \E_{B_1}\act{\Pi^{A_1A_2}}\mem \E_{A_2} \wedge \E_{A_1} \wedge \Pi^{B_1B_2}\mem \E_{B_2}
			\\&
				- \Pi^{A_1A_2}\mem \Om^C{}_{A_2B_1}\mem \E_C
				\wedge \E_{A_1} \wedge \Pi^{B_1B_2}\mem \E_{B_2}
			\\&
				+ \Pi^{A_1A_2}\mem \E_{A_2}\act{\Pi^{B_1B_2}}\, 
					\E_{B_2} \wedge \E_{A_1} \wedge \E_{B_1}
			\\&
				- \Pi^{B_1B_2}\mem \E_{B_2}\act{\Pi^{A_1A_2}}\,
					\E_{A_2} \wedge \E_{A_1} \wedge \E_{B_1}
			\\&
				+ \Pi^{A_1A_2}\mem \Pi^{B_1B_2}\mem 
					\Om^C{}_{A_2B_2}\mem \E_C \wedge \E_{A_1} \wedge \E_{B_1}
		\end{aligned}
		\,}
		\,,
	\end{split}
	\end{align}
	which shows that the components of the trivector $- \tfrac{1}{2}\: \comm{\Pi}{\Pi}$ are
	\begin{align}
		\label{SNB.noncoord}
		- \tfrac{1}{2}\:
			\comm{\Pi}{\Pi}^{A_1A_2A_3}
		\,=\,
			3\mem \BB{
				\E_D\act{\Pi^{[A_1A_2|}}\mem \Pi^{D|A_3]}
				- \Om^{[A_1|}{}_{BC}\mem \Pi^{B|A_2|}\mem \Pi^{C|A_3]}
			}
		\,.
	\end{align}
	As compared to \eqref{SNB.coord},
	an additional ``algebraic'' part arises due to anholonomy.
	In the symplectic vielbein formalism, one finds
	\begin{align}
		\label{SNB.sv}
		- \tfrac{1}{2}\:
			\comm{\Pi}{\Pi}^{A_1A_2A_3}
		\,=\,
			- 3\mem \Om^{[A_1A_2A_3]}
		\,,
	\end{align}
	reproducing the statement of symplecticity in \eqref{asw}.
	
\enter
\item
	For (C\textsubscript{3}),
	one uses
	the first line of \eqref{Jac3} and \eqref{A1.frame},
	or
	the second line of \eqref{Jac3} and \eqref{SNB.noncoord},
	to find
	\begin{align}
		\label{plug.noncoord.form}
		&\Jac{f}{g}{h}
		\\
		&=\,
			3\mem \BB{
				\E_D\act{\Pi^{[AB|}}\mem \Pi^{D|C]}
				- \Om^{[A|}{}_{EF}\mem \Pi^{E|B|}\mem \Pi^{F|C]}
			}\,
			\E_{A}\act{f}\,
			\E_{B}\act{g}\,
			\E_{C}\act{h}
		\,.\nonumber
	\end{align}
	
\end{itemize}

\end{itemize}
\hrule

\enter
\para{Covariant Hamiltonian Vector Fields}

Regarding the analysis on
(B\textsubscript{1}), (B\textsubscript{2}), and (C\textsubscript{1})
given above,
our discussions in the main text suggest that
a more physically meaningful version 
will examine
a generalized variant of Hamiltonian vector fields.

Suppose an almost-symplectic manifold $(\ps,\omega)$.
For any $\a\in\Omega^1(\ps)$ and $v\in\Gamma(T\ps)$,
we denote
\begin{align}
	\a^\sharp
	\,:=\,
		\omega^{-1}(\blank,\alpha)
	\,,\quad
	v^\flat
	\,:=\,
		\omega(\blank,v)
	\,.
\end{align}
In particular, $\a^\sharp$ will be referred to as
a \define{covariant Hamiltonian vector field}
if $\a$ describes a covariant one-form on $\ps$
as a phase space.
Since the $\omega$ will be stipulated to be gauge-invariant,
$\a^\sharp$ is gauge-covariant if $\a$ is gauge-covariant and vice versa.
If $\a$ is exact, then $\a^\sharp$ is simply an ordinary Hamiltonian vector field.
For instance,
\begin{align}
	(df)^\sharp
	\,=\,
		\Xvec_f
	\qiq
		\comm{(df)^\sharp}{(dg)^\sharp}
		\,=\,
			\comm{\Xvec_f}{\Xvec_g}
		\,=\,
			-\Xvec_{\pb{f}{g}}
	\,.
\end{align}

In this context,
we consider the formalism put forward by
Abraham and Marsden
\cite{Marsden:1978FoM}.
Note first that
\begin{align}
	\label{cartan-alpha}
	\pounds_{\a^\sharp}\mem \omega
	\,=\,
		\i_{\a^\sharp}\mem d\omega - d\alpha
	\,.		
\end{align}
The Lie bracket between 
covariant Hamiltonian vector fields
$\a^\sharp,\b^\sharp \in \Gamma(T\ps)$
satisfy
\begin{align}
\begin{split}
	\comm{\a^\sharp}{\b^\sharp}^\flat
	\,=\,
		- \i_{\comm{\a^\sharp}{\b^\sharp}}\mem \omega
	\,&=\,
		\i_{\b^\sharp}\mem \pounds_{\a^\sharp}\mem \omega
		- \pounds_{\a^\sharp}\mem \i_{\b^\sharp}\mem \omega
	\,,\\
	\,&=\,
		\i_{\b^\sharp}\mem \BB{
			\i_{\a^\sharp}\mem d\omega -d\alpha
		}
		+ \pounds_{\a^\sharp}\mem \beta
	\,,\\
	\,&=\,
		\i_{\b^\sharp}\mem \i_{\a^\sharp}\mem d\omega
		- \i_{\b^\sharp}\mem d\a 
		+ \i_{\a^\sharp}\mem d\b
		- d\mem \cont{\a}{\b^\sharp}
	\,.
\end{split}
\end{align}
As a result, it holds that
\begin{align}
	\label{id-marsden}
	\comm{\a^\sharp}{\b^\sharp}
	\,=\,
		-\Xvec_{\langle\a,\b^\sharp\rangle}
		+ \BB{
			\i_{\b^\sharp}\mem \i_{\a^\sharp}\mem d\omega
			+ \i_{\a^\sharp}\mem d\b
	  		- \i_{\b^\sharp}\mem d\a 
		}^{\nem\sharp}
	\,.
\end{align}

\hrule
\enter
\begin{itemize}
\item
	From \eqref{cartan-alpha},
	we see that
	the Liouville theorem in (A\textsubscript{1})
	is genuinely violated
	when $d\a \neq 0$.
	In particular,
	taking $\a = \bme^A$ and $d\omega = 0$ gives
	\begin{align}
		\pounds_{(\bme^A)^\sharp}\mem \omega
		\,=\,
			- d\bme^A
		\,=\,
			\tfrac{1}{2}\:
				\Om^A{}_{BC}\,
				\bme^B \swedge \bme^C
		\,.
	\end{align}

\item
	Meanwhile, \eqref{id-marsden}
	describes that the homomorphism property in (C\textsubscript{1}) is also obstructed.
	In particular,
	taking $\a = \bme^A$, $\b = \bme^B$, and $d\omega = 0$ gives
	\begin{align}
		\comm{(\bme^A)^\sharp}{(\bme^B)^\sharp}^C
		\,=\,
			- \Xvec_{(\omega^{-1})^{AB}}^C
			- 2\mem \Om^{[A}{}_{EF}\mem 
				(\omega^{-1})^{B]E}\mem 
				(\omega^{-1})^{CF}
		\,,
	\end{align}
	where $(\omega^{-1})^{AB} = \omega^{-1}(\bme^A,\bme^B)$ will describe the covariant Poisson bracket.
	The anholonomy coefficients on the right-hand side
	arises by the non-vanishing $d\bme^A$ and $d\bme^B$.
	A slight caveat is that $\Xvec_{(\omega^{-1})^{AB}}$
	can still be non-covariant (e.g., in gravity).

\end{itemize}

\enter
\hrule

\newpage

\newpage
\noindent\textbf{Acknowledgements.}
	J.-H. K. is supported by the Department of Energy (Grant No.~DE-SC0011632) and by the Walter Burke Institute for Theoretical Physics.
\medskip

\let\c\oldc
\let\i\oldi

\bibliography{references.bib}

\end{document}